\documentclass[twocolumn,times,trackchanges,tighten]{aastex62}

\usepackage{amsmath,amstext}
\usepackage[figure,figure*]{hypcap}
\usepackage{newtxmath} 

\defcitealias{worseck16}{W16}
\defcitealias{davies17}{D17}

\def \lya  {Ly$\alpha$}

\def\mteff {\tau_\mathrm{eff}}
\def\teff {$\mteff$}
\def\mgheii {\Gamma_\mathrm{HeII}}
\def\gheii {$\mgheii$}
\def\mghi {\Gamma_\mathrm{HI}}
\def\ghi {$\mghi$}
\def\mxheii {x_{\rm HeII}}
\def\xheii {$\mxheii$}

\received{2018 August 15}
\revised{2019 February 28}
\accepted{2019 March 5}

\shorttitle{The Evolution of the \ion{He}{2}-Ionizing Background}
\shortauthors{Worseck et al.}

\begin{document}

\title{The Evolution of the \ion{He}{2}-Ionizing Background at Redshifts $2.3<z<3.8$ Inferred from a Statistical Sample of 24 \textit{HST}/COS \ion{He}{2} Ly$\alpha$ Absorption Spectra
\footnote{Based on observations made with the NASA/ESA \textit{Hubble Space Telescope}, obtained at
the Space Telescope Science Institute, which is operated by the Association of Universities
for Research in Astronomy, Inc., under NASA contract NAS5-26555. These observations are
associated with Programs 13013 and 13875. Archival \textit{Hubble Space Telescope} data 
(Programs 7575, 9350, 11528, 12033, 12178, 12249, 12816) were obtained from the Mikulski Archive for
Space Telescopes (MAST). 
}
}

\correspondingauthor{G\'abor Worseck}
\email{gworseck@uni-potsdam.de}

\author[0000-0003-0960-3580]{G\'abor Worseck}
\affiliation{Max-Planck-Institut f\"ur Astronomie, K\"onigstuhl 17, D-69117 Heidelberg, Germany}
\affiliation{Institut f\"ur Physik und Astronomie, Universit\"at Potsdam, Karl-Liebknecht-Str.\ 24/25, D-14476 Potsdam, Germany}

\author{Frederick B. Davies}
\affiliation{Max-Planck-Institut f\"ur Astronomie, K\"onigstuhl 17, D-69117 Heidelberg, Germany}
\affiliation{Department of Physics, University of California, Santa Barbara, CA 93106, USA}

\author{Joseph~F. Hennawi}
\affiliation{Max-Planck-Institut f\"ur Astronomie, K\"onigstuhl 17, D-69117 Heidelberg, Germany}
\affiliation{Department of Physics, University of California, Santa Barbara, CA 93106, USA}

\author{J.~Xavier Prochaska}
\affiliation{Department of Astronomy and Astrophysics, University of California, 1156 High Street, Santa Cruz, CA 95064, USA}
\affiliation{University of California Observatories, Lick Observatory, 1156 High Street, Santa Cruz, CA 95064, USA}

\begin{abstract}

We present measurements of the large-scale ($\approx 40$ comoving Mpc) effective optical depth of
\ion{He}{2} \lya\ absorption, \teff, at $2.54<z<3.86$ toward 16 \ion{He}{2}-transparent quasars observed with the
Cosmic Origins Spectrograph (COS) on the \textit{Hubble Space Telescope} (\textit{HST}), 
to characterize the ionization state of helium in the intergalactic medium (IGM).
We provide the first statistical sample of \teff\ measurements in six
signal-to-noise ratio $\ga 3$ \ion{He}{2} sightlines at $z>3.5$,
and study the redshift evolution and sightline-to-sightline variance of \teff\ in 24 \ion{He}{2} sightlines.
We confirm an increase of the median \teff\ from $\simeq 2$ at $z=2.7$ to $\mteff\ga 5$ at $z>3$,
and a scatter in \teff\ that increases with redshift.
The $z>3.5$ \ion{He}{2} absorption is predominantly saturated, but isolated narrow
($\Delta v<650$\,km\,s$^{-1}$) transmission spikes indicate patches of reionized helium. 
We compare our measurements to predictions for a range of UV background models applied to outputs
of a large-volume (146 comoving Mpc)$^3$ hydrodynamical simulation by forward-modeling our sample's quality and size.
At $z>2.74$ the variance in \teff\ significantly exceeds expectations for a spatially uniform
UV background, but is consistent with a fluctuating radiation field sourced by variations
in the quasar number density and the mean free path in the post-reionization IGM.
We develop a method to infer the approximate median \ion{He}{2} photoionization rate \gheii\ of a fluctuating
UV background from the median \teff, finding a factor $\simeq 5$ decrease in \gheii\ between
$z\simeq 2.6$ and $z\simeq 3.1$.
At $z\simeq 3.1$ a $\mgheii=\left[9.1^{+1.1}_{-1.2}\,\mathrm{(stat.)}\,^{+2.4}_{-3.4}\,\mathrm{(sys.)}\right]\times 10^{-16}$\,s$^{-1}$
corresponds to a median \ion{He}{2} fraction of $\simeq 2.5$\%, indicating that our data probe the tail end
of \ion{He}{2} reionization.
\end{abstract}


\section{Introduction}

The epoch of helium reionization marked the final baryonic phase transition to substantially
influence the thermal and ionization state of the IGM.
While hydrogen reionization occurred at $z\ga 6$ \citep[e.g.,][]{fan06,becker15,davies18b,planckcollab18},
the completion of \ion{He}{2} reionization was likely delayed to $z\sim 3$ when quasars became
numerous enough to supply the required hard ($E=h_\mathrm{P}\nu>54.4$\,eV) UV photons
\citep[e.g.][]{madau94,miralda-escude00,mcquinn09a,compostella13,compostella14}.
The current picture of a quasar-driven \ion{He}{2} reionization process extending over $\sim 1$\,Gyr
is supported by most measurements of the $z>3$ quasar luminosity function
\citep[e.g.,][]{hopkins07,mcgreer13,jiang16,kulkarni18}, which yield a total quasar emissivity sufficient
to complete \ion{He}{2} reionization by $z\sim 3$
\citep{madau99,miralda-escude00,wyithe03,furlanetto08,haardt12,laplante16,khaire17,puchwein19,kulkarni18}.
However, such photon budget arguments only provide rough constraints on the \ion{He}{2} reionization
history due to their simplified treatment of the gas density distribution, in particular the gradual
reionization of optically thick absorbers near the end of reionization \citep{bolton09,madau17}.

Semianalytic models and detailed cosmological radiative transfer simulations that self-consistently
include the physics governing \ion{He}{2} reionization both predict that the bulk of intergalactic
\ion{He}{2} was reionized by the emerging quasar population at $z\la 5$, and ended with the
percolation of the \ion{He}{3} zones around quasars at $z\sim 3$
\citep{fardal98,miralda-escude00,sokasian02,gleser05,furlanetto08,tittley07,furlanetto09,faucher09,mcquinn09a,furlanetto10,tittley12,compostella13,compostella14}.
However, there remains significant uncertainty in the precise timing and morphology of \ion{He}{2} reionization,
as the detailed conditions of the intergalactic gas during and after \ion{He}{2} reionization
depend on several poorly constrained parameters of the high-redshift quasar population
(e.g.\ their duty cycle, spectral energy distribution and opening angle), and the frequency
and structure of self-shielding absorbers. Typically, several generations of quasars are required
to fully reionize a given region, resulting in a rich thermal and ionization structure
of the gas \citep{compostella13,compostella14}.

Over the last two decades much theoretical and observational work has focused on the thermal state
of the IGM during and after \ion{He}{2} reionization. During \ion{He}{2} reionization supersonic quasar ionization fronts 
impulsively heat the IGM which subsequently relaxes to a tight post-reionization power-law temperature-density relation
governed by adiabatic (Hubble) cooling and photoheating by a quasi-homogeneous UV background
\citep[e.g.][]{hui97,haehnelt98b,furlanetto08b,bolton09b,mcquinn09b,becker11,compostella13,compostella14,puchwein15,mcquinn16,laplante17}.
The exact amount of injected heat ($\Delta T= 5,000$--$10,000$\,K) depends on the duration of \ion{He}{2} reionization,
the spatial clustering of the sources, and their typical spectral energy distribution
\citep{tittley07,bolton09,mcquinn09b,compostella13,compostella14,puchwein15}.
Information on the thermal state of the IGM has been extracted from the \ion{H}{1} Ly$\alpha$ forest by decomposing it into
individual absorption lines \citep{schaye00,ricotti00,bryan00,mcdonald01b,rudie12,bolton14,rorai18,hiss18}
or by treating it as a continuous field using various transmission statistics, i.e.\ the probability distribution function
\citep{bolton08,viel09,calura12,lee15,rorai17a}, the power spectrum \citep{zaldarriaga01,walther18,walther19},
the transmission curvature \citep{becker11,boera14}, wavelet decomposition \citep{theuns02a,theuns02c,lidz10,garzilli12},
and the quasar pair phase angle distribution \citep{rorai17b}.
Despite large statistical errors and some remaining tension between the measurements, these studies indicate
an extended heating of the IGM from $z\simeq 6$ \citep{bolton10,bolton12} to $z\simeq 2.8$
\citep{schaye00,becker11,boera14,hiss18} expected from an extended \ion{He}{2} reionization process.
Very recent results from \citet{walther19} show a rise in the IGM temperature from $z=5$, peaking at $z\sim 3.4$,
and subsequent cooling to $z=1.8$, which can only be explained as being a result of extended \ion{He}{2} reionization.
The \ion{He}{2} reionization epoch can be studied directly via spectroscopy of intergalactic
\ion{He}{2} Ly$\alpha$ absorption ($\lambda_\mathrm{rest}=303.78$\,\AA) toward far-UV
(FUV)-bright quasars at $z>2$. The discovery and systematic study of the \ion{He}{2}
Ly$\alpha$ forest at $z\ga 2.7$ has been a science driver for \textit{HST}
since its inception \citep{bahcall79}.
However, despite extensive efforts during the first 15 years of \textit{HST} operations,
only seven \ion{He}{2} sightlines had been successfully probed, because for most
$z_\mathrm{em}>2.7$ quasars the spectral range covering the \ion{He}{2} Lyman series
is blacked out by optically thick \ion{H}{1} Lyman limit systems in the foreground IGM \citep{picard93,worseck11}.
Early \textit{HST} spectra of varying spectral resolution and quality revealed strong evolution of the
\ion{He}{2} Ly$\alpha$ effective optical depth \teff\ from Gunn-Peterson troughs ($\mteff\ga 3$)
measured toward four $z_\mathrm{em}>3$ quasars
\citep{jakobsen94,hogan97,anderson99,heap00,zheng04b,zheng08}
to fluctuating \ion{He}{2} absorption at $2.7\la z\la 2.9$ \citep{reimers97,reimers05,heap00,smette02}.
High-resolution spectra taken with the \textit{Far Ultraviolet Spectroscopic Explorer}
(\textit{FUSE}, $R=\lambda/\Delta\lambda\approx 20,000$) resolved the \ion{He}{2} Ly$\alpha$ forest in
two sightlines \citep{kriss01,shull04,zheng04,fechner06,fechner07}.
Inferences on the \ion{He}{2} reionization history, however, were severely limited by sample variance
and data quality \citep{mcquinn09a,furlanetto10}.

The first panoramic FUV imaging surveys by the \textit{Galaxy Evolution Explorer}
\citep[\textit{GALEX},][]{martin05,morrissey07} enabled the efficient photometric selection of likely
\ion{He}{2}-transparent quasar sightlines \citep{syphers09a,syphers09b,worseck11}.
\textit{HST} follow-up spectroscopy with the FUV-sensitive Cosmic Origins Spectrograph
\citep[COS,][]{green12} yielded a sample of 11 new science-grade (signal-to-noise ratio S/N$\ga 3$)
\ion{He}{2} Ly$\alpha$ absorption spectra covering $2.7\la z\la 3.8$ \citep{worseck11b,syphers12,zheng15,worseck16}.
Together with high-quality COS spectra of the previously known sightlines \citep{shull10,syphers13,syphers14},
these new sightlines enabled statistical studies of the intergalactic \ion{He}{2} Ly$\alpha$ opacity and constraints on the
\ion{He}{2} reionization history. The diminishing sightline-to-sightline variance in the large-scale ($\sim 40$ comoving Mpc)
\ion{He}{2} effective optical depth indicates that \ion{He}{2} reionization completed at $z\simeq 2.7$ \citep{worseck11b}. 

In \citet[][hereafter \citetalias{worseck16}]{worseck16} we presented first results of the Helium Reionization Survey (HERS),
a comprehensive campaign to study the epoch of \ion{He}{2} reionization with \textit{HST} \ion{He}{2} absorption spectra.
Our analysis of 17 sightlines revealed a gradual increase in the \ion{He}{2} effective optical depth from $z=2.3$
($\mteff\simeq 1$) to $z=3.4$ ($\mteff\ga 4.5$) with strong sightline-to-sightline variance at $z>2.7$,
consistent with a predominantly ionized IGM with large UV background fluctuations \citep[][hereafter \citetalias{davies17}]{davies14,davies17}.
Numerical simulations of quasar-driven \ion{He}{2} reionization struggle to reproduce the observed distribution
of \ion{He}{2} effective optical depths at $z\simeq 3.4$ \citep{mcquinn09a,compostella13},
suggesting a very extended epoch of \ion{He}{2} reionization \citepalias{worseck16}.
However, this apparent tension may be due to the limited number of quasar models run in limited simulation
volumes \citep{daloisio17}, or due to the small number of four \ion{He}{2} sightlines covering $z\simeq 3.4$ \citep{compostella14}.
A large population of faint quasars may accomplish \ion{He}{2} reionization at $z>4$ \citep{madau15},
but underpredicts the observed \ion{He}{2} effective optical depths at $2.7<z<3$\citep{worseck16,mitra18,puchwein19,garaldi19}
unless the post-reionization UV background is strongly fluctuating on large scales (\citealt{furlanetto10,mcquinn14,davies14}; \citetalias{davies17}).
Likewise, an early \ion{He}{2} reionization would prematurely heat the IGM \citep{daloisio17,mitra18,puchwein19,garaldi19}.
Similarly, the \ion{H}{1} effective optical depths display a large sightline-to-sightline variance at $z>5.5$ \citep{becker15},
possibly persisting to $z\simeq 5.2$ \citep{bosman18,eilers18}. These may be explained by relic temperature fluctuations from patchy
\ion{H}{1} reionization \citep{daloisio15,keating18} and/or large UV background fluctuations that are either sourced
by clustered galaxies and a short mean free path to ionizing photons \citep{davies16,davies18a,daloisio18,becker18}
or rare bright sources, such as quasars \citep{chardin15,chardin17}.
While quasars are likely not abundant enough to substantially contribute to the $z\ga 5$ \ion{H}{1}-ionizing UV background
and \ion{H}{1} reionization \citep[e.g.][but see \citealt{giallongo15}]{jiang16,ricci17,mcgreer18,parsa18,matsuoka18,kulkarni18,wang18},
it may be necessary to include them in reionization models in order to explain the fluctuations in the effective optical depths
of \ion{H}{1} at $z>5.5$ \citep[although see \citealt{kulkarni18b}]{chardin15,chardin17} and of \ion{He}{2} at $z\simeq 3.4$ \citep{compostella14}.

In light of this heated debate on the duration of \ion{He}{2} reionization, the sources of reionization,
and the high-redshift UV background, it is timely to extend the sample of \ion{He}{2} absorption measurements,
particularly at $z>3$. In this manuscript we present \textit{HST}/COS spectra of eight additional
\ion{He}{2}-transparent quasars (Section~\ref{sect:obsdatared}) that more than double the redshift pathlength
sensitive to $\mteff\sim 5$ at $3<z<3.5$, and provide a statistically meaningful sample of six \ion{He}{2} sightlines
probing $z>3.5$ (Section~\ref{sect:he2zpath}). With an enlarged sample of measured \ion{He}{2} effective optical depths
(Section~\ref{sect:he2tauevol}) and using realistic forward-modeled mock spectra from a large-volume hydrodynamical
simulation \citep{lukic15}, we test recent models of the \ion{He}{2}-ionizing background \citep[\citetalias{davies17};][]{puchwein19},
and provide measurements of the evolving \ion{He}{2} photoionization rate and the corresponding \ion{He}{2} fraction
at $2.3<z<3.86$ (Sections~\ref{sect:he2uvb} and \ref{sect:discussion}). We conclude in Section~\ref{sect:conclusions}.

We adopt a flat cold dark matter cosmology with dimensionless Hubble constant
$h=0.685$ ($H_0=100h$\,km\,s$^{-1}$\,Mpc$^{-1}$) and density parameters
$\left(\Omega_\mathrm{m},\Omega_\mathrm{b},\Omega_\Lambda\right)=(0.3,0.047,0.7)$
for total matter, baryons, and cosmological constant, consistent with \citet{planckcollab18}.
In this cosmology $\Delta z=0.04$ -- the standard interval that we will use for our \teff\ measurements -- corresponds
to a distance interval of $\approx 40$ comoving Mpc (hereafter cMpc) at $z=3$.
The object designations of quasars discovered by SDSS will be abbreviated to
SDSS~J\texttt{HHMM$\pm$DDMM}.


\section{Observations and data reduction}
\label{sect:obsdatared}

\subsection{Discovery of FUV-Bright $z_\mathrm{em}>3.1$ Quasars}

Our sample includes six \ion{He}{2}-transparent sightlines toward FUV-bright $z_\mathrm{em}>3.1$ quasars
discovered in our dedicated ground-based survey, dubbed the `\ion{He}{2} Quasar Survey'
(HE2QS, Worseck et al.\ in preparation), which we will briefly summarize here.
Our survey was motivated by the dearth of UV-bright $z_\mathrm{em}\ga 3$ quasars that is due to a
combination of physical effects (i.e.\ the declining space density of bright quasars and the incidence of
optically thick \ion{H}{1} Lyman limit systems in the IGM) and optical color selection effects \citep{worseck11}.
While matching \textit{GALEX} photometry to existing quasar catalogs had led to many tens of candidates
\citep{syphers09a,syphers09b,worseck11}, only 11 of these were either bright enough for efficient science-grade
(S/N$\ga 3$) spectroscopic follow-up with \textit{HST}/COS \citep[FUV$\la 21.5$;][]{worseck11b,syphers12,worseck16}
or probed $z>3.5$, justifying exposure times $\ga 20$\,ks \citep{zheng15,worseck16}. 

Target selection was performed mainly via applying color cuts to combined catalogs
of wide-field multi-wavelength photometry from \textit{GALEX} \citep[FUV
  and NUV;][]{morrissey07}, SDSS \citep[$ugri$; e.g.][]{aihara11} or
Pan-STARRS \citep[$gri$;][]{chambers16}, and \textit{WISE} \citep[W1
  and W2;][]{wright10}.  Typically we required significant flux in the
FUV (S/N$>3$, FUV$<21.5$) and the optical ($i<19.5$) to ensure
efficient spectroscopic follow-up with \textit{HST}/COS and
ground-based telescopes, respectively. On the SDSS footprint we
also applied extreme deconvolution techniques \citep{bovy11b,bovy11}
specifically trained on SDSS$+$\textit{GALEX} mock photometry
\citep{worseck11} to select FUV-bright high$-z$ quasar candidates.
The quasar candidates were verified with low-resolution ($R\sim 200$) optical spectroscopy,
mainly with the Calar Alto Faint Object Spectrograph (CAFOS) at the
Calar Alto 2.2\,m telescope and the Kast spectrometer at the Lick 3\,m
Shane Telescope during several campaigns (2010--2015).  Unambiguous
redshifts were measured from broad emission lines (typically
\ion{H}{1} Ly$\alpha$ and \ion{C}{4}).

\subsection{\textit{HST}/COS Follow-up Spectroscopy}

In \textit{HST} Cycles~20 and 22 we used COS \citep{green12} to obtain follow-up spectroscopy of eight
$3.1<z_\mathrm{em}<3.8$ quasars discovered in HE2QS (Programs 13013 and 13875, PI Worseck; Table~\ref{tab:qsosample}).
The UV brightness of our targets enabled efficient, simultaneous verification of quasar flux at \ion{He}{2} Ly$\alpha$
in the quasar rest frame and science-grade spectroscopy of the \ion{He}{2} absorption along their sightlines in single visits
of 3--4 \textit{HST} orbits per target. We employed the COS grating G140L in the 1105\,\AA\ setup ($\lambda\lambda$ 1110--2150\,\AA)
at COS Lifetime Positions (LPs) 2 and 3, resulting in a spectral resolution $R\simeq 1400$ (LP2) and $R\simeq 1600$ (LP3) at 1150\,\AA.
The 20--30\% lower spectral resolution compared to COS LP1 is inconsequential to our science goal to quantify
the \ion{He}{2} absorption on scales $\Delta z\ge 0.01$ ($\ga 4$ G140L resolution elements). The Cycle~20 observations were taken
between 2013 April and December, i.e.\ in a period of high solar activity, while the Cycle~22 observations were taken between
2015 August and November, i.e.\ past solar maximum.
In our Cycle~22 program we also obtained science-grade COS G140L spectra of two $z_\mathrm{em}\simeq 3.8$
SDSS quasars that had been verified to show flux at \ion{He}{2} Ly$\alpha$ in earlier surveys with STIS \citep{zheng05}
and the ACS prism \citep{syphers09a}. Being a factor $\ga 2$ fainter than the quasars discovered in HE2QS,
these two targets needed significantly longer exposures of $2\times 5$ orbits to yield S/N$=3$--4 at \ion{He}{2} Ly$\alpha$.
Individual visits were scheduled within two months of another to minimize complications in the coaddition of very low S/N data
due to quasar variability.

\tabletypesize{\footnotesize}
\begin{deluxetable*}{lcclllrrrrrrrr}
\tablewidth{0pt}
\setlength{\tabcolsep}{1ex}
\renewcommand{\arraystretch}{1.0}
\tablecaption{\label{tab:qsosample}Sample of UV-bright $z_\mathrm{em}>2.7$ Quasars}
\tablehead{
\colhead{Object} &
\colhead{R.\,A. (J2000)}&
\colhead{Decl. (J2000)} &
\colhead{$z_\mathrm{em}$} &
\colhead{PID\tablenotemark{a}} &
\colhead{Instrument} &
\colhead{$R$\tablenotemark{b}}&
\colhead{$t_\mathrm{exp}$ (s)}&
\colhead{S/N\tablenotemark{c}}&
\colhead{$f_{1500\mathrm{\AA}}\tablenotemark{d}$}&
\colhead{$\alpha$\tablenotemark{e}}&
\colhead{$z_\mathrm{abs}$}&
\colhead{$\log N_\mathrm{HI}\tablenotemark{f}$}&
\colhead{Ref.\tablenotemark{g}}}
\startdata
HS~1700$+$6416    &$17^\mathrm{h}01^\mathrm{m}00\fs61$ &$+64\degr12\arcmin09\farcs1$ &$2.751$ &11528 &COS G140L &2000 &15705 &15 &$21.590$ &$-1.756$ &$0.8648$ &$16.05$        &1\\
                  &                                    &                             &        &      &          &     &      &   &         &         &$0.7222$ &$16.17$        & \\
                  &                                    &                             &        &      &          &     &      &   &         &         &$0.5528$ &$15.87$        & \\
HS~1024$+$1849    &$10^\mathrm{h}27^\mathrm{m}34\fs13$ &$+18\degr34\arcmin27\farcs5$ &$2.860$ &12816 &COS G130M &15000&28689 &5  &$2.935$  &$-0.560$ &\nodata  &\nodata        &2\\
Q~1602$+$576      &$16^\mathrm{h}03^\mathrm{m}55\fs92$ &$+57\degr30\arcmin54\farcs4$ &$2.862$ &12816 &COS G130M &15000&15613 &7  &$6.188$  &$-2.463$ &\nodata  &\nodata        &2\\
HE~2347$-$4342    &$23^\mathrm{h}50^\mathrm{m}34\fs21$ &$-43\degr25\arcmin59\farcs6$ &$2.887$ &11528 &COS G140L &2000 &11557 &14 &$20.325$ &$-2.690$ &$0.5766$ &$<15.8$        &1\\
                  &                                    &                             &        &      &COS G130M &16000&28458 &19 &         &         &$0.4215$ &$<15.8$        & \\
PC~0058$+$0215    &$01^\mathrm{h}00^\mathrm{m}58\fs39$ &$+02\degr31\arcmin31\farcs4$ &$2.89$  &11742 &COS G140L &2000 &6212  &4  &$1.415$  &$-1.289$ &\nodata  &\nodata        &1\\
SDSS~J0936$+$2927 &$09^\mathrm{h}36^\mathrm{m}43\fs50$ &$+29\degr27\arcmin13\farcs6$ &$2.930$ &11742 &COS G140L &2000 &4739  &4  &$1.086$  &$-2.314$ &$0.2121$ &blend          &1\\
SDSS~J0818$+$4908 &$08^\mathrm{h}18^\mathrm{m}50\fs01$ &$+49\degr08\arcmin17\farcs0$ &$2.957$ &11742 &COS G140L &2000 &7598  &4  &$1.246$  &$-2.295$ &$0.2015$ &$\lesssim 17.0$&1\\
HS~1157$+$3143    &$12^\mathrm{h}00^\mathrm{m}06\fs24$ &$+31\degr26\arcmin30\farcs8$ &$2.989$ &9350  &STIS G140L&1000 &26820 &11 &$0.541$  &$-7.346$ &\nodata  &\nodata        &1\\
SDSS~J0924$+$4852 &$09^\mathrm{h}24^\mathrm{m}47\fs35$ &$+48\degr52\arcmin42\farcs8$ &$3.027$ &11742 &COS G140L &2000 &7598  &8  &$2.432$  &$-2.085$ &$0.4570$ &$<16.0$        &1\\
                  &                                    &                             &        &      &          &     &      &   &         &         &$0.2280$ &blend          & \\
SDSS~J1101$+$1053 &$11^\mathrm{h}01^\mathrm{m}55\fs74$ &$+10\degr53\arcmin02\farcs3$ &$3.029$ &11742 &COS G140L &2000 &7157  &4  &$1.028$  &$-2.953$ &$0.3177$ &$\sim 16.5$    &1\\
                  &                                    &                             &        &      &          &     &      &   &         &         &$0.1358$ &$21.13$        & \\
HE2QS~J2157$+$2330&$21^\mathrm{h}57^\mathrm{m}43\fs63$ &$+23\degr30\arcmin37\farcs3$ &$3.143$ &13013 &COS G140L &1400 &8074  &4  &$5.772$  &$-3.185$ &$0.8092$ &$17.27$        &2\\
                  &                                    &                             &        &      &          &     &      &   &         &         &$0.7112$ &$16.96$        & \\
                  &                                    &                             &        &      &          &     &      &   &         &         &$0.5545$ &$16.52$        & \\
SDSS~J1237$+$0126 &$12^\mathrm{h}37^\mathrm{m}48\fs99$ &$+01\degr26\arcmin07\farcs0$ &$3.154$ &11742 &COS G140L &2000 &6212  &4  &$1.401$  &$-2.290$ &\nodata  &\nodata        &1\\
HE2QS~J2321$+$1558&$23^\mathrm{h}21^\mathrm{m}54\fs98$ &$+15\degr58\arcmin34\farcs2$ &$3.212$ &13013 &COS G140L &1400 &7427  &0  &$6.427$  &$-1.165$ &$0.7572$ &$17.11$        &2\\
                  &                                    &                             &        &      &          &     &      &   &         &         &$0.4900$ &$\sim 18.2$    & \\
HE2QS~J1706$+$5904&$17^\mathrm{h}06^\mathrm{m}21\fs74$ &$+59\degr04\arcmin06\farcs4$ &$3.248$ &13013 &COS G140L &1400 &15300 &4  &$0.809$  &$-1.152$ &$0.6220$ &$16.29$        &2\\
                  &                                    &                             &        &      &          &     &      &   &         &         &$0.4040$ &$\sim 18.5$    & \\
HE2QS~J2149$-$0859&$21^\mathrm{h}49^\mathrm{m}27\fs77$ &$-08\degr59\arcmin03\farcs6$ &$3.259$ &13013 &COS G140L &1400 &7561  &3  &$0.928$  &$-2.116$ &$0.3530$ &$<16.5$        &2\\
Q~0302$-$003      &$03^\mathrm{h}04^\mathrm{m}49\fs85$ &$-00\degr08\arcmin13\farcs5$ &$3.286$ &12033 &COS G130M &16000&24870 &3  &$2.816$  &$-3.534$ &\nodata  &\nodata        &2\\
HE2QS~J0233$-$0149&$02^\mathrm{h}33^\mathrm{m}06\fs01$ &$-01\degr49\arcmin50\farcs5$ &$3.314$ &13013 &COS G140L &1400 &7307  &4  &$1.885$  &$-2.074$ &$0.3825$ &$<16.6$        &2\\
                  &                                    &                             &        &      &          &     &      &   &         &         &$0.3186$ &$<16.8$        & \\
HS~0911$+$4809    &$09^\mathrm{h}15^\mathrm{m}10\fs01$ &$+47\degr56\arcmin58\farcs8$ &$3.350$ &12816 &COS G130M &15000&26864 &6  &$3.034$  &$-0.475$ &$0.3028$ &$<16.8$        &2\\
                  &                                    &                             &        &      &          &     &      &   &         &         &$0.1827$ &$\sim 18.5$    & \\
HE2QS~J0916$+$2408&$09^\mathrm{h}16^\mathrm{m}20\fs85$ &$+24\degr08\arcmin04\farcs6$ &$3.440$ &13013 &COS G140L &1400 &7954  &4  &$3.241$  &$-2.117$ &$0.6638$ &$16.79$        &2\\
SDSS~J1253$+$6817 &$12^\mathrm{h}53^\mathrm{m}53\fs71$ &$+68\degr17\arcmin14\farcs2$ &$3.481$ &12249 &COS G140L &2000 &14095 &7  &$1.834$  &$-2.858$ &$0.6930$ &$16.09$        &2\\
SDSS~J2346$-$0016 &$23^\mathrm{h}46^\mathrm{m}25\fs66$ &$-00\degr16\arcmin00\farcs4$ &$3.512$ &12249 &COS G140L &2000 &20737 &8  &$2.053$  &$-1.771$ &\nodata  &\nodata        &2\\
HE2QS~J2311$-$1417&$23^\mathrm{h}11^\mathrm{m}45\fs46$ &$-14\degr17\arcmin52\farcs1$ &$3.705$ &13875 &COS G140L &1600 &11084 &4  &$2.025$  &$-2.600$ &$0.7515$ &$16.37$        &2\\
                  &                                    &                             &        &      &          &     &      &   &         &         &$0.4779$ &$<17.2$        & \\
SDSS~J1137$+$6237 &$11^\mathrm{h}37^\mathrm{m}21\fs72$ &$+62\degr37\arcmin07\farcs2$ &$3.779$ &13875 &COS G140L &1600 &50993 &4  &$0.509$  &$-2.980$ &\nodata  &\nodata        &2\\
HE2QS~J1630$+$0435&$16^\mathrm{h}30^\mathrm{m}56\fs34$ &$+04\degr35\arcmin59\farcs4$ &$3.788$ &13013 &COS G140L &1400 &7908  &4  &$3.397$  &$-1.158$ &\nodata  &\nodata        &2\\
SDSS~J1614$+$4859 &$16^\mathrm{h}14^\mathrm{m}26\fs81$ &$+48\degr59\arcmin58\farcs8$ &$3.812$ &13875 &COS G140L &1600 &28369 &3  &$0.559$  &$-2.472$ &\nodata  &\nodata        &2\\
SDSS~J1711$+$6052 &$17^\mathrm{h}11^\mathrm{m}34\fs41$ &$+60\degr52\arcmin40\farcs3$ &$3.834$ &12249 &COS G140L &2000 &23951 &4  &$1.467$  &$-4.803$ &$0.7750$ &$16.53$        &2\\
                  &                                    &                             &        &      &          &     &      &   &         &         &$0.4370$ &$\lesssim 18.0$& \\
SDSS~J1319$+$5202 &$13^\mathrm{h}19^\mathrm{m}14\fs20$ &$+52\degr02\arcmin00\farcs1$ &$3.930$ &12249 &COS G140L &2000 &26643 &2  &$0.876$  &$-5.767$ &$0.7026$ &$17.27$        &2\\
\enddata
\tablenotetext{a}{\textit{HST} Program Number.}
\tablenotetext{b}{Spectral resolution $R\equiv\lambda/\mathrm{FWHM}$ at $\lambda=1150$\,\AA.}
\tablenotetext{c}{Signal-to-noise ratio per pixel (COS G140L: $\simeq 0.24$\,\AA\,pixel$^{-1}$, COS G130M $R\simeq 16000$: $\simeq 0.03$\,\AA\,pixel$^{-1}$, COS G130M $R\simeq 15000$: $\simeq 0.04$\,\AA\,pixel$^{-1}$, STIS G140L: $0.6$\,\AA\,pixel$^{-1}$) near \ion{He}{2} Ly$\alpha$ in the quasar rest frame. HE2QS~J2321$+$1558 and HE2QS~J1706$+$5904 are not considered further due to strong intervening \ion{H}{1} Lyman limit systems.}
\tablenotetext{d}{Continuum flux density at 1500\,\AA\ in $10^{-16}$\,erg\,cm$^{-2}$\,s$^{-1}$\,\AA$^{-1}$.}
\tablenotetext{e}{Power-law spectral index $\alpha$ for $f_\lambda=f_{1500\mathrm{\AA}}\left(\lambda/1500\mathrm{\AA}\right)^{\alpha}$ including a correction for identified \ion{H}{1} Lyman continuum absorption.}
\tablenotetext{f}{Logarithmic column density of identified intervening \ion{H}{1} absorber in cm$^{-2}$.}
\tablenotetext{g}{Data reference. 1: \citet{worseck16}, 2: This paper.}
\end{deluxetable*}

\subsection{Archival Data}

We supplemented this data set with \textit{HST} spectra of 17 \ion{He}{2}-transparent quasars studied in \citetalias{worseck16},
but using more recent higher-quality COS G130M spectra for four of these (Table~\ref{tab:qsosample}).
Q~0302$-$003 \citep{syphers14} was reobserved in the COS GTO Cycle~18 Program 12033 (PI Green) at
COS LP1 ($R\simeq 16,000$ at 1150\,\AA). Three UV-bright \ion{He}{2}-transparent quasars discovered in a COS G140L survey by
\citet{syphers12} were followed up with G130M in its 1222\,\AA\ setup between 2013 March and July (Cycle~20 Program 12816, PI Syphers).
Compared to the redder G130M setups, the 1222\,\AA\ setup ($\lambda\lambda$ 1066--1367\,\AA) provides extended coverage
of the \ion{He}{2} Ly$\alpha$ absorption down to $z=2.51$ at lower spectral resolution ($R\simeq 15,000$ at 1150\,\AA,
declining toward shorter and longer wavelengths). Visits were scheduled within a few days of each other to minimize the adverse effects
of quasar variability. For the highest-redshift target HS~0911$+$4809 at $z_\mathrm{em}=3.350$ a 1-orbit G140L spectrum was
taken together with the G130M spectrum to aid continuum definition and to gauge the level of quasar variability between the
G140L survey observations in Cycle~18 and the G130M follow-up in Cycle~20.

\subsection{\textit{HST}/COS Data Reduction}

All raw data were retrieved from the \textit{HST} archive and were reduced with the CALCOS pipeline (v2.21)
and custom software that preserves Poisson counts in coadded spectra and offers reliable background subtraction
(dark current, quasi-diffuse sky emission and scattered light) at the flux limit of \textit{HST}/COS
($f_\lambda\la 10^{-17}$\,erg\,cm$^{-2}$\,s$^{-1}$\,\AA$^{-1}$; \citetalias{worseck16}).
The only significant update to the procedure detailed in \citetalias{worseck16} is that we routinely limited the
COS detector pulse height amplitude (PHA) to the range encountered in the science data, which effectively excludes
a part of the background and therefore maximizes the sensitivity to high \teff\ at $z>3.3$ where the COS G140L sensitivity
significantly declines \citepalias{worseck16}. Pulse height screening was employed for all G130M spectra and for
G140L spectra obtained in Programs 12249, 13013 and 13875. With the reanalyzed data from Program 12249
we confirmed that for G140L \ion{He}{2} spectra, pulse height screening significantly enhances the data quality only
at $z>3.3$, so we took 8 lower-redshift sightlines from \citetalias{worseck16} without reprocessing them (Table~\ref{tab:qsosample}). 

The data reduction parameters were customized according to the LP, the grating setting, and the changing PHA distribution
of the COS detector. Boxcar source extraction was performed on rectangular apertures that include $\simeq 100$\%
of the flux of a point source in the covered wavelength range (i.e.\ accounting for COS's astigmatism in the cross-dispersion direction)
while minimizing the background contribution. For G140L spectra we chose a 25-pixel aperture at every LP, which for LP3
conveniently excludes the gain-sagged trace at the close-by LP1\footnote{We refrained from using the new
\texttt{TWOZONE} extraction algorithm in CALCOS due to its inaccurate dark current estimation from offset windows with different
gain sag characteristics. Moreover, \texttt{TWOZONE} is currently not applicable to dark exposures, and to data obtained at COS LPs 1 and 2.}. 
For the G130M spectra of Q~0302$-$003 taken in the 1291\,\AA\ and 1318\,\AA\ setups we deemed a 23-pixel aperture
as adequate, while the 1222\,\AA\ setup used in Program 12816 required 35 pixels due to its wider cross-dispersion profile.

The PHA range of the science data was determined from the two-dimensional count distribution
in the trace (G130M) or from the fraction of geocoronal Ly$\alpha$ counts at a given PHA value (G140L),
and is listed for every (re)analyzed target in Table~\ref{tab:obsdetail}. The PHA range determined from
geocoronal Ly$\alpha$ extends to the lowest values encountered in the science data due to locally stronger
detector gain sag, and includes $>99$\% of the total Ly$\alpha$ flux in the extraction aperture.

The dark current in the science aperture was estimated from post-processed dark monitoring exposures
taken within $\pm 1.5$ months around the date of observation \citepalias{worseck16} at the same detector
voltage and using the PHA range determined for the science exposures. Different time windows were adopted
for Q~0302$-$003 and Q~1602$+$576 that were observed shortly before or after an increase of the detector
voltage that interferes with the long-term dark monitoring. As detailed in \citetalias{worseck16}, we coadded
dark exposures taken in similar environmental conditions, as determined from the PHA distribution in two
50--60-pixel wide unilluminated windows above and below the science aperture. These included previous
gain-sagged COS LPs, but that is inconsequential to our calibration efforts. The pulse-height-screened
dark current extracted in the science aperture was smoothed by a 500-pixel running average and scaled
to the total dark current expected in the science exposure using the PHA calibration windows.
Validation tests with dark exposures following \citetalias{worseck16} confirmed that the chosen 500-pixel
averaging scale captured the large-scale spatial variations (up to a factor $2.8$) in the Segment A dark current
occurring around solar maximum. Together with more frequent dark monitoring beginning in Cycle~20
($65\times1330$\,s in the considered 3-month window) this ensured percent-level accurate dark estimation in
highly variable environmental conditions.

\tabletypesize{\footnotesize}
\begin{deluxetable}{lllcc}
\tablewidth{0pt}
\setlength{\tabcolsep}{1ex}
\renewcommand{\arraystretch}{1.0}
\tablecaption{\label{tab:obsdetail}PHA Ranges Adopted for (Re)analyzed Quasars}
\tablehead{
\colhead{Object} &
\colhead{LP} &
\colhead{Dates}&
\colhead{Segment A} &
\colhead{Segment B}
}
\startdata
SDSS~J2346$-$0016	&1	&2010 Nov 29		&2--15	&\nodata\\
					&1	&2010 Dec 4		&2--15	&\nodata\\
SDSS~J1711$+$6052	&1	&2011 Apr 27--28	&2--15	&\nodata\\
					&1	&2011 Oct 9		&1--14	&\nodata\\
SDSS~J1319$+$5202	&1	&2011 May 4, 8		&2--15	&\nodata\\
SDSS~J1253$+$6817	&1	&2011 May 5		&2--15	&\nodata\\
Q~0302$-$003			&1	&2012 Mar 8, 9		&1--15	&1--15\\
HS~1024$+$1849		&2	&2013 Mar 25, 26	&3--14	&3--14\\
HE2QS~J1706$+$5904	&2	&2013 Apr 11		&3--14	&\nodata\\
HE2QS~J1630$+$0435	&2	&2013 Apr 12		&3--14	&\nodata\\
HE2QS~J2149$-$0859	&2	&2013 Apr 27		&3--14	&\nodata\\
HS~0911$+$4809		&2	&2013 May 3, 5		&3--14	&3--14\\
HE2QS~J2321$+$1558	&2	&2013 May 9		&3--14	&\nodata\\
HE2QS~J2157$+$2330	&2	&2013 Jul 19		&2--13	&\nodata\\
Q~1602$+$576			&2	&2013 Jul 20		&3--14	&4--15\\
HE2QS~J0233$-$0149	&2	&2013 Aug 7		&2--13	&\nodata\\
HE2QS~J0916$+$2408	&2	&2013 Dec 4		&2--13	&\nodata\\
SDSS~J1614$+$4859	&3	&2015 Aug 28, 31	&2--14	&\nodata\\
SDSS~J1137$+$6237	&3	&2015 Sep 17--18	&2--14	&\nodata\\
					&3	&2015 Nov 10--11	&2--14	&\nodata\\
HE2QS~J2311$-$1417	&3	&2015 Nov 7		&2--14	&\nodata\\
 \enddata
\end{deluxetable}

Open-shutter backgrounds were treated as in \citetalias{worseck16}. In short, the quasi-continuous FUV
sky background was taken from \citet{murthy14}. Earthshine was significant only for HE2QS~J1706$+$5904,
so for this target we restricted the limb angle to $>21\degr$ in the affected wavelength range 1142--1181\,\AA.
Likewise, geocoronal emission lines (\ion{N}{1}\,$\lambda 1134$\,\AA, \ion{N}{1}\,$\lambda 1200$\,\AA, \ion{H}{1}\,$\lambda 1216$\,\AA,
\ion{N}{1}\,$\lambda 1243$\,\AA, \ion{O}{1}\,$\lambda 1304$\,\AA, \ion{O}{1}]\,$\lambda 1356$\,\AA)
were suppressed using data restricted in solar altitude and/or limb angle. Residual geocoronal emission was flagged,
in particular the remaining core of \ion{H}{1} Ly$\alpha$ and the region around \ion{O}{1}\,$\lambda 1304$\,\AA\ that
has very little sensitivity to high \ion{He}{2} absorption due to the reduced exposure time in orbital night.
Scattered geocoronal Ly$\alpha$ emission in the G140L spectra was estimated from the recorded geocoronal
Ly$\alpha$ flux and an empirical model describing its amplitude along the dispersion axis \citepalias{worseck16}.
For HE2QS~J2149$-$0859 and SDSS~J1614$+$4859 that were observed mostly during orbital day
(daytime fractions $79.8$\% and $83.7$\%, respectively) the uncertainty in our scattered light model dominates
the total background uncertainty and results in appreciable negative flux on large scales, limiting the sensitivity
to strong \ion{He}{2} absorption.
Comparing data taken during orbital day and night we verified the presence of scattered geocoronal Ly$\alpha$
emission in the G130M spectra of Q~0302$-$003 and HS~0911$+$4809, and used only nighttime data in the
affected wavelength range ($\simeq 10$\,\AA\ around geocoronal Ly$\alpha$). For HS~0911$+$4809 the cuts
were more extensive to exclude strong geocoronal \ion{N}{1}\,$\lambda 1200$\,\AA\ and very weak but present
\ion{N}{1}\,$\lambda 1243$\,\AA\ emission. Beyond that, saturated \ion{He}{2} absorption confirms that
scattered light is insignificant in G130M spectra.

Subexposures were coadded by summing the detected Poisson counts and the post-processed background
on the CALCOS wavelength grid, accounting for varying pixel exposure times due to offsets in dispersion direction,
detector blemishes, and geocoronal emission. Absorption lines in the quasar continuum and \ion{He}{2}
transmission features were used to verify the alignment of each individual grating setup across multiple visits.
However, subexposures taken at different central wavelengths showed local distortions in the wavelength solution
by up to several resolution elements \citep[e.g.][]{wakker15} that are probably caused by inaccuracies in the
geometric distortion correction and detector walk \citep{sahnow16}. These local wavelength calibration errors
do not affect our measurements of the mean \ion{He}{2} absorption on scales $\Delta z\ge 0.01$, corresponding
to $\ga 4$ G140L and $\ga 25$ G130M resolution elements, respectively.

The coadded counts were flux-calibrated using the time-dependent flux calibration curve determined by CALCOS.
For targets observed in multiple visits or at several central wavelengths we used an exposure-time-weighted
average calibration curve to preserve the Poisson count statistics. The G140L spectra were binned by a factor
of three to yield a sampling of 2--3 $0.24$\,\AA\ pixels per resolution element in the wavelength range of interest
(1100--1500\,\AA). The G130M spectra were binned to two pixels per resolution element accounting for the lower
resolution of the 1222\,\AA\ setup, yielding dispersions of $\simeq 0.03$\,\AA\,pixel$^{-1}$ for Q~0302$-$003,
and $\simeq 0.04$\,\AA\,pixel$^{-1}$ for the three quasars observed in Program 12816, respectively.
The asymmetric Poisson error was calculated accounting for the background \citep{feldman98}.
For plotting purposes we computed an approximate $1\sigma$ error array by adding in quadrature the larger
of the asymmetric Poisson error and the propagated background error. Almost all G140L spectra obtained in our
Cycle~20 and 22 programs reach a S/N$\simeq 4$ per $0.24$\,\AA\ pixel near \ion{He}{2} Ly$\alpha$ in the
quasar rest frame (Table~\ref{tab:qsosample}). The G130M spectra have a higher continuum S/N per \AA\ due
to their higher dispersion.

\subsection{Continuum Definition}

Since the intrinsic FUV flux of \ion{He}{2}-transparent quasars is partially attenuated by low-redshift \ion{H}{1}
Lyman limit absorbers, many of which are in the NUV  \citep[e.g.][]{worseck11}, one can only fit quasi-continua
in the covered FUV spectral range that do not represent the intrinsic quasar spectral energy distributions.
Their full reconstruction would require contiguous and near-simultaneous spectral coverage from the FUV
to the optical to locate Lyman limit breaks while minimizing uncertainties due to quasar variability
\citep{syphers13,syphers14}. Moreover, due to the strong unresolved \ion{He}{2} absorption in the G140L
spectra it is necessary to extrapolate a continuum from redward of \ion{He}{2} Ly$\alpha$, including
identified Lyman limit breaks in the covered spectral range.

We modeled the continuum as a power-law $f_\lambda\propto\lambda^\alpha$, accounting for Galactic extinction,
low-redshift IGM and ISM absorption, weak quasar emission lines apparent in a few spectra, and residual
geocoronal contamination.
All spectra were corrected for Galactic extinction with their respective line of sight selective extinction $E(B-V)$
from \citet{schlegel98} and the \citet{cardelli89} extinction curve assuming the Galactic average ratio between
total $V$ band extinction and selective extinction $R_V=3.1$. Each G140L spectrum was searched for (partial)
Lyman limit absorbers identified by their Lyman series transitions, whose column densities were added as
additional free parameters to the power-law continuum if the spectral range redward of the break improved the
continuum fit. 
The power-law including Lyman limit breaks was fitted to select regions deemed free of emission and absorption
lines redward of \ion{He}{2} Ly$\alpha$ via a maximum likelihood routine, with continuum errors estimated by
refitting Poisson deviates of the inferred continuum counts 10,000 times. Table~\ref{tab:qsosample} lists the
continuum parameters and the identified Lyman limit systems for all quasars considered in this work,
including nine quasars from \citetalias{worseck16} for completeness. For absorbers with inferred optically thin
Lyman limit breaks overlapping with the \ion{He}{2} absorption we obtained upper limits on the \ion{H}{1}
column densities from \ion{He}{2} IGM transmission features. These potential breaks were not considered
in the continuum fit, such that the \ion{He}{2} absorption at shorter wavelengths may be overestimated by
up to $\delta\mteff\simeq 1$ (HE2QS~J2311$-$1417 in Table~\ref{tab:qsosample}). Two quasars observed
in our Cycle~20 survey have their FUV spectra truncated by intervening optically thick Lyman limit systems
shortly redward (HE2QS~J2321$+$1558) or blueward (HE2QS~J1706$+$5904) of \ion{He}{2} Ly$\alpha$
in the quasar rest frame (Table~\ref{tab:qsosample}), and are not considered for further analysis\footnote{Due
to the confusion of saturated \ion{He}{2} absorption and optically thick \ion{H}{1} Lyman continuum absorption
one requires S/N$\ga 3$ survey spectra to identify Lyman limit systems via their Lyman series.}.
The different PHA cuts adopted for the quasars from Program 12249 did not significantly change their continuum
fits reported in \citetalias{worseck16}. 

Since most G130M spectra have insufficient spectral coverage
redward of \ion{He}{2} Ly$\alpha$ for a reliable continuum extrapolation
we adopted their respective G140L continua after correcting for quasar variability between the G140L
and the G130M observations. Specifically, we measured the average fluxes in 20\,\AA\ bins in the
overlapping spectral range redward of \ion{He}{2} Ly$\alpha$ and determined their ratio to check for
variations in the flux normalization and the spectral slope. The two $z_\mathrm{em}=2.86$ quasars
from Program 12816 had enough spectral coverage redward of \ion{He}{2} Ly$\alpha$ to rescale their
Cycle~18 G140L spectra that were analyzed in \citetalias{worseck16}. As the 1-orbit G140L exposures
of Q~0302$-$003 and HS~0911$+$4809 taken together with the G130M spectra were of insufficient
quality for a continuum fit, we used them only to determine the level of quasar variability compared to
their earlier STIS and COS G140L spectra \citepalias{worseck16}. All four quasars were consistent with
a constant flux ratio redward of \ion{He}{2} Ly$\alpha$, so for all quasars we adopted its mean to rescale
the G140L continuum normalization, adding its standard error in quadrature to the statistical continuum error
(Table~\ref{tab:contscale}). For HS~0911$+$4809 the 45\,\AA\ continuum overlap between the G130M
and the G140L spectrum revealed a $\simeq 3$\% inconsistency in the absolute flux calibration that was
corrected by rescaling the G130M flux calibration curve.

\tabletypesize{\footnotesize}
\begin{deluxetable}{lllc}
\tablewidth{0pt}
\setlength{\tabcolsep}{1ex}
\renewcommand{\arraystretch}{1.0}
\tablecaption{\label{tab:contscale}Flux Ratio between FUV Spectra at different Epochs}
\tablehead{
\colhead{Object} &
\colhead{1st epoch} &
\colhead{2nd epoch}&
\colhead{Mean flux ratio}\\
& & &
\colhead{(2nd/1st epoch)}
}
\startdata
Q~0302$-$003			&1997 Dec 2, 10	&2012 Mar 8, 9		&$0.90\pm 0.02$\\
HS~0911$+$4809		&2010 Oct 6--7		&2013 May 3, 5		&$0.78\pm 0.01$\\
HS~1024$+$1849		&2011 Mar 14		&2013 Mar 25, 26	&$0.59\pm 0.01$\\
Q~1602$+$576			&2011 Mar 4		&2013 Jul 20		&$1.14\pm 0.01$\\
 \enddata
\end{deluxetable}

\section{Eight New Science-grade \ion{He}{2} Quasar Sightlines at $z>3.1$}
\label{sect:he2zpath}

\begin{figure*}[t]\centering
\includegraphics[width=0.9\textwidth]{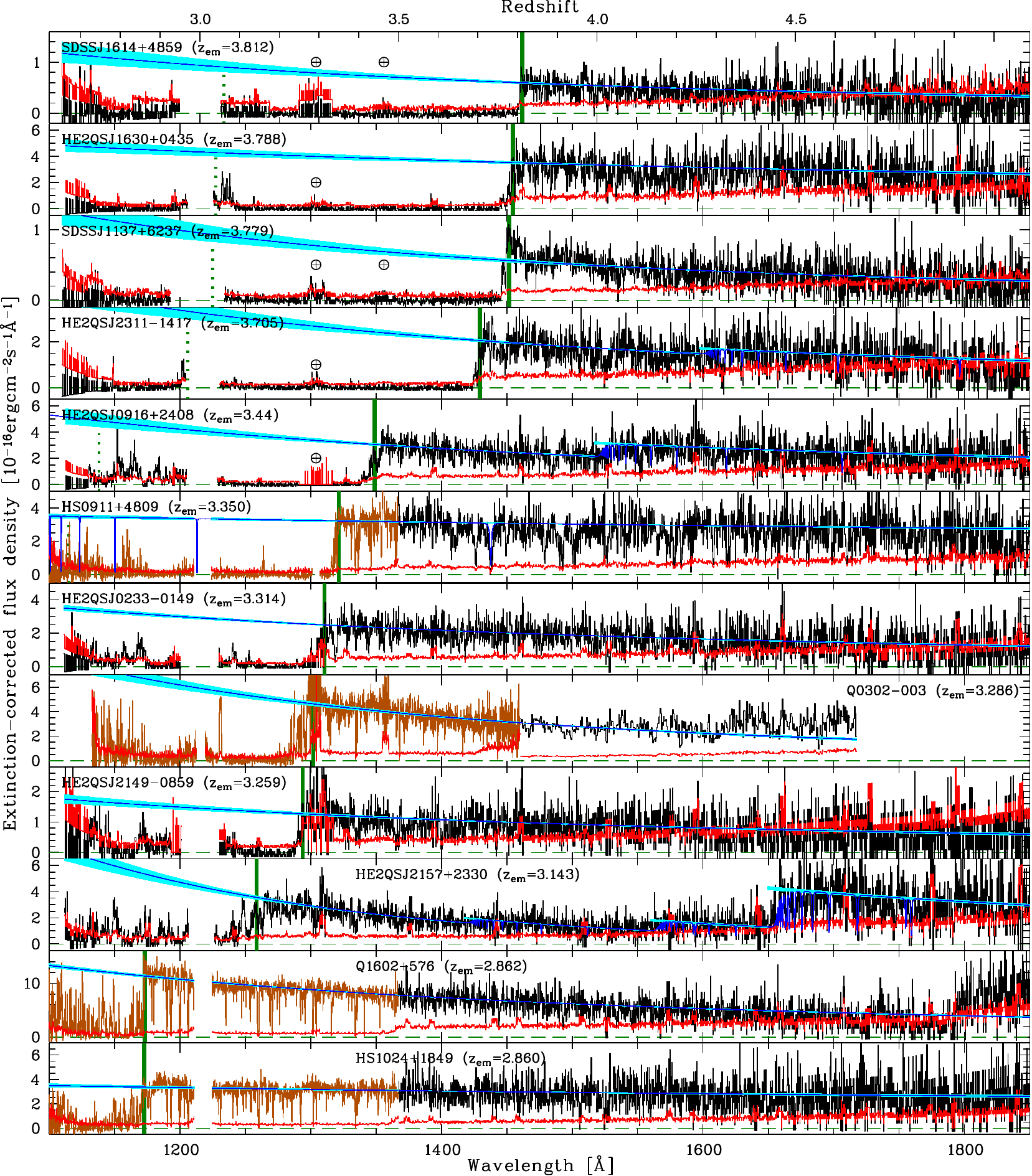}
\caption{\label{fig:he2newspc}
Extinction-corrected \textit{HST} FUV spectra (black/brown) and corresponding $1\sigma$ error arrays (red)
of 12 \ion{He}{2}-transparent quasar sightlines with either new or significantly deeper data than presented in \citetalias{worseck16}.
The redshift axis (top) is for \ion{He}{2} Ly$\alpha$. Low-resolution spectra (black) have been taken with COS
(G140L, $R= 1400$--2000 at 1150\,\AA\ depending on the COS LP, binned to $0.24$\,\AA\,pixel$^{-1}$) with the exception
of Q~0302$-$003 for which we show its higher-quality STIS spectrum (G140L, $R\sim 1000$, 0.6\,\AA\,pixel$^{-1}$).
COS G130M spectra are plotted in brown ($R=15,000$--$16,000$ at 1150\,\AA, binned to $0.12$\,\AA\,pixel$^{-1}$ for display purposes),
and the corresponding G140L spectra taken at different epochs have been rescaled to correct for quasar variability.
The spectral region contaminated by geocoronal Ly$\alpha$ is not shown, and regions with residual \ion{O}{1}
emission have been marked (Earth symbols). The green dashed lines mark the zero level, whereas the green
vertical lines mark \ion{He}{2} Ly$\alpha$ (solid) and \ion{He}{2} Ly$\beta$ (dotted) in the quasar rest frame.
The blue lines show power-law continuum fits to absorption-free regions redward of \ion{He}{2} Ly$\alpha$
and the corresponding $1\sigma$ error (cyan shaded).
For four quasars (HE2QS~J2311$-$1417, HE2QS~J0916$+$2408, HS~0911$+$4809, and HE2QS~J2157$+$2330)
the power-law continua include identified Lyman series and continuum absorption from intervening low-redshift
\ion{H}{1} Lyman limit systems (Table~\ref{tab:qsosample}), convolved to COS resolution.
}
\end{figure*}

\begin{figure}[t]
\includegraphics[width=\linewidth]{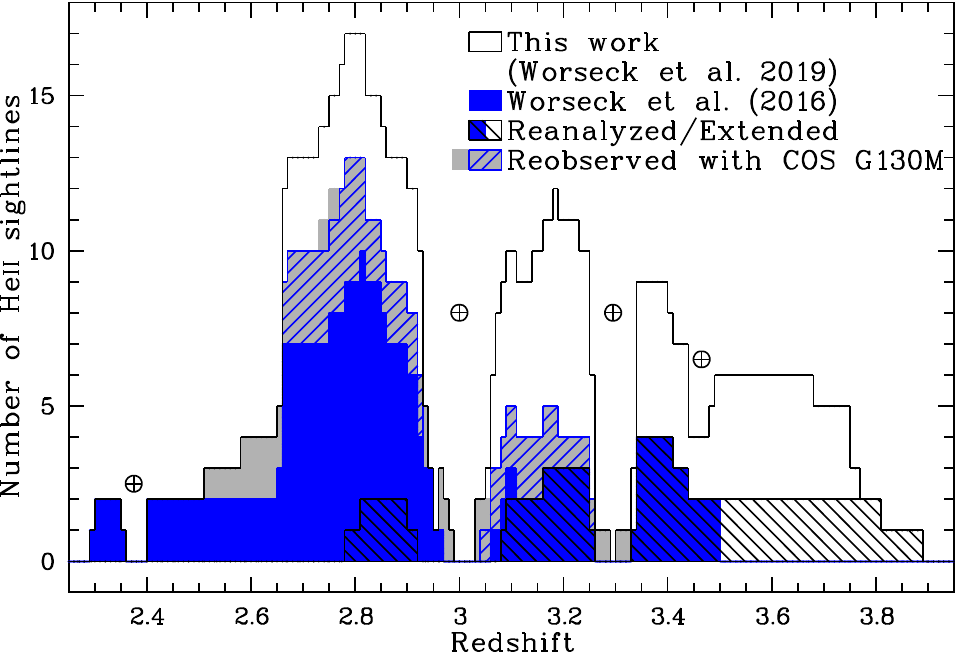}
\caption{\label{fig:he2zpath}
Redshift coverage of the present \ion{He}{2} quasar sample (black) compared to our previous study at $z<3.5$
(blue; \citetalias{worseck16}). The black hatched histogram shows the redshift coverage of four $z_\mathrm{em}>3.4$
quasars with reanalyzed \textit{HST}/COS data. Four quasars have been reobserved with the
\textit{HST}/COS G130M grating, extending their redshift coverage (gray vs.\ blue hatched).
Earth symbols mark redshift ranges impacted by geocoronal emission.
}
\end{figure}

Six out of the eight targeted FUV-bright $z_\mathrm{em}>3.1$ quasars discovered in HE2QS do not show
optically thick low-$z$ \ion{H}{1} Lyman limit absorbers, and qualify for a quantitative analysis of the
\ion{He}{2} Ly$\alpha$ absorption along their sightlines. Figure~\ref{fig:he2newspc} shows their
\textit{HST}/COS G140L spectra together with our G140L follow-up spectra of two known \ion{He}{2}-transparent
SDSS quasars and our reductions of recent archival G130M data of four \ion{He}{2} sightlines.
Blueward of \ion{He}{2} Ly$\alpha$ and the proximity zone of the background quasar \citep[e.g.][]{zheng15,khrykin16}
we observe the redshift evolution of the intergalactic \ion{He}{2} Ly$\alpha$ absorption.
At $z>3.5$ the \ion{He}{2} absorption is ubiquitously saturated, with the exception of an isolated \ion{He}{2}
transmission spike at $z=3.582$ in the sightline toward HE2QS~J1630$+$0435. These spikes become more
frequent and prominent at $3<z<3.3$, evolving into coherent structures of alternating patchy absorption at $z<3$.
At $z\la 2.8$ the G130M spectra probe the emerging \ion{He}{2} Ly$\alpha$ forest at the end of \ion{He}{2} reionization. 

Figure~\ref{fig:he2zpath} compares the \ion{He}{2} Ly$\alpha$ redshift coverage of the eight new science-grade
\ion{He}{2} sightlines reported here to the 17 sightlines presented in \citetalias{worseck16}. Wavelength ranges
with residual geocoronal emission, higher-order \ion{He}{2} Lyman series absorption, interloping saturated
\ion{H}{1} Lyman continuum absorption, and the proximity zones of the background quasars have been excluded.
Together with the two $z>3.5$ archival sightlines discussed in \citetalias{worseck16} and reanalyzed here,
our \textit{HST} Cycle~20 and 22 programs provide the first statistically meaningful sample of 6 science-grade
(continuum S/N$\simeq 4$) \ion{He}{2} sightlines at $z>3.5$.
The total redshift pathlength $g\left(z>3.5\right)=\sum_{z>3.5}\Delta z$ is $1.65$. In \citetalias{worseck16}
we focused on $z<3.5$ due to the limited sample size and decreasing instrument sensitivity at higher redshifts,
providing the first statistical analysis of \ion{He}{2} absorption at $3<z<3.5$ and a sizeable sample of
$\sim 10$ \ion{He}{2} sightlines at $2.7<z<3$. With four new $z_\mathrm{em}>3.5$ quasars and a quasar at
$z_\mathrm{em}=3.44$ we more than double the redshift pathlength at $3.3<z<3.5$ from $0.50$ to $1.16$.
At redshifts $3<z<3.3$ the increase is even larger ($g\left(3<z<3.3\right)=2.04$ vs.\ $0.83$ in \citetalias{worseck16})
due to the two additional $z_\mathrm{em}\simeq 3.3$ \ion{He}{2}-transparent quasars discovered in HE2QS.

The COS G130M spectra of the four previously known \ion{He}{2} sightlines provide higher sensitivity to probe
saturated \ion{He}{2} absorption than their existing low-resolution COS and STIS spectra.
At G130M resolution ($R\sim 15,000$) most \ion{He}{2} transmission spikes are resolved, enabling detailed studies
of the small-scale patchiness of the \ion{He}{2} absorption. Moreover, the eight times higher dispersion of the G130M
compared to the G140L results in an eight times narrower spectral range contaminated by geocoronal emission,
allowing one to probe \ion{He}{2} absorption in otherwise inaccessible redshift ranges at $z\sim 3$ and $z\sim 3.3$.
In addition, the G130M 1222\,\AA\ setup extends the spectral coverage to $z<2.66$, i.e.\ outside the spectral range
of the G140L 1105\,\AA\ setup.

\section{The Redshift Evolution of the \ion{He}{2} Effective Optical Depth}
\label{sect:he2tauevol}

\subsection{Measurement Technique}
\label{sect:he2tautech}

To quantify the level of \ion{He}{2} absorption in our mostly Poisson-count-limited spectra we employed
the same technique as in \citetalias{worseck16}, which we briefly summarize here.
We computed the effective optical depth $\mteff=-\ln{\left<f_\lambda/E_\lambda\right>}$,
where $f_\lambda$ is the observed flux density, $E_\lambda$ is the extrapolated quasar continuum,
and $\left<\,\right>$ denotes the average taken over a redshift range $\Delta z$.
As in \citetalias{worseck16} we adopted regular redshift bins of common size $\Delta z=0.04$,
corresponding to $\simeq 44$ ($\simeq 30$) cMpc at $z=2.7$ ($z=3.8$) in our assumed cosmology.
The chosen bin size mainly preserves the sensitivity to high \teff\ values at $z>3$, while the fixed
bin centers ensure an objective comparison of the sightlines without focusing on individual features.
Moreover, large-scale structures are not highly correlated on $\simeq 40$\,cMpc scales,
such that the measurements along a sightline can be treated as
independent\footnote{
On smaller scales the density field and possibly also the radiation field exhibit progressively larger correlations.
Repeating the statistical comparison to models of the \ion{He}{2}-ionizing background (Section~\ref{sect:he2uvb})
with a scale $\Delta z=0.01$ -- but accounting for the expected correlations by sampling sightlines with four
consecutive $\Delta z=0.01$ bins per $\Delta z=0.04$ segment -- does not change our results.}.
For each sightline, redshift bins with incomplete spectral coverage $g\left(z\right)$ were excluded
(Figure~\ref{fig:he2zpath}), yielding a sample of 206 $\Delta z=0.04$ redshift bins.

In each redshift bin, \teff\ was computed by maximizing the Poisson likelihood function
\begin{equation}\label{eq:likelihood}
L = \prod_{j=1}^{k}\frac{\left(S_j+B_j\right)^{N_j}e^{-\left(S_j+B_j\right)}}{N_j!}
\end{equation}
of $k$ contiguous pixels with an integer number of registered counts $N_j$, the post-processed 
multi-component background $B_j$, and the unknown signal $S_j$.
The signal was modeled as a constant in \ion{He}{2} transmission,
converted to non-integer source counts via the pixel exposure time $t_j$,
the extinction-corrected flux calibration curve $C_j$, and the continuum $E_j$ as
\begin{equation}
S_j=t_jC_jE_je^{-\tau_\mathrm{eff}}\quad.
\end{equation}
For our rebinned COS spectra $\Delta z=0.04$ corresponds to $k=51$ G140L
pixels, and 406 (305) pixels in the standard (1222\,\AA) G130M settings, respectively.
Confidence intervals ($1\sigma$, $68.26$\% confidence) were computed via
ordering the Poisson likelihood ratio \citep{feldman98}.
If the \ion{He}{2} transmission was formally negative or if the upper confidence limit included $\mteff=\infty$
we adopted the $1\sigma$ lower limit on \teff\ as a proxy for the sensitivity limit of our measurement,
computed from 200,000 Poisson deviates of the background at zero source flux.
Likewise, we estimated the significance of each \teff\ measurement by computing the probability
\begin{equation}\label{eq:prob}
P\left(>N|B\right) = 1-\sum_{k=0}^{N}\frac{B^k e^{-B}}{k!}
\end{equation}
of having measured more than $N=\sum_j N_j$ counts just from the total background $B=\sum_j B_j$.
Very small $P$ values correspond to a statistically significant signal ($N\gg B$), whereas downward
Poisson background fluctuations with $N\ll B$ result in $P\rightarrow 1$.
The $1\sigma$ lower limit on \teff\ corresponds to $P=0.1587$.

\begin{figure*}[t]
\includegraphics[width=\textwidth]{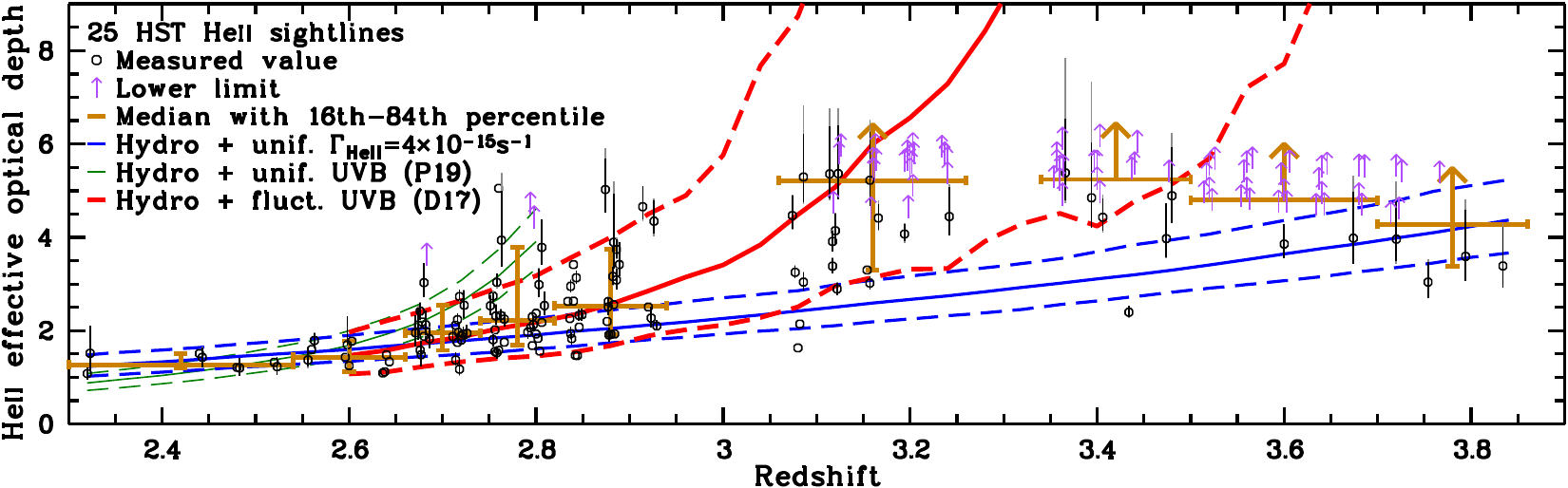}
\caption{\label{fig:he2tau_model}
\ion{He}{2} effective optical depth \teff\ vs.\ redshift for 25 \ion{He}{2} sightlines in identical redshift bins of
$\Delta z=0.04$ ($\approx 40$\,cMpc at $z=3$), measured in the newly observed spectra (Fig.~\ref{fig:he2newspc})
or previously reported in \citetalias{worseck16}.
Measurements are plotted as black circles with error bars distinguishing statistical errors due to Poisson count
statistics (black, double-sided $1\sigma$ errors corresponding to a confidence level of $68.26$\%) and estimated
systematic errors from background uncertainties (gray). For clarity, the data are plotted slightly offset with
respect to the identical bin centers and total error bars smaller than the symbol size have been omitted.
If the upper confidence limit includes $\mteff=\infty$ or if the mean transmission is formally negative ($P>0.1587$),
we adopt the $1\sigma$ lower limit on \teff\ given by our instrumental sensitivity limit (arrow symbols).
We also show the median \teff\ in subsamples in larger redshift intervals (brown), with the error bar indicating
the 16th and the 84th percentile of the \teff\ distribution in the subsample.
Overplotted are predictions from three UV background models applied to the Nyx hydrodynamic simulation output
in the same $\Delta z=0.04$ redshift bins (solid: median \teff; dashed: 16th and 84th percentile):
A model with a constant \ion{He}{2} photoionization rate $\mgheii=4\times 10^{-15}$\,s$^{-1}$ (thick blue),
an evolving but uniform $\Gamma_\mathrm{HeII}\left(z\right)$ applicable at $z\la 2.8$ \citep[thin green,][]{puchwein19},
and a fluctuating UV background model \citep[thick red,][]{davies17}.
}
\end{figure*}

For each \teff\ measurement, background subtraction errors were incorporated as a systematic error
estimated by Monte Carlo simulations, drawing 200,000 Gaussian deviates from the background
including its estimated relative uncertainty of 2--14\% depending on the orbit parameters and the
environmental conditions during the observations \citepalias{worseck16}. With the inferred \teff\ for
the modified background we generated a mock data sample by drawing from the Poisson distribution
of the background and the inferred signal. Measuring \teff\ on these mock samples yielded an estimate
of the total error from statistical Poisson shot noise and systematic background error.
Accounting for the background uncertainties marginally increases the fraction of sensitivity lower limits
on \teff\ from $34.5$\% to $36.9$\%\ in the sample. Significant detections with $P\ll 0.01$ are not
affected by background uncertainties, and because of the object-to-object variations in the background
uncertainties we did not compute formal errors on $P$. Given the strong \ion{He}{2} absorption
($\mteff\ga 1.5$), absorption lines from species other than \ion{He}{2} and the statistical continuum
error do not significantly add to the error budget, while systematic continuum error was minimized
by excluding regions with uncertain interloping low-$z$ \ion{H}{1} Lyman continuum absorption.

\subsection{Observational Results}
\label{sect:obsresults}

Figure~\ref{fig:he2tau_model} presents the 206 \teff\ measurements or limits thereof in the 25 \ion{He}{2}-transparent
sightlines of our combined sample (Table~\ref{tab:qsosample}). The 149 values for the 16 sightlines with either new
or reanalyzed spectra are reported in Appendix~\ref{sect:taueffapp}, while the remainder 
(9 sightlines marked in Table~\ref{tab:qsosample})
has been taken from \citetalias{worseck16}.
With the eight new sightlines from our \textit{HST} programs in Cycle~20 and 22, and the extended spectral coverage
of the recent COS G130M spectra we increase the $z<3.5$ sample from \citetalias{worseck16} by 63\% (168 vs.\ 103 \teff\ values).
Thirty-eight newly reported \teff\ values provide the first statistical sample of quantified \ion{He}{2} absorption at $z>3.5$.

The \ion{He}{2} effective optical depth increases with redshift, but with significant sightline-to-sightline variance at $z>2.7$
that increases with redshift \citepalias{worseck16}. To quantify this, we computed the median \teff\ and its scatter
(estimated from the range between the 16th and the 84th percentile) in larger redshift ranges, each having
$n\ga 20$ \teff\ values except at the lowest and highest redshifts. Note that for all following statistical analyses and data
modeling (Section~\ref{sect:he2uvb}) we excluded the single remaining STIS spectrum of our sample (HS~1157$+$3143),
since the STIS background uncertainties have been characterized to a lesser extent than for COS.
Given the large redshift path, the exclusion of its four $z<3$ \teff\ values does not change our results.

The results are plotted in Figure~\ref{fig:he2tau_model} in brown and are listed in Table~\ref{tab:taueffmedian}.
Table~\ref{tab:taueffmedian} also lists the error on the median that has been determined from the 16th and the 84th
percentile of the distribution of the median \teff\ in 1000 bootstrap samples.
In Table~\ref{tab:taueffstat} we give the empirical cumulative distributions $F(<\mteff)$ in these redshift ranges for the 24
\textit{HST}/COS sightlines. Lower limits on \teff\ are included, such that the cumulative distributions represent a lower
limit on the actual spread.

Redshifts $z<2.66$ are covered by a maximum of four sightlines, limiting the robustness of the estimated scatter in
\teff\ and the error on its median. At $2.66<z<2.74$ we observe a narrow \teff\ distribution around $\mteff\simeq 2$
sampled by 24 $\Delta z=0.04$ bins in 13 sightlines.
Given that the subsample contains only one \teff\ sensitivity limit, the scatter in \teff\ is well determined.
At $z>2.74$ the median \teff\ gradually increases and its scatter doubles with respect to the data at $z\simeq 2.7$.
Our new \ion{He}{2} sightlines have the lowest \teff\ measured in 13 COS sightlines at $z=2.84$
($\mteff=1.47\pm 0.05$ toward HE2QS~J0916$+$2408 and $\mteff=1.47^{+0.07}_{-0.06}$ toward HE2QS~J0233$-$0149).
At the other extreme we obtain $\mteff>4.19$ at $z=2.80$ in the HE2QS~J2149$-$0859 sightline, confirming the result of
\citetalias{worseck16} that the sightline-to-sightline variance in \ion{He}{2} absorption increases between $z\simeq 2.7$ and $z\simeq 2.8$.

\tabletypesize{\footnotesize}
\begin{deluxetable}{cccrrr}
\renewcommand{\arraystretch}{1.0}
\tablecaption{\label{tab:taueffmedian}50th, 16th and 84th percentiles of the \teff\ distribution in subsamples of $n$ values
in $m$ \textit{HST}/COS \ion{He}{2} sightlines}
\tablehead{
\colhead{\hspace{0.1in}$\Delta z$}\hspace{0.1in} &
\colhead{\hspace{0.1in}$n$}\hspace{0.1in} &
\colhead{\hspace{0.1in}$m$}\hspace{0.1in} &
\colhead{\hspace{0.1in}$\tau_\mathrm{eff,50}$}\hspace{0.1in} &
\colhead{\hspace{0.1in}$\tau_\mathrm{eff,16}$}\hspace{0.1in} &
\colhead{\hspace{0.1in}$\tau_\mathrm{eff,84}$}\hspace{0.1in}
}
\startdata
$2.30$--$2.54$	&8	&2	&$1.27^{+0.10}_{-0.06}$	&$1.20$	&$1.51$\\
$2.54$--$2.66$	&11	&4	&$1.43^{+0.17}_{-0.10}$	&$1.12$	&$1.78$\\
$2.66$--$2.74$	&24	&13	&$1.95^{+0.09}_{-0.06}$	&$1.58$	&$2.54$\\
$2.74$--$2.82$	&28	&15	&$2.21^{+0.10}_{-0.15}$	&$1.68$	&$3.78$\\
$2.82$--$2.94$	&28	&13	&$2.53^{+0.10}_{-0.20}$	&$1.92$	&$3.74$\\
$3.06$--$3.26$	&42	&13	&$>5.22$				&$3.30$	&\nodata\\
$3.34$--$3.50$	&23	&9	&$>5.24$				&$>4.68$	&\nodata\\
$3.50$--$3.70$	&29	&6	&$>4.80$				&$>4.47$	&\nodata\\
$3.70$--$3.86$	&9	&5	&$>4.28$				&$3.38$	&\nodata\\
 \enddata
\end{deluxetable}

\begin{figure*}[t]
\includegraphics[width=\textwidth]{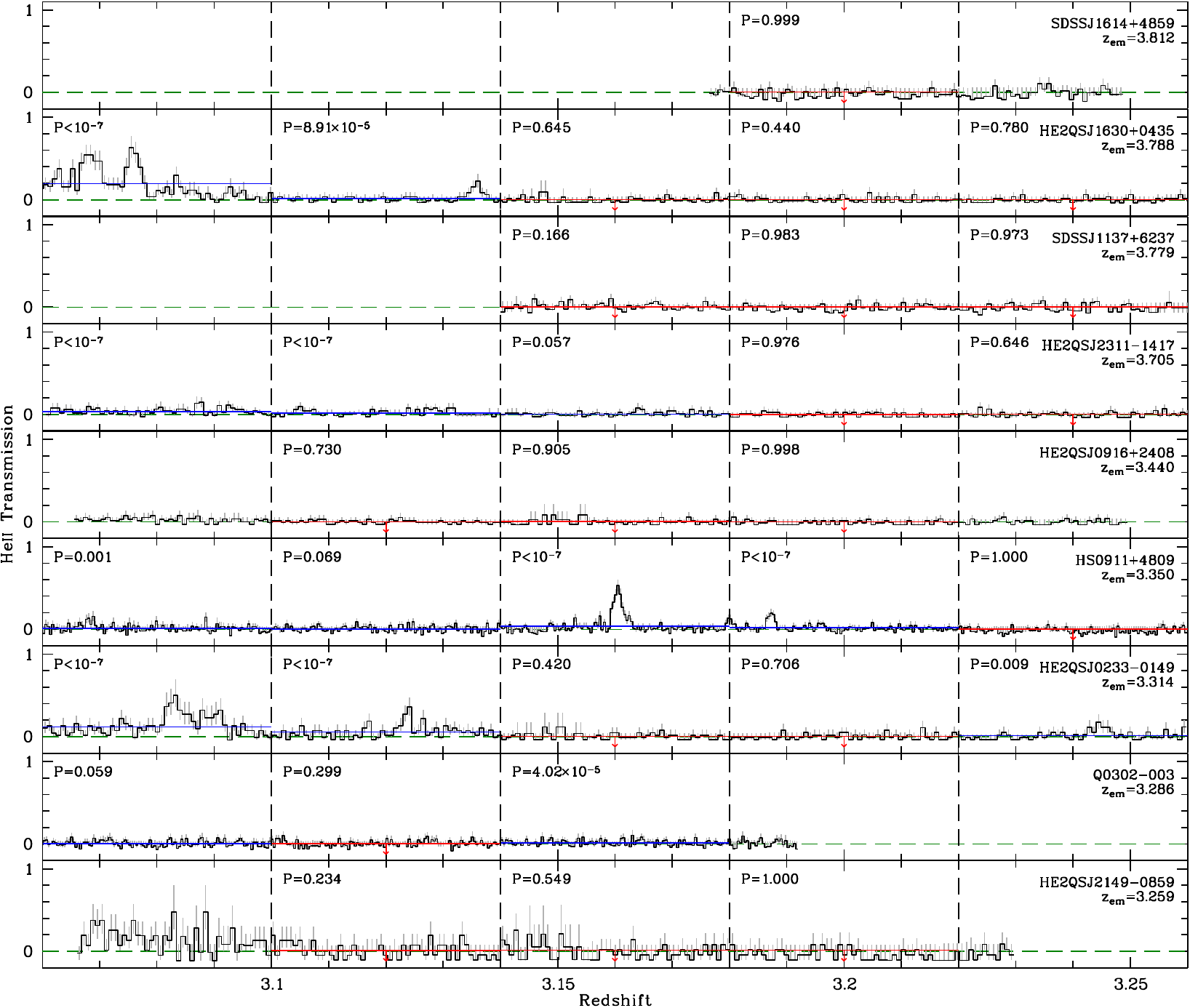}
\caption{\label{fig:he2spczoom1}
\ion{He}{2} Ly$\alpha$ transmission spectra of nine quasar sightlines at $z\simeq 3.16$ (black)
with Poisson errors for individual pixels (gray). We only show spectra that have not been analyzed
in \citetalias{worseck16} (Table~\ref{tab:qsosample}).
The COS G130M spectra of HS~0911$+$4809 and Q~0302$-$003 have been binned to
$\simeq 0.12$\,\AA\,pixel$^{-1}$ ($0.7$--$0.9$ pixels per resolution element).
The COS G140L spectra of the remaining 7 sightlines have been binned to $\simeq 0.24$\,\AA\,pixel$^{-1}$
(3 pixels per resolution element). The proximity zones of the background quasars and regions contaminated
by geocoronal Ly$\alpha$ and residuals of geocoronal \ion{O}{1} $\lambda 1304$\,\AA\ emission have been omitted.
The dashed horizontal lines mark the zero level, while the vertical lines indicate our regular $\Delta z=0.04$ bins
for our \teff\ measurements. The \teff\ measurements are overplotted, converted to the mean \ion{He}{2} transmission,
distinguishing robustly measured values (blue; measurement error comparable to the line thickness) and sensitivity
limits (red with arrow symbols). Incompletely covered redshift bins were not considered. For every redshift bin we
indicate the probability $P$ that the measured mean \ion{He}{2} transmission is consistent with a
Poisson fluctuation of the background.
}
\end{figure*}

At $z>3$ the \teff\ distribution is affected by frequent sensitivity limits to high values $\mteff\simeq 5$.
In each $z>3$ subsample the median \teff\ corresponds to a sensitivity limit, and therefore the true median
\teff\ is underestimated. At $3.06<z<3.26$ the 16th percentile of the distribution is still well determined due to
the remaining tail toward low \teff\ and the large sample size. The tail toward low \teff\ is more than a factor of
three larger than at $z\simeq 2.9$, indicating a significant increase in the \teff\ scatter at $z>3$.
Figure~\ref{fig:he2spczoom1} shows the spectra of nine \ion{He}{2} sightlines covering $3.06<z<3.26$,
seven of which are new, while two (Q~0302$-$003 and HS~0911$+$4809) have significantly deeper COS G130M
data than the low-resolution STIS and COS data analyzed in \citetalias{worseck16}. Several sightlines show
significant \ion{He}{2} transmission spikes that are much narrower ($\Delta z\la 0.01$ or $\Delta v\la 720$\,km\,s$^{-1}$)
than our chosen bin size $\Delta z=0.04$, but limited S/N prevents a robust characterization of their size distribution.
Since we observe strong saturation between them, these transmission spikes are robust to residual background
subtraction systematics. In the HS~0911$+$4809 sightline the new G130M data confirm the three most significant $z>3$
transmission features ($P<0.01$) detected in the G140L data \citepalias{worseck16}, although the $z\simeq 3.07$ spike
is weaker, probably due Poisson noise in the G140L data. In our new sightlines we find
the three lowest \teff\ values at $3.06<z<3.26$ (HE2QS~J1630$+$0435: $\mteff=1.63\pm 0.06$ at $z=3.08$;
HE2QS~J0233$-$0149: $\mteff=2.14^{+0.08}_{-0.09}$ at $z=3.08$ and $\mteff=2.89^{+0.14}_{-0.13}$ at $z=3.12$).
HE2QS~J2149$-$0859 and SDSS~J1614$+$4859 -- the two spectra that were observed mostly during orbital
day -- show significantly negative transmission on large scales, indicating that the simple model developed in
\citetalias{worseck16} may overestimate the scattered geocoronal Ly$\alpha$ emission by $\sim 20$\%,
but our results remain unaffected by this uncertainty\footnote{We tested this by varying the estimated scattered light
for the 22 redshift bins at $3.06<z<3.26$ covered by COS G140L data that have insignificant \ion{He}{2} transmission
(i.e.\ $P\ge 0.1587$ and an adopted \teff\ sensitivity limit). A 20\% reduction in the scattered light results in nominally
positive transmission in about half of them (12/22), but in only three of them the detection is significant ($P<0.1587$).
This confirms that our procedure to adopt the $1\sigma$ lower limit on \teff\ guards against possible background systematics.}.

\begin{figure*}[t]
\includegraphics[width=\textwidth]{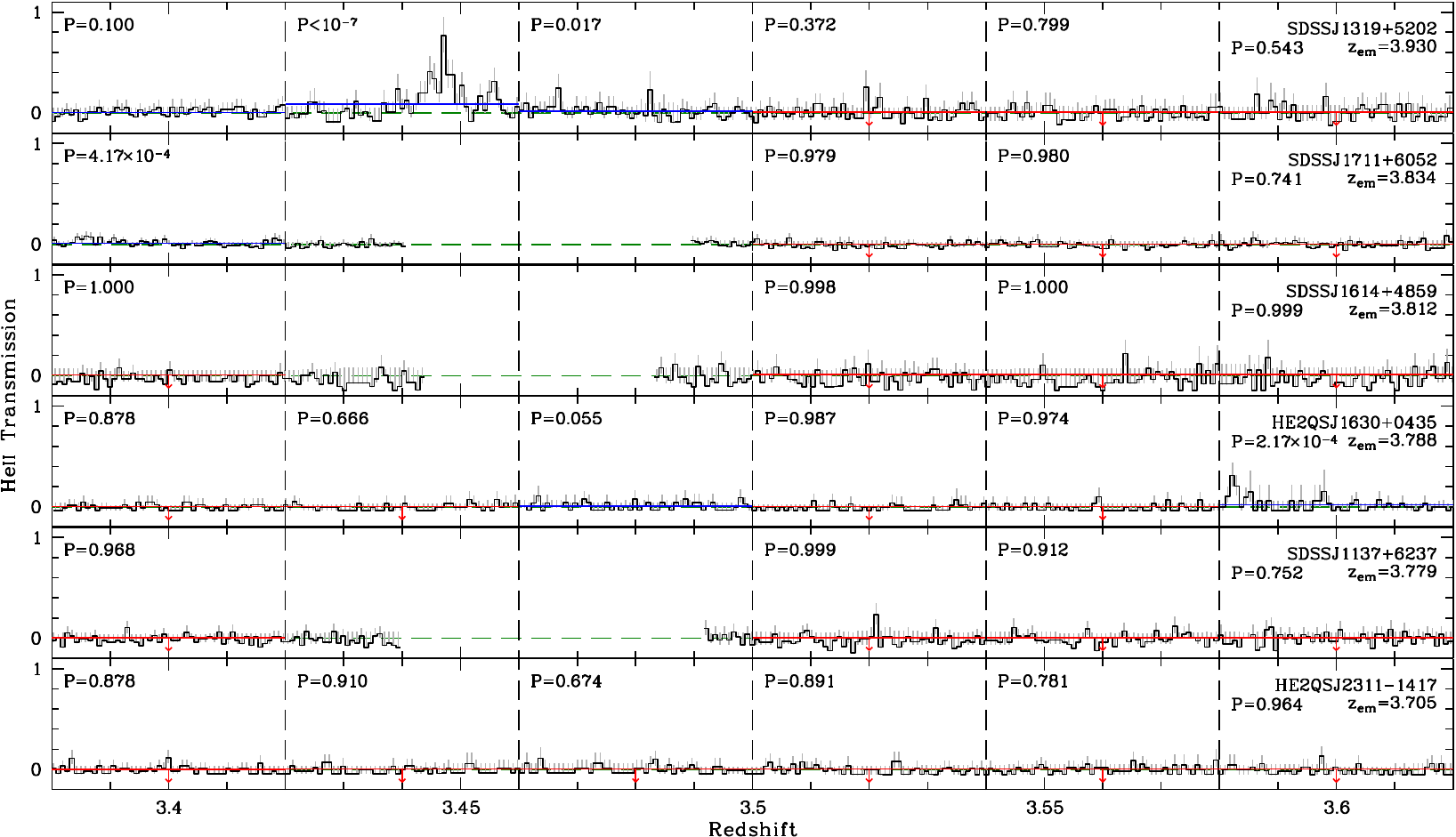}
\includegraphics[width=\textwidth]{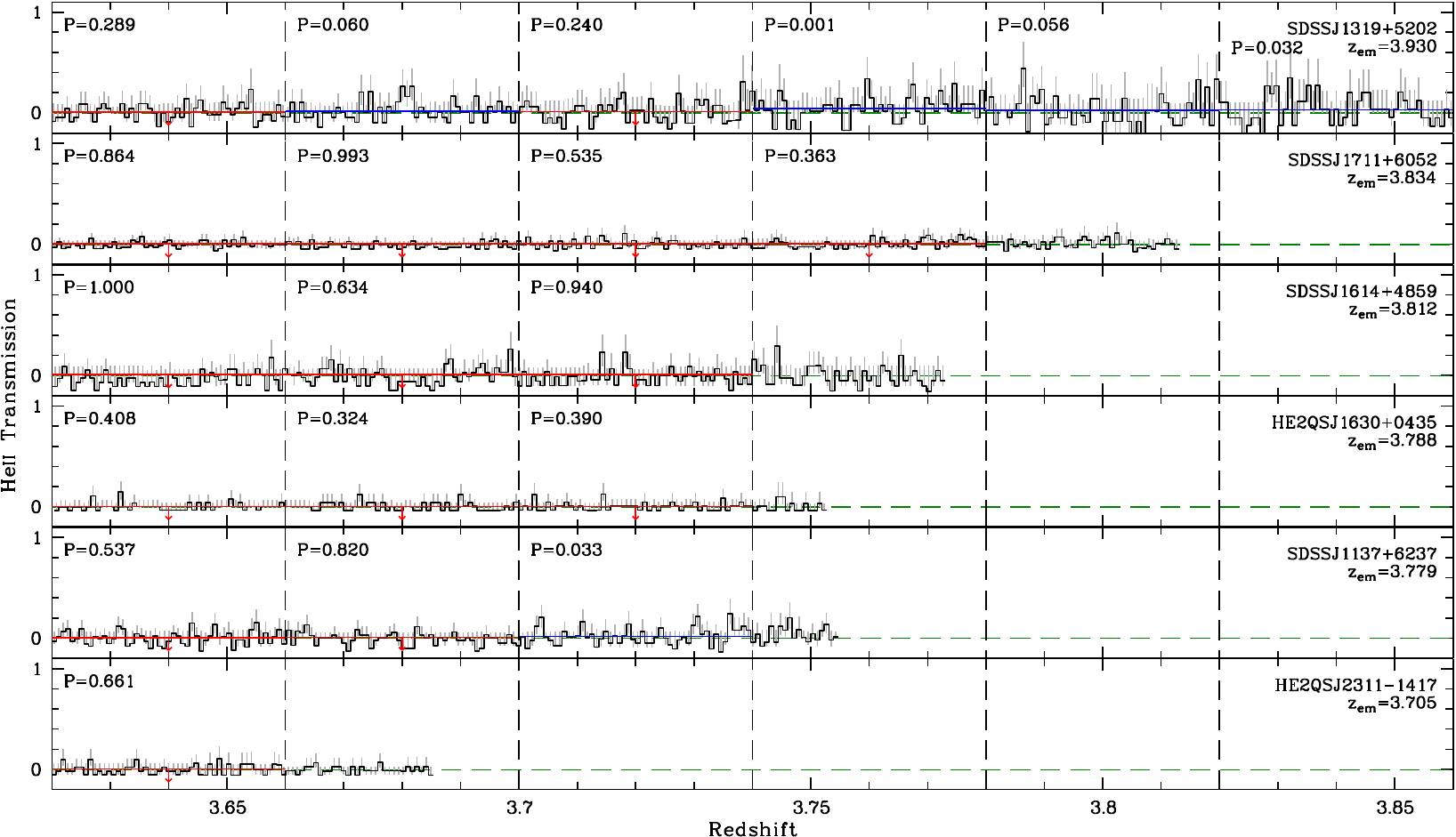}
\caption{\label{fig:he2spczoom2}
Similar to Figure~\ref{fig:he2spczoom1}, but showing the redshift ranges $3.36<z<3.62$ (upper panels)
and $3.62<z<3.86$ (lower panels) for the 6 $z_\mathrm{em}>3.5$ quasars (Table~\ref{tab:qsosample}).
The spectra have been binned to 2--3 pixels per resolution element ($\simeq 0.24$\,\AA\,pixel$^{-1}$).
Occasional geocoronal \ion{O}{1}] $\lambda 1356$\,\AA\ emission has been omitted.
}
\end{figure*}

At $z>3.3$ the tail toward well determined low \teff\ values gradually disappears. The fraction of robust ($P<0.1587$)
\teff\ measurements decreases from $42.9$\% at $3.06<z<3.26$ to $26.1$\% at $3.34<z<3.5$ at similar sensitivity,
suggesting that the actual median \teff\ is significantly higher than our sensitivity limit. With more than twice
the pathlength than in \citetalias{worseck16} we obtain a significantly lower fraction of robust \teff\ measurements
at $3.34<z<3.5$ ($26.1$\% vs.\ 50\%), indicating strong sightline-to-sightline variance in the \ion{He}{2} absorption.
Figure~\ref{fig:he2spczoom2} focuses on the $z>3.3$ \ion{He}{2} absorption spectra of the 6 $z_\mathrm{em}>3.5$
quasars including the reanalyzed sightlines of SDSS~J1319$+$5202 and SDSS~J1711$+$6052.
In \citetalias{worseck16} we reported on the strong transmission spike at $z\simeq 3.45$ in the SDSS~J1319$+$5202 sightline.
A narrower and weaker, but still significant ($P=2.17\times 10^{-4}$) spike occurs in the sightline to
HE2QS~J1630$+$0435 at $z=3.582$. Note that $P$ is even smaller ($P<10^{-7}$ eventually limited by background uncertainty)
across the observed width of the spike ($3.58<z\la 3.59$, $\Delta v\la 650$\,km\,s$^{-1}$).
At $z>3.4$ several sightlines show unresolved narrow transmission features of single $0.24$\,\AA\ pixels.
Although formally significant ($P\la 0.001$ for single pixels), these could also be rare background events,
and more or higher-resolution data are required to confirm them. Otherwise most spectra are saturated on
large scales or appear to be affected by background oversubtraction (SDSS~J1614$+$4859).
At $z>3.5$ we obtain sensitivity limits $4.3\la\mteff\la 5.4$ in $84.2$\% of the pathlength, mostly due to the
declining sensitivity of the G140L grating (Figure~\ref{fig:he2tau_model}).

\tabletypesize{\footnotesize}
\begin{deluxetable}{ccccc}
\tablecaption{\label{tab:taueffstat} Cumulative distributions of \teff\ in redshift ranges}
\tablehead{
\colhead{\hspace{0.13in}$z_\mathrm{min}$}\hspace{0.13in} &
\colhead{\hspace{0.13in}$z_\mathrm{max}$}\hspace{0.13in} &
\colhead{\hspace{0.13in}\teff}\hspace{0.13in} &
\colhead{\hspace{0.13in}$F(<\mteff)$}\hspace{0.13in} &
\colhead{\hspace{0.13in}Flag\tablenotemark{a}}\hspace{0.13in}
}
\startdata
$2.30$ &$2.54$ &$1.08$ &$0.000$ &$0$\\
$2.30$ &$2.54$ &$1.20$ &$0.125$ &$0$\\
$2.30$ &$2.54$ &$1.21$ &$0.250$ &$0$\\
$2.30$ &$2.54$ &$1.23$ &$0.375$ &$0$\\
$2.30$ &$2.54$ &$1.32$ &$0.500$ &$0$\\
\enddata
\tablenotetext{a}{Flag indicating whether the \teff\ value is a measurement (0) or a $1\sigma$ upper limit (1).}
\tablecomments{Table~\ref{tab:taueffstat} is published in its entirety in the machine-readable format.
      A portion is shown here for guidance regarding its form and content.}
\end{deluxetable}

An exception to this trend is the SDSS~J1319$+$5202 sightline, in which low-level \ion{He}{2} transmission
is detected in 3/4 $\Delta z=0.04$ bins at $z>3.7$. Combining these three redshift bins, the probability that the
transmission at $3.74<z<3.86$ arises from a Poisson background fluctuation is very small ($P=1.2\times 10^{-4}$).
Large-scale background undersubtraction in this particular sightline is unlikely due to insignificant or slightly negative
transmission in 5/9 $z>3.5$ redshift bins. Specifically, a factor of three increase in the modeled scattered geocoronal
Ly$\alpha$ emission or a 17\% increase in the dark current would render the transmission at $3.74<z<3.86$
insignificant ($P=0.16$), but would result in unphysical strongly negative transmission in adjacent redshift bins.
Also, the required increases in these background components are much larger than their estimated relative uncertainty
($\sim 18$\% for the scattered light and $1.2$\% for the dark current). The shadow data of SDSS~J1319$+$5202
($58.6$\% of the total exposure at a scattered light level reduced by $\simeq 80$\%) give consistent results within
the necessarily larger errors ($P=3.6\times 10^{-3}$ at $3.74<z<3.86$). We conclude that the low-level \ion{He}{2}
transmission in the SDSS~J1319$+$5202 sightline is likely real. Similar to lower redshifts, this transmission may be
dominated by narrow spikes, but the data quality is particularly low due to the degrading sensitivity of the COS G140L
grating and the low continuum flux of the background quasar. Note that the detected transmission at $z<3.86$
is unlikely to be affected by the background quasar, because the proximity zones of $z=3.9$ quasars rarely
exceed 50\,cMpc for a range of quasar lifetimes and ambient IGM \ion{He}{2} fractions \citep{khrykin16,khrykin19}.

In summary, the \ion{He}{2} effective optical depth increases from $\mteff\simeq 2$ at $z\simeq 2.7$ to $\mteff>5$
at $z>3$ with an intrinsic scatter that increases with redshift and indications of large-scale sightline-to-sightline variance.
At $z>3.5$ the 6 robust values at $\mteff\la 4$ probe the low tail of the intrinsic \teff\ distribution, but our results are
still limited by sample size and sightline variance. Narrow $\Delta z\la 0.01$ \ion{He}{2} transmission patches
may exist even at $z>3.5$, only a few of which are robustly detected at the current data quality.

\section{The \ion{He}{2}-Ionizing Background}
\label{sect:he2uvb}

After having measured the redshift evolution of the \ion{He}{2} effective optical depth and its scatter,
we seek to compare our measurements to realistic mock \teff\ distributions from outputs of numerical
hydrodynamical IGM simulations. Considering the large fraction of robust \teff\ measurements out to $z=3.8$
and the saturation limit of the data at \ion{He}{2} fractions of a few percent,
we take the simplifying approach that our measurements probe the post-reionization IGM that is in
photoionization equilibrium with a \ion{He}{2}-ionizing background that may exhibit spatial fluctuations on large scales
($\ga 20$\,cMpc). Comparisons to detailed numerical radiative transfer simulations
of quasar-driven \ion{He}{2} reionization are left for future work.

\subsection{Nyx Cosmological Hydrodynamical Simulation}

To compare our \teff\ measurements to predictions for a range of UV background models we used outputs of a cubic
($100h^{-1}$\,cMpc)$^3$ hydrodynamical simulation performed with the Nyx code \citep{almgren13,lukic15}.
Eulerian hydrodynamics of the baryons was computed on a fixed Cartesian grid with 4096$^3$ cells, and the evolution of 
dark matter was followed with 4096$^3$ Lagrangian particles. The spatial resolution of $25h^{-1}$\,ckpc
is sufficient to resolve the \ion{H}{1} and \ion{He}{2} Ly$\alpha$ forest, and to reach convergence in \ion{H}{1} Ly$\alpha$
transmission statistics to percent-level accuracy \citep{lukic15}. The simulation was run without adaptive mesh refinement,
star formation or thermal feedback prescription, and is therefore optimized to resolve underdense regions in the IGM
where much of the \ion{He}{2} Ly$\alpha$ absorption originates \citep{croft97,theuns98,mcquinn09a,compostella13}.
Convergence tests on ($20h^{-1}$\,cMpc)$^3$ volumes indicate that our fiducial simulation overestimates the \ion{He}{2} effective optical depth by $\sim 5$\% (Appendix~\ref{sect:convapp}).
Primordial ideal gas chemistry (mass fractions $X=0.76$ for H and $Y=0.24$ for He) was computed assuming
photoionization equilibrium in the optically thin limit, using the spatially uniform but redshift-dependent photoionization
and photoheating rates by \citet{haardt12}. 

We used the density, velocity and temperature fields of the seven outputs in the redshift range of interest
($z_\mathrm{sim}=2.2$, $2.4$, $2.5$, $2.6$, 3, $3.5$, 4). We extracted $\Delta z=0.08$
skewers along one axis of the simulation volume, starting from random positions and applying periodic boundary conditions.
To match the regular $\Delta z=0.04$ grid of the \teff\ measurements we rescaled the density field along each skewer as
\begin{equation}
\rho\left(z\right)=\rho_\mathrm{sim}\left(\frac{1+z}{1+z_\mathrm{sim}}\right)^3\quad,
\end{equation}
and extracted 1000 skewers from the closest two Nyx outputs spaced by $\Delta z_\mathrm{sim}$. For each redshift
bin centered on $z_\mathrm{bin}$ the number of skewers from the two considered Nyx outputs was weighted with a
factor $w=1-\left|z_\mathrm{sim}-z_\mathrm{bin}\right|/\Delta z_\mathrm{sim}$.
Ly$\alpha$ optical depths of \ion{H}{1} and \ion{He}{2} were computed accounting for peculiar velocities and thermal
broadening (see e.g.\ \citealt{lukic15} for details). Each $\Delta z=0.08$ skewer comprised 2431 pixels, corresponding
to a pixel scale of 3\,km\,s$^{-1}$ at $z=2.3$ and 2\,km\,s$^{-1}$ at $z=3.8$. To approximately account for the lack of
strong \ion{H}{1} Lyman limit systems in the observed \ion{He}{2} sightlines (\citealt{compostella14}; \citetalias{worseck16})
we discarded the 4--7\% of skewers in which pixels reached $\tau_\mathrm{HI}>3000$, corresponding to \ion{H}{1}
column densities $N_\mathrm{HI}\ga 10^{17.1}$\,cm$^{-2}$. In the observed \ion{He}{2} sightlines the $N_\mathrm{HI}$
limit may be somewhat higher if there are no strong Lyman limit systems at $z\ll z_\mathrm{bin}$ along the lightcone.
Eventually, 900 of the 1000 skewers at each $z_\mathrm{bin}$ were kept.

\subsection{UV Background Models Applied to the Nyx Skewers}

\subsubsection{Uniform UV Background Models}
A posteriori, we linearly rescaled the \ion{He}{2} Ly$\alpha$ optical depths to match the observed range in \teff.
This rescaling adjusts the \ion{He}{2} photoionization rate \gheii\ in the optically thin limit ($\tau_\mathrm{HeII}\propto\mgheii^{-1}$),
but in reality also depends on the thermal state of the gas during and after \ion{He}{2} reionization that the Nyx simulation
does not follow self-consistently. We explore the impact of this approximation in Section~\ref{sect:uncertainties} below.
Changes in the ionization rate by a factor $\sim 2$ result in percent level errors in the transmission probability
distribution and power spectrum at $z>2$ \citep{lukic15,onorbe17}. The much larger systematic effects of a fixed thermal
history under photoionization equilibrium are explored in Section~\ref{sect:uncertainties}. Across the redshift range
of interest we applied a range of constant \ion{He}{2} photoionization rates $\mgheii=10^{-15.3}$--$10^{-14}$\,s$^{-1}$
to the $\Delta z=0.08$ skewers. Additionally, we mimicked a totally opaque IGM by multiplying all
\ion{He}{2} optical depths by a factor of 1000 (i.e.\ $\mgheii\approx 0$).

We also considered the spatially uniform but redshift-dependent UV background synthesis model by \citet{puchwein19}
that consistently traces the overall thermal and ionization history via one-dimensional radiative transfer in
a three-phase medium. Since the Nyx simulation assumes photoionization equilibrium, we used the \citet{puchwein19}
equivalent-equilibrium \ion{He}{2} photoionization rates, interpolated onto our redshift bins.
For simplicity we did not rescale the Nyx gas temperatures to the to those predicted by \citet{puchwein19},
such that $\mteff(z)$ is not predicted self-consistently. We mainly aim to check whether a redshift-dependent uniform
\gheii\ is consistent with our data.

\subsubsection{The \citetalias{davies17} Fluctuating UV Background Model}

Recently, \citetalias{davies17} presented a three-dimensional model of a spatially fluctuating
\ion{He}{2}-ionizing background produced by scarce bright $z\sim 3$ quasars and a spatially varying mean free path
of \ion{He}{2}-ionizing photons. The model broadly reproduces the large \teff\ variations reported in \citetalias{worseck16}.
Hence, spatial fluctuations of the \ion{He}{2}-ionizing UV background in a post-reionization IGM may be a viable
alternative explanation to still ongoing \ion{He}{2} reionization at $2.7<z<3$ (\citealt{worseck11b}; \citetalias{worseck16}).

We used the default model parameters from \citetalias{davies17}: A cubic (500\,cMpc)$^3$ volume was randomly
populated with quasars according to the \citet{hopkins07} luminosity function, each emitting isotropically at a
constant luminosity for a fixed lifetime of 50\,Myr. Approximate three-dimensional radiative transfer of
\ion{He}{2}-ionizing photons was calculated in grid cells of ($7.8$\,cMpc)$^3$, 
modestly improved compared to the ($10$\,cMpc)$^3$ resolution in \citetalias{davies17}, with outputs every 5\,Myr,
accounting for a spatially varying mean free path computed assuming local photoionization equilibrium.
The IGM was approximated as an ensemble
of randomly distributed discrete absorbers drawn from the \citet{prochaska14} column density distribution function.
While this model certainly simplifies the absorber physics and captures UV background fluctuations only on scales
larger than the mean free path ($\ga 20$\,cMpc), \citetalias{davies17} estimate that volumes $\ga$(500\,cMpc)$^3$
are required to reach convergence in the distribution of \teff\ on scales $\Delta z=0.04$ to better than 10\%\ at $z\la 3$.
Current cosmological simulations with more accurate radiative transfer still lack the required box size to capture
the large-scale fluctuations and/or the spatial resolution to resolve transmission in the \ion{He}{2} Ly$\alpha$ forest
\citep{compostella13,compostella14,laplante17}.

We applied the fluctuating UV background rates to the Nyx skewers by drawing $2.5<z<4$ lightcone skewers
from the simulation volume, and rescaling the \ion{He}{2} Ly$\alpha$ optical depths from the Nyx skewers
($\tau_\mathrm{HeII}\propto\mgheii^{-1}$) in the overlapping redshift range without interpolating
$\mgheii\left(z\right)$ between time steps.

\subsubsection{The Predicted Redshift Evolution of the \ion{He}{2} Effective Optical Depth}

In Figure~\ref{fig:he2tau_model} we compare our \teff\ measurements to predictions from the 900 skewers
per rescaled Nyx simulation snapshot, applying three UV background models described above: (1) a uniform
constant $\mgheii=4\times 10^{-15}$\,s$^{-1}$, (2) the uniform but redshift-dependent UV background model
from \citet{puchwein19}, and (3) the fluctuating UV background model from \citetalias{davies17}.
We show the redshift evolution of the median \teff\ and its scatter, estimated from the 16th and 84th percentiles
of the \teff\ distribution on our adopted scale $\Delta z=0.04$. 

Among the uniform UV background models with a constant \ion{He}{2} photoionization rate, the one with
$\mgheii=4\times 10^{-15}$\,s$^{-1}$ roughly matches the \teff\ measurements at $z\la 2.6$, i.e.\ in the
post-reionization \ion{He}{2} \lya\ forest. In this model, the shallow redshift evolution of the median \ion{He}{2}
effective optical depth from $\tau_\mathrm{eff,50}\simeq 1.3$ at $z=2.3$ to $\tau_\mathrm{eff,50}\simeq 4$
at $z=3.8$ is entirely due to density evolution in the IGM. Density fluctuations on the scale $\Delta z=0.04$
give rise to a relative scatter of $\simeq 18$\% around the median \teff\ that only slightly evolves with redshift.
The observed scatter in \teff\ is much larger, indicating redshift evolution and/or spatial fluctuations in \gheii.
Nevertheless, the predicted range in \teff\ matches the locus of the lowest measured \teff\ values at all redshifts.
These likely correspond to fully reionized regions in the IGM. However, at $z>3.5$ it is increasingly difficult to
distinguish such regions from saturated \ion{He}{2} Gunn-Peterson troughs due to density evolution in the IGM
and the sensitivity limit of the data.

\citet{puchwein19} use a quasar emissivity model and an empirical IGM absorber model to predict the redshift
evolution of the average \gheii\ and the corresponding \teff\ in the post-reionization IGM. In their model \ion{He}{2}
reionization completes at $z\simeq 2.8$, and \gheii\ increases by a factor $\sim 6$ from $z=2.75$ to $z=2.3$,
corresponding to a strong decrease in \teff\ from $\simeq 3$ to $\simeq 0.8$. Applying their $\mgheii(z)$ to the
Nyx skewers we obtain a strongly evolving $\mteff(z)$ that agrees very well (maximum relative deviation 5\%)
with the prediction from the empirical IGM absorber model used by \citet{puchwein19}.
In addition, we use the Nyx skewers to predict the scatter of \teff\ on our adopted scale $\Delta z=0.04$.
A comparison to our \teff\ measurements in Figure~\ref{fig:he2tau_model} reveals that the redshift-dependent
but spatially uniform $\mgheii(z)$ from \citet{puchwein19} significantly overpredicts the observed median \teff\
and underpredicts its scatter at $2.7\la z\la 2.8$., suggesting that \gheii\ is actually higher on average and
spatially varying. Moreover, our ubiquitously low \teff\ at $z\simeq 2.9$ (see also \citetalias{worseck16}) indicates
that \ion{He}{2} reionization finishes too late in the \citet{puchwein19} model.

\begin{figure*}[t]
\includegraphics[width=\textwidth]{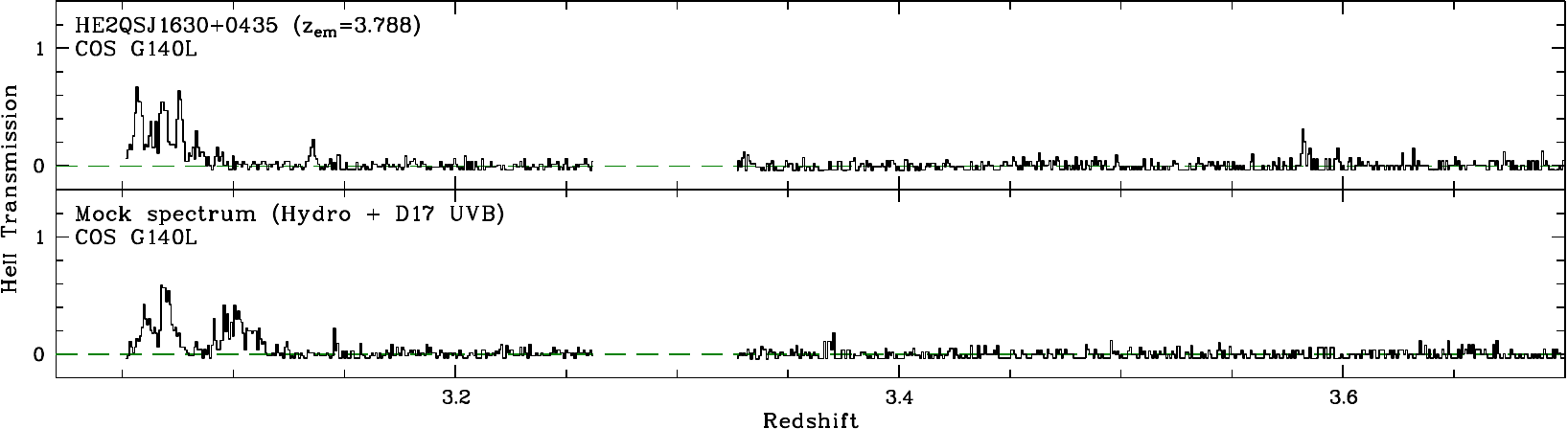}
\caption{\label{fig:he2mockreal}
Comparison of an observed COS G140L \ion{He}{2} transmission spectrum (top) and a mock spectrum (bottom).
The mock spectrum was generated by applying a lightcone realization of the fluctuating \ion{He}{2}-ionizing
background from \citetalias{davies17} to concatenated skewers from the Nyx hydrodynamical simulation at different redshifts,
and simulating the observed continuum and noise characteristics.
}
\end{figure*}

In contrast, the redshift evolution of the \ion{He}{2} effective optical depth predicted by the \citetalias{davies17}
fluctuating UV background model is in very good agreement with our measurements. At $z<3$ both the median
and the scatter in the observed \teff\ distribution are well reproduced.
Applying the spatially fluctuating \gheii\ from \citetalias{davies17} to the realistic IGM density field in the Nyx simulation result
in a consistent range in \teff\ on the considered scale $\Delta z=0.04$. In the \citetalias{davies17} model the tails of the
\teff\ distribution are driven by the volume of space far away from and very close to quasars, corresponding to regions with
low and high \gheii, respectively. At $z>3$ the model predicts a large scatter in \teff\ due to the decreasing space density
of quasars and increasing spatial fluctuations in the mean free path to \ion{He}{2}-ionizing photons.
The lowest \teff\ values at these redshifts are produced by intersected isolated quasar proximity zones
(\citealt{davies14}; \citetalias{davies17}). The rest of the volume may have substantially higher \ion{He}{2} fractions,
such that the assumption of photoionization equilibrium breaks down and a proper treatment of \ion{He}{2}
reionization becomes necessary. Because saturation in \ion{He}{2} \lya\ at $z>3$ limits constraints on the
\ion{He}{2} fraction to a few percent at most \citep{mcquinn09b,khrykin16,worseck16}, our data cannot
discriminate between ongoing \ion{He}{2} reionization and residual UV background fluctuations in a
post-reionization IGM at $z>3$. For a detailed statistical comparison, however, it is necessary to forward
model the predicted $\mteff\left(z\right)$ distribution to account for data quality and sample size.

\subsection{Generation of Realistic Mock COS Spectra}

To produce realistic mock COS spectra from the rescaled Nyx skewers we took a forward-modeling approach
similar to \citetalias{worseck16}. In each $\Delta z=0.04$ bin the contributing \textit{HST}/COS \ion{He}{2} sightlines
have a somewhat different sensitivity to high \teff\ values due to their varying exposure time, continuum level,
spectral dispersion, and time-dependent background conditions. First, the same number of skewers was drawn
randomly from the set of 900 Nyx skewers for the considered \gheii\ model and centered on the redshift bin.
The Nyx skewers were convolved with the line spread functions of the actual \textit{HST}/COS \ion{He}{2} spectra,
accounting for the different central wavelengths and COS LPs. 
From the initial $\Delta z=0.08$ skewers only the
central $\Delta z=0.04$ part was kept to account for the wings of the COS G140L line spread function.
With the specific parameters of the \textit{HST}/COS spectra (quasar continuum, grating sensitivity, binning,
pixel exposure time, background) the convolved Nyx skewers were converted to expected COS counts per pixel.
Realistic COS counts were simulated by drawing from a Poisson distribution with a mean equal to the expected
COS counts. For each spectrum a systematic background subtraction error across the $\Delta z=0.04$ bin was
incorporated by adding to the background a Gaussian random number with zero mean and standard deviation
equal to the estimated background error for the considered spectral region. Then the Poisson counts were
converted to \ion{He}{2} transmission with the modified background. Finally, \teff\ was evaluated in the mock
COS spectra as described in Section~\ref{sect:he2tautech}.
In particular, this ensured a proper treatment of the sightline-specific \teff\ sensitivity limits as in the observed
spectra.

For each \gheii\ model and $\Delta z=0.04$ bin, the above procedure was repeated 2000 times to generate a
statistical ensemble of subsample realizations. To illustrate its fidelity, we plot in Figure~\ref{fig:he2mockreal}
the \textit{HST}/COS \ion{He}{2} transmission spectrum of HE2QS~J1630$+$0435 and a mock spectrum
using a lightcone skewer from the \citetalias{davies17} fluctuating \ion{He}{2}-ionizing background model.
At $3.05<z<3.12$ the skewer intersects a \ion{He}{3} zone with a peak value of $\mgheii\simeq 6\times 10^{-15}$\,s$^{-1}$,
giving rise to large-scale \ion{He}{2} transmission that closely mimics the prominent \ion{He}{2} transmission region
toward HE2QS~J1630$+$0435. Small-scale structure in the mock transmission region is impacted by Poisson noise,
but still reflects variations in the density field drawn from the Nyx skewers. The generally strong \ion{He}{2} absorption
masks large-scale correlations and small discontinuities in the density field that arise from simply concatenating
independent $\Delta z=0.04$ Nyx skewers.

\subsection{Statistical Comparison of Observed and Mock Data}
\label{sect:statcomp}

\begin{figure*}[t]
\includegraphics[width=\textwidth]{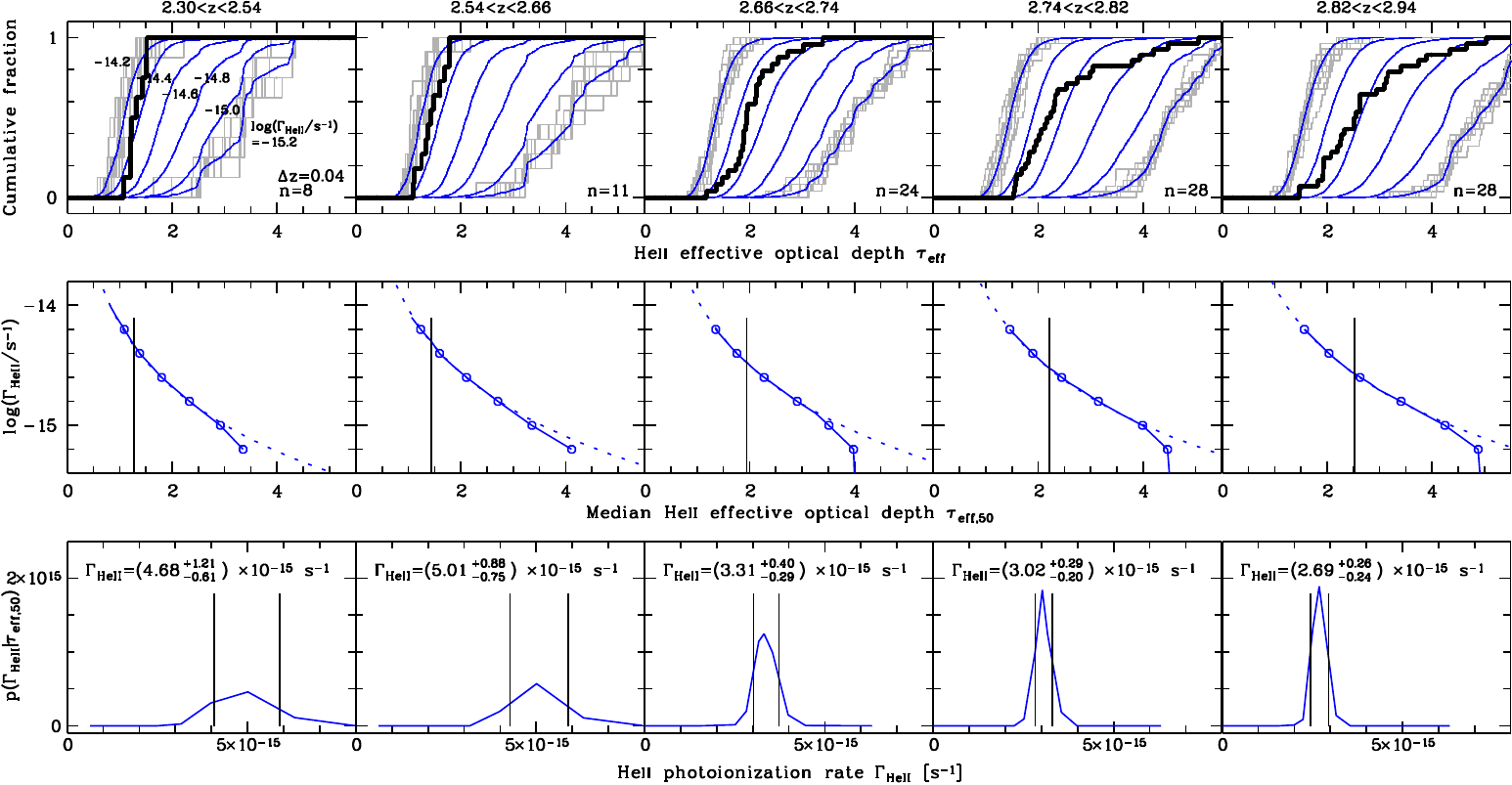}
\caption{\label{fig:he2gammavar1}
\emph{Top: } Cumulative \teff\ distributions on a scale $\Delta z=0.04$ for observed COS \ion{He}{2} spectra
(thick black, Table~\ref{tab:taueffstat})
and for realistic mock COS spectra from the Nyx hydrodynamical simulation for a uniform
UV background assuming six different constant \ion{He}{2} photoionization rates \gheii\ (labeled) in
five redshift ranges at $z<3$ defined in Table~\ref{tab:taueffmedian}.The blue curves show the merged
samples of 2000 realizations of mock data samples with $n$ values each, and sample variance is indicated
for the minimum and maximum considered \gheii\ by plotting 10 representative realizations (gray).
\emph{Middle: } \gheii\ corresponding to the median \teff\ in the merged mock samples (solid curves).
The values of the specific models in the top panel are indicated with circles. In noise-free simulated
spectra the median \teff\ is well described by a power-law $\tau_\mathrm{eff,50}=a\mgheii^b$ that varies
with redshift (dotted curves). The measured $\tau_\mathrm{eff,50}$ from Table~\ref{tab:taueffmedian}
is indicated by the vertical lines, yielding inferred values of \gheii.
\emph{Bottom: } Normalized posterior of \gheii\ given the mesured $\tau_\mathrm{eff,50}$ in the considered
redshift ranges. The 16th and the 84th percentile of the integral of the posterior (vertical lines) yield a
$1\sigma$ confidence interval on \gheii\ inferred from the middle panels (labeled).
}
\end{figure*}

In the mock spectra, the primary measure of interest is the \teff\ distribution that closely mimics the sensitivity limit
of the \textit{HST}/COS spectra. In the top panels of Figure~\ref{fig:he2gammavar1} we compare the observed
cumulative \teff\ distributions in the five $z<3$ redshift ranges
(Table~\ref{tab:taueffstat})
to the respective cumulative \teff\ distributions of mock data assuming several constant \ion{He}{2} photoionization
rates on a logarithmic grid. Due to sample variance, individual realizations of mock subsamples with $n$ mock
\teff\ values scatter around the distribution obtained from the total 2000 realizations. Nevertheless, the mock
distributions can be clearly distinguished from another, allowing us to constrain \gheii\ (see below).
The mock distributions have characteristic shapes due to \teff\ sensitivity limits which occur more frequently at low \gheii. 

For a quantitative comparison of the observed and the mock data we use the Kolmogorov-Smirnov summary statistic
\begin{equation}\label{eq:dstat}
D=\max_k\left|F\left(<\tau_\mathrm{eff}^k\right)-H\left(<\tau_\mathrm{eff}^k|\mgheii\right)\right|\quad, 
\end{equation}
where $F\left(<\tau_\mathrm{eff}^k\right)$ is the empirical cumulative distribution function to the $k$th highest
out of $n$ \teff\ values in the subsample
(Table~\ref{tab:taueffstat}),
and $H$ is the hypothesized cumulative distribution function of
\teff\ given \gheii\ that is computed from the set of 2000 realizations. Computing the summary statistic for the
individual mock subsamples we obtain an estimate of the distribution of $D$ due to sample variance
(i.e.\ the difference between the gray histograms and the blue curves in Figure~\ref{fig:he2gammavar1}),
while the observed data yield a value $D_\mathrm{obs}$. The fraction of mock realizations with
$D>D_\mathrm{obs}$ yields an estimate of the probability $P\left(>D_\mathrm{obs}\right)$
that the model (i.e.\ a given constant \gheii)
is consistent with the data, with small values of this probability indicating inconsistency.

\begin{figure}[t]
\includegraphics[width=\linewidth]{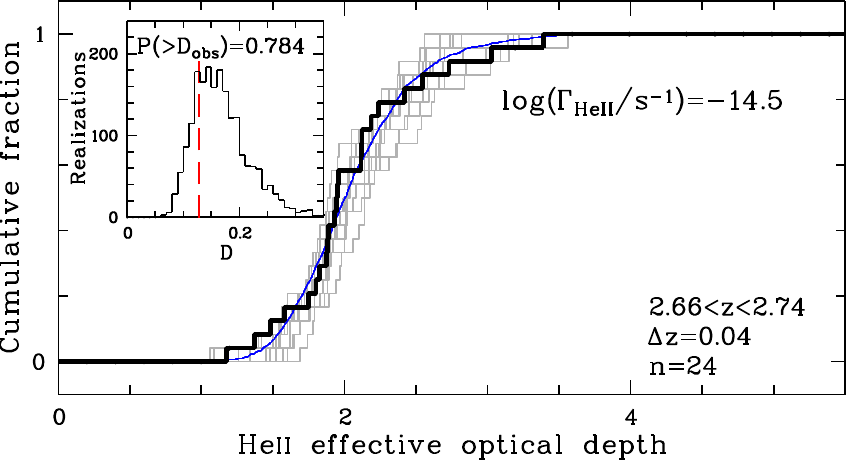}
\caption{\label{fig:he2gammalowz}
Observed cumulative \teff\ distribution on a scale $\Delta z=0.04$ at $2.66<z<2.74$ (thick black) compared
to mock distributions from the Nyx simulation with $\mgheii=10^{-14.5}$\,s$^{-1}$
(blue: all 2000 realizations with $n=24$ values each, gray: 10 representative realizations to indicate sample variance).
The inset shows the distribution of the maximum distance $D$ between the cumulative \teff\ distributions of
individual realizations and all mock data (Equation~\ref{eq:dstat}).
The dashed line marks the value $D_\mathrm{obs}=0.1280$ of the actual data and all mock data.
A high $P\left(>D_\mathrm{obs}\right)=0.784$ indicates that the model is consistent with the data.
}
\end{figure}

\begin{figure*}[t]
\includegraphics[width=\textwidth]{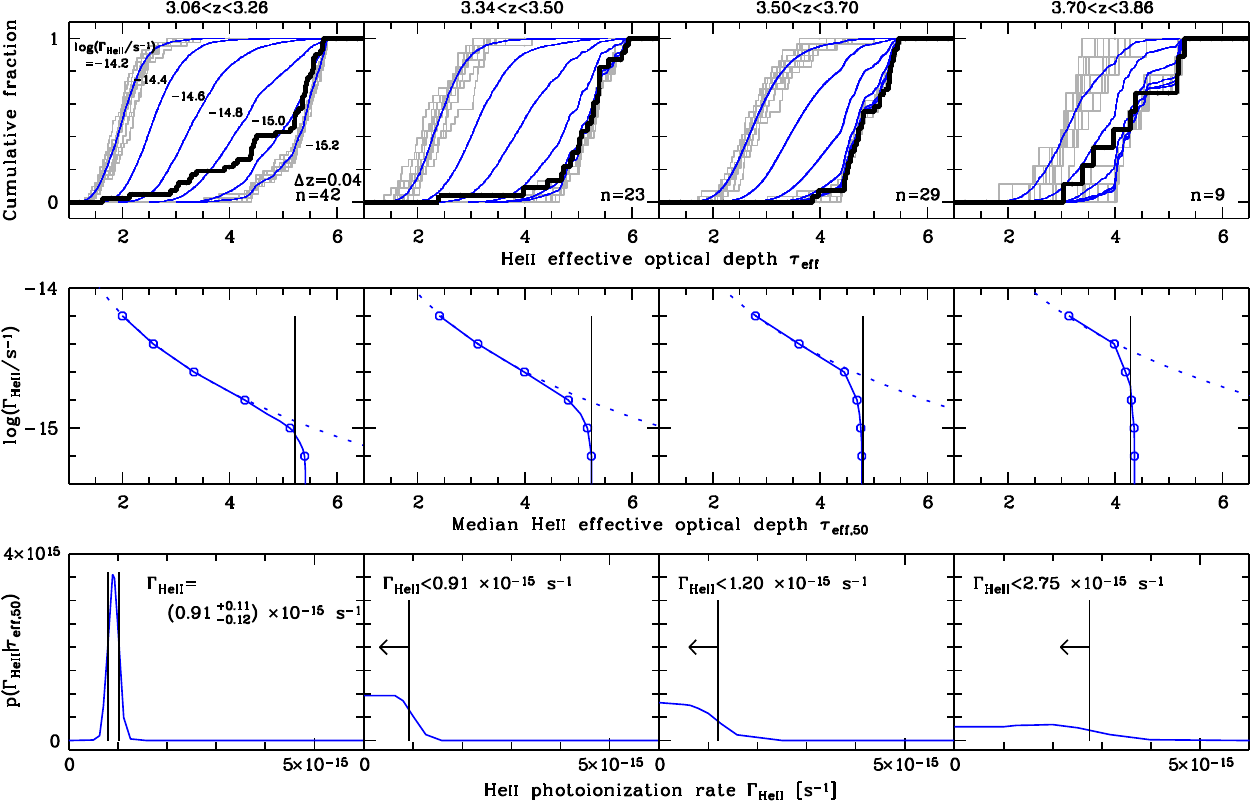}
\caption{\label{fig:he2gammavar2}
Similar to Figure~\ref{fig:he2gammavar1}, but showing the four subsamples at $z>3$.
}
\end{figure*}

The spread in the model distributions in the top panels of Figure~\ref{fig:he2gammavar1} reveals that
each subsample is consistent with a narrow range in \gheii. At $z<2.66$ the data set is sparse, but only
the model with $\mgheii=10^{-14.4}$\,s$^{-1}$ is consistent with the data ($P\left(>D_\mathrm{obs}\right)=0.205$
at $2.30<z<2.54$ and $P\left(>D_\mathrm{obs}\right)=0.250$ at $2.54<z<2.66$).
We will give confidence intervals of \gheii\ below.
At $2.66<z<2.74$, the models with
$\mgheii=10^{-14.4}$\,s$^{-1}$ and $\mgheii=10^{-14.6}$\,s$^{-1}$ are inconsistent with the data
($P\left(>D_\mathrm{obs}\right)=0.004$ and $0.007$, respectively), but bracket the observed \teff\ distribution.
To investigate this further, we simulated the \teff\ distribution for $\mgheii=10^{-14.5}$\,s$^{-1}$,
finding a very good match to the data ($P\left(>D_\mathrm{obs}\right)=0.784$), as illustrated in Figure~\ref{fig:he2gammalowz}.
The agreement between the model and the data in Figure~\ref{fig:he2gammalowz} indicates that the
\ion{He}{2}-ionizing background at $z\simeq 2.7$ is quasi-uniform at an amplitude
$\mgheii\simeq 10^{-14.5}$\,s$^{-1}$, and that the scatter in \teff\ is consistent with density
fluctuations on the chosen scale $\Delta z=0.04$ that are reproduced in the Nyx simulation.

At higher redshifts none of the uniform UV background models are consistent with the data. As shown in
the top panels of Figure~\ref{fig:he2gammavar1} the observed \teff\ distributions at $z>2.74$ show
a much larger spread than the mock distributions for any constant \gheii.
Consequently, also any spatially uniform but redshift-dependent \ion{He}{2}-ionizing background
\citep[e.g.][]{faucher09,haardt12,puchwein19} will not reproduce the observed \teff\ distribution at $z>2.74$.
Instead, the tail toward high \teff\ likely indicates fluctuations in the \ion{He}{2}-ionizing background that
are either due to the sparse source population (\citealt{davies14}; \citetalias{davies17}) or ongoing \ion{He}{2}
reionization (\citealt{worseck11b}; \citetalias{worseck16}).

Figure~\ref{fig:he2gammavar2} shows the cumulative \teff\ distributions in the four $z>3$ subsamples
(Table~\ref{tab:taueffstat}).
At $3.06<z<3.26$ the observations are still sensitive to a range in \gheii,
but uniform UV background models do not reproduce the large spread in the observed \teff\ distribution.
The steep increase in the observed cumulative \teff\ distribution at $\mteff>5$ is due to the very similar
sensitivity limits of the contributing spectra.
At $z>3.3$ the frequent \teff\ sensitivity limits affect our constraints on \gheii\ and possible UV background fluctuations.
Nevertheless, based on the few low \teff\ measurements and the trend from lower redshifts, a non-uniform
\ion{He}{2}-ionizing background is favored. The apparent agreement of the data at $3.50<z<3.70$ with a
uniform UV background with $\mgheii\la 10^{-15}$\,s$^{-1}$ is a mere coincidence due to the frequent sensitivity limits.
For such low \gheii\ values the \teff\ distribution is saturation-limited, so the apparent agreement just indicates that our
model for noise and systematics is reasonable, rather than the existence of a uniform low-amplitude UV background.

\subsection{The Median \ion{He}{2} Photoionization Rate}
\label{sect:he2ionrate}

Given the very good agreement of the data with a uniform UV background at $z\la 2.7$ (Figure~\ref{fig:he2gammalowz})
and the sufficient dynamic range in \gheii\ from our mock spectra, we investigated how to infer \gheii\ as a function of redshift.
We chose the median \teff\ as a representative value in each redshift range (Table~\ref{tab:taueffmedian}),
since it is well defined and robust to sample variance at $2.66<z<2.94$. The tails of the \teff\ distributions are due to
IGM density fluctuations and fluctuations in \gheii, but these cannot be readily disentangled.
As we will show below, the measured median \teff\ allows us to obtain an estimate of the characteristic \gheii\ value of a
mildly fluctuating UV background, whereas our summary statistic $D$ is highly sensitive to the shape
of the \teff\ distribution that is not reproduced by a uniform \gheii\ model at $z>2.74$.

The middle panels in Figures~\ref{fig:he2gammavar1} and \ref{fig:he2gammavar2} show the relation between the median
\teff\ and \gheii\ from the Nyx simulation output. In noise-free simulated spectra the median \teff\ is well described by a
power-law $\tau_\mathrm{eff,50}=a\mgheii^b$ with parameters $a$ and $b$ that slightly vary with redshift
(dotted lines in Figures~\ref{fig:he2gammavar1} and \ref{fig:he2gammavar2}).
At low \gheii\ the realistic mock data depart from the power-law relation due to the characteristic sensitivity limit
of the observed data. At $z<3$ the measured $\tau_\mathrm{eff,50}$ is well below the sensitivity limit, such that
\gheii\ can be inferred. At $z>3$, however, the measured $\tau_\mathrm{eff,50}$ is close to or at the sensitivity limit
characterized by a sharp turnover of \gheii\ as a function of $\tau_\mathrm{eff,50}$, implying a sensitivity limit to
\gheii\ at $z>3.3$ (Figure~\ref{fig:he2gammavar2}).

To obtain a confidence interval for \gheii\ we ran 2000--5000 mock sample realizations per input \gheii\ value on
a dense grid in \gheii\ around the value implied by the data. On the mock sample realizations we performed
Gaussian kernel density estimation to describe the likelihood $L\left(\tau_\mathrm{eff,50}|\mgheii\right)$.
Kernel widths and number of realizations were chosen to accurately reproduce multimodal distributions arising
from sightly different sensitivity limits of the contributing spectra.
According to Bayes' Theorem the posterior probability distribution $p\left(\mgheii|\tau_\mathrm{eff,50}\right)$ is
related to the likelihood $L\left(\tau_\mathrm{eff,50}|\mgheii\right)$ via
\begin{equation}
p\left(\mgheii|\tau_\mathrm{eff,50}\right)=\frac{L\left(\tau_\mathrm{eff,50}|\mgheii\right)p\left(\mgheii\right)}{p\left(\tau_\mathrm{eff,50}\right)}\quad,
\end{equation}
with the prior probability distribution $p\left(\mgheii\right)$ and the evidence
$p\left(\tau_\mathrm{eff,50}\right)=\int L\left(\tau_\mathrm{eff,50}|\mgheii\right)p\left(\mgheii\right)\,\mathrm{d}\mgheii$
that is the normalization of the posterior probability distribution.
With the grid in \gheii\ and the measured $\tau_\mathrm{eff,50}$ we constructed the posterior $p\left(\mgheii|\tau_\mathrm{eff,50}\right)$,
assuming a linear prior. 
While a logarithmic prior might be preferable to explore the range of \gheii\ during \ion{He}{2} reionization,
such a prior is improper due to the saturation limit of the data at $z>3.3$. Given our limited sensitivity to
high \teff\ there would be an arbitrary large prior volume at low $\log\mgheii$, and even for a restricted range in
$\log\mgheii$ the confidence intervals would depend on the lower and upper limit of \gheii. 
For this reason we chose a linear prior, since $\mgheii=0$ can be represented in a non-infinite parameter volume,
but we caution that our limits on \gheii\ obtained from \teff\ sensitivity limits are sensitive to this choice. 

The resulting posterior distributions are shown in the bottom panels of Figures~\ref{fig:he2gammavar1} and \ref{fig:he2gammavar2}.
Equal-tailed $1\sigma$ confidence intervals were obtained by integrating the normalized posteriors to
$0.16$ and $0.84$, respectively. At $z<3.3$ the posteriors are strongly peaked and their widths are primarily given by sample size.
At higher redshifts the posteriors are non-zero at $\mgheii=0$, indicating the sensitivity limit of the data.
For each redshift range where the posterior at $\mgheii=0$ is at or close to its peak value
(in practice we choose $p\left(\mgheii=0|\tau_\mathrm{eff,50}\right)>10^{14}$ in Figure~\ref{fig:he2gammavar2})
we obtained a $1\sigma$ upper limit on \gheii\ by integrating the posterior to $0.84$.
Our inferred values and limits of \gheii\ are listed in Table~\ref{tab:gammahe2}.

\begin{figure}[t]
\includegraphics[width=\linewidth]{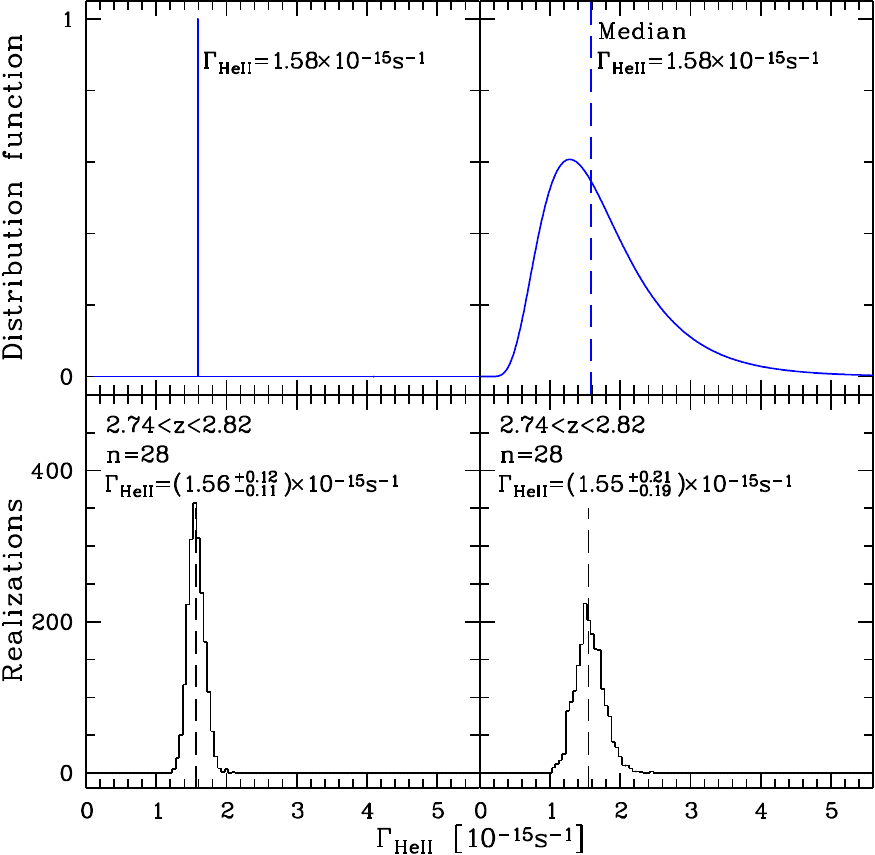}
\caption{\label{fig:he2gammavalid}
\emph{Top: } Two input distributions of \gheii\ (left: uniform, right: lognormal but constant on scales $\Delta z=0.04$)
with indicated median value (dashed and labeled).
\emph{Bottom: } Distributions of \gheii\ values inferred from 2000 mock data sets at $2.74<z<2.82$.
Dashed lines mark the median values. The median \teff\ approximately recovers the median value of the input \gheii\ distribution.
}
\end{figure}

The above inferences assume a uniform \gheii\ across the redshift ranges of the subsamples.
Since uniform models fail to reproduce the data at $z>2.74$ we may question the meaning of the inferred
\gheii\ values at these redshifts. We explored this by producing mock spectra for several \gheii\ distributions
in the representative redshift range $2.74<z<2.82$. For each of the 28 $\Delta z=0.04$ bins from
15 contributing COS \ion{He}{2} sightlines we randomly drew a \gheii\ value from the chosen distribution,
assumed it to be constant across the $\Delta z=0.04$ bin, and generated a mock spectrum as before.
From 2000 mock data sets we obtained the distribution of $\tau_\mathrm{eff,50}$ and the inferred \gheii.
Figure~\ref{fig:he2gammavalid} shows the results for a uniform and a lognormal \gheii\ distribution.
For a uniform \gheii\ the median \teff\ recovers the input value, with the $1\sigma$ scatter dominated
by sample variance. For an underlying lognormal \gheii\ distribution the median \teff\ approximately
yields the median \gheii. The variance in the inferred \gheii\ increases due to the sampling from the
\gheii\ distribution. Tests with several other skewed or bimodal distributions gave similar results close
to the median of the input distribution. While we cannot rigorously show that this holds in general,
we note that the statistical error inferred from the output \gheii\ distribution is much larger than
systematic deviations between the assumed and the revealed median \gheii\ even for a uniform \gheii.
Therefore we conclude that our measurements approximately recover the median \gheii\ of
a fluctuating \ion{He}{2}-ionizing background.

\tabletypesize{\footnotesize}
\begin{deluxetable}{ccc}
\renewcommand{\arraystretch}{1.0}
\tablecaption{\label{tab:gammahe2}Median \ion{He}{2} photoionization rate \gheii\ and corresponding
median \ion{He}{2} fraction \xheii\ in redshift ranges $\Delta z$}
\tablehead{
\colhead{\hspace{0.2in}$\Delta z$}\hspace{0.2in} &
\colhead{\hspace{0.2in}\gheii~$\left(10^{-15}\,s^{-1}\right)$}\hspace{0.2in} &
\colhead{\hspace{0.2in}\xheii~(\%)}\hspace{0.2in}
}
\startdata
$2.30$--$2.54$	&$4.68^{+1.21}_{-0.61}$ &$0.29^{+0.04}_{-0.06}$\\
$2.54$--$2.66$	&$5.01^{+0.88}_{-0.75}$ &$0.32^{+0.05}_{-0.05}$\\
$2.66$--$2.74$	&$3.31^{+0.40}_{-0.29}$ &$0.52^{+0.05}_{-0.06}$\\
$2.74$--$2.82$	&$3.02^{+0.29}_{-0.20}$ &$0.61^{+0.04}_{-0.05}$\\
$2.82$--$2.94$	&$2.69^{+0.26}_{-0.24}$ &$0.73^{+0.07}_{-0.06}$\\
$3.06$--$3.26$	&$0.91^{+0.11}_{-0.12}$ &$2.54^{+0.26}_{-0.37}$\\
$3.34$--$3.50$	&$<0.91$                        &$>3.15$\\
$3.50$--$3.70$	&$<1.20$                        &$>2.71$\\
$3.70$--$3.86$	&$<2.75$                        &$>1.47$\\
 \enddata
\end{deluxetable}

\subsection{Fluctuations in the \ion{He}{2}-Ionizing Background}

\begin{figure*}[t]
\includegraphics[width=\textwidth]{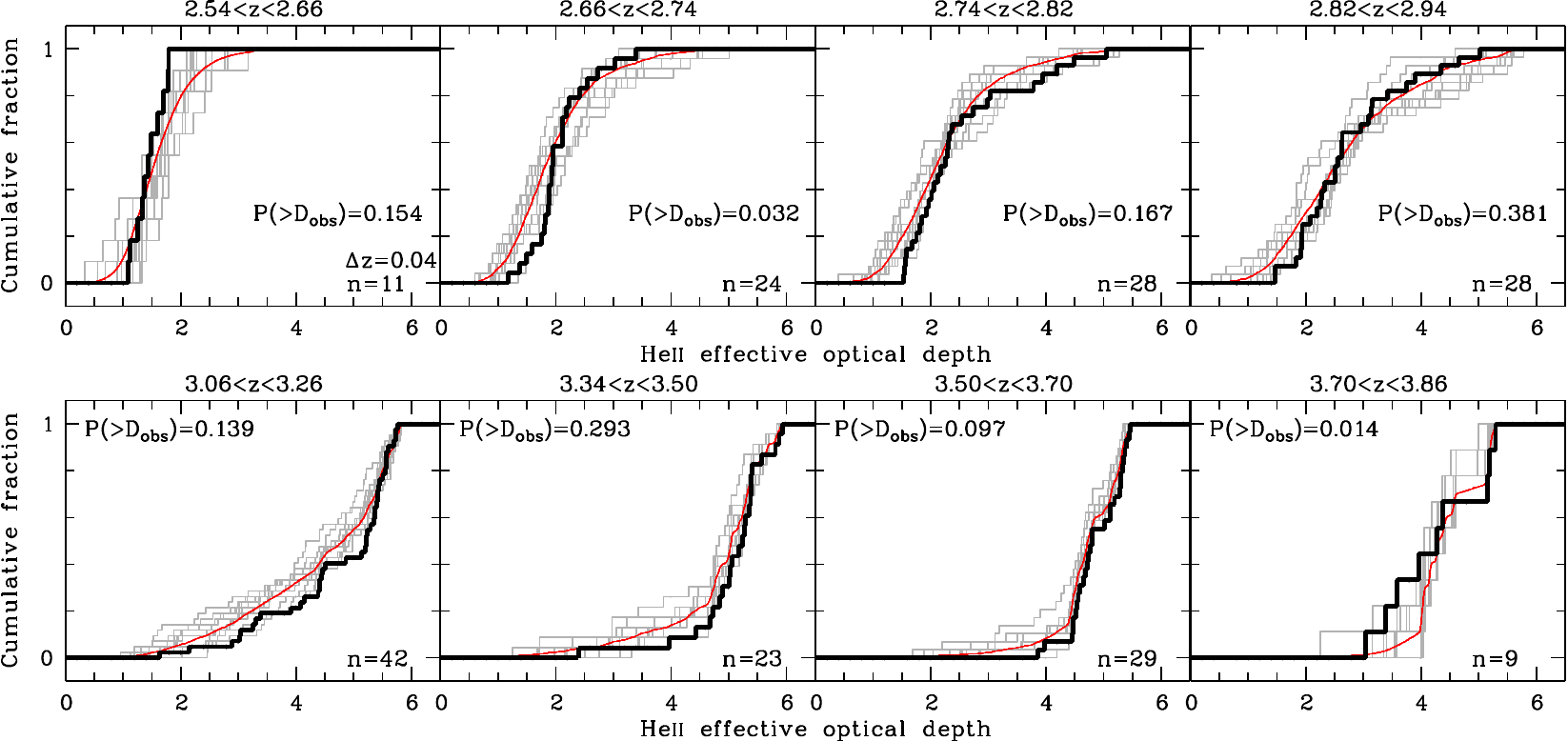}
\caption{\label{fig:he2fluctcomp}
Cumulative \teff\ distributions on a scale $\Delta z=0.04$ for observed COS \ion{He}{2} spectra
(thick black, Table~\ref{tab:taueffstat})
and for realistic mock COS spectra from the Nyx hydrodynamical simulation applying the fluctuating
UV background model from \citet{davies17} in eight of the redshift ranges defined in Table~\ref{tab:taueffmedian}.
The red curves show the merged samples of 2000 realizations of mock data samples with $n$ values each,
and sample variance is indicated by plotting 10 representative realizations (gray). For each redshift range
we list the estimated probability $P\left(>D_\mathrm{obs}\right)$ that the model is consistent
with the data (Section~\ref{sect:statcomp}).
}
\end{figure*}

Analogously to the uniform UV background models presented in Section~\ref{sect:statcomp}, we show in
Figure~\ref{fig:he2fluctcomp} the comparison of the observed cumulative \teff\ distributions to the respective
mock \teff\ distributions assuming the fluctuating UV background model from \citetalias{davies17}.
We reiterate that we have not tuned the \citetalias{davies17} model parameters
(quasar luminosity function, quasar spectral energy distribution, quasar lightcurve and opening angle, IGM absorption)
to fit our \teff\ distributions.
Further exploration of the partially degenerate parameter space constrained by recent observational data
and inferences \citep[e.g.][]{inoue14,khaire17,schmidt18,khrykin19,kulkarni18} is left for future work.

High values of $P\left(>D_\mathrm{obs}\right)$ indicate that the model is in remarkable agreement with
the data in most of the considered redshift ranges. In particular, in contrast to the uniform UV background models,
the \citetalias{davies17} model reproduces the observed skewed \teff\ distributions at $z>2.74$ that are due to
the combined effect of fluctuations in the density field and the UV background.

Minor mismatches of the data and the model occur at the lowest and the highest redshifts
($2.66<z<2.74$ and $3.70<z<3.86$), which may be due to the particular choice of model parameters
(i.e.\ quasar and IGM absorber properties) or remaining cosmic variance due to the small number of
bright $z>3$ quasars in the (500\,cMpc)$^3$ simulation volume \citepalias{davies17}.
Since the \teff\ distribution at $2.66<z<2.74$ is consistent with a uniform UV background
(Figure~\ref{fig:he2gammalowz}) we tested whether the minor inconsistency at $2.66<z<2.74$ is due to
diminishing UV background fluctuations (the \citetalias{davies17} model predicts variations by a factor $\simeq 1.5$
around the median \gheii\ at $z=2.7$). Comparing mock samples generated with a uniform
UV background to mock samples assuming the \citetalias{davies17} fluctuating UV background, we found
that a narrow range of constant \gheii\ values is consistent with the fluctuating model.
We conclude that at $z\simeq 2.7$ our data cannot distinguish between a uniform UV background
and a mildly fluctuating one, and that the minor discrepancy in the \teff\ distribution can be resolved by
an insignificant $-0.05$\,dex adjustment of the median \gheii\ in the \citetalias{davies17} model at $z\simeq 2.7$.

At $z>3$ the \citetalias{davies17} model reproduces the tail toward low \teff\ values, indicating that these are
likely due to intersected proximity zones of quasars. The model predicts a somewhat stronger tail than observed,
which may require either adjustments in the model parameters or proper modeling of the \ion{He}{2}
reionization process that becomes relevant at $z>3$ (\citealt{davies14}; \citetalias{davies17}).
The overall good agreement indicates that the \citetalias{davies17} model successfully captures the percolation
of \ion{He}{3} zones around quasars at $z\ga 3$. For saturated regions in our \ion{He}{2} spectra, however,
we cannot tell whether they correspond to downward UV background fluctuations in a post-reionization IGM
($\mgheii\lesssim10^{-15}\,\mathrm{s}^{-1}$) or to not yet reionized \ion{He}{2} patches ($\mgheii\rightarrow 0$).
The observed excess toward low \teff\ values at $z>3.7$ that is inconsistent with the model despite the
small sample ($P\left(>D_\mathrm{obs}\right)=0.014$) may be due to large-scale correlations in the radiation field
in the peculiar SDSS~J1319$+$5202 sightline (Section~\ref{sect:obsresults}) or due to a particularly small number
of $z\sim 3.8$ quasars in the (500\,cMpc)$^3$ simulation volume. Analysis of a second (500\,cMpc)$^3$
volume yields similar results at $z>3$, indicating that the predicted \teff\ distribution is not strongly affected by
cosmic variance given our sample size and limited \teff\ sensitivity.

\section{Discussion}
\label{sect:discussion}

\subsection{Redshift Evolution of the \ion{He}{2}-Ionizing Background}

\begin{figure}[t]
\includegraphics[width=\linewidth]{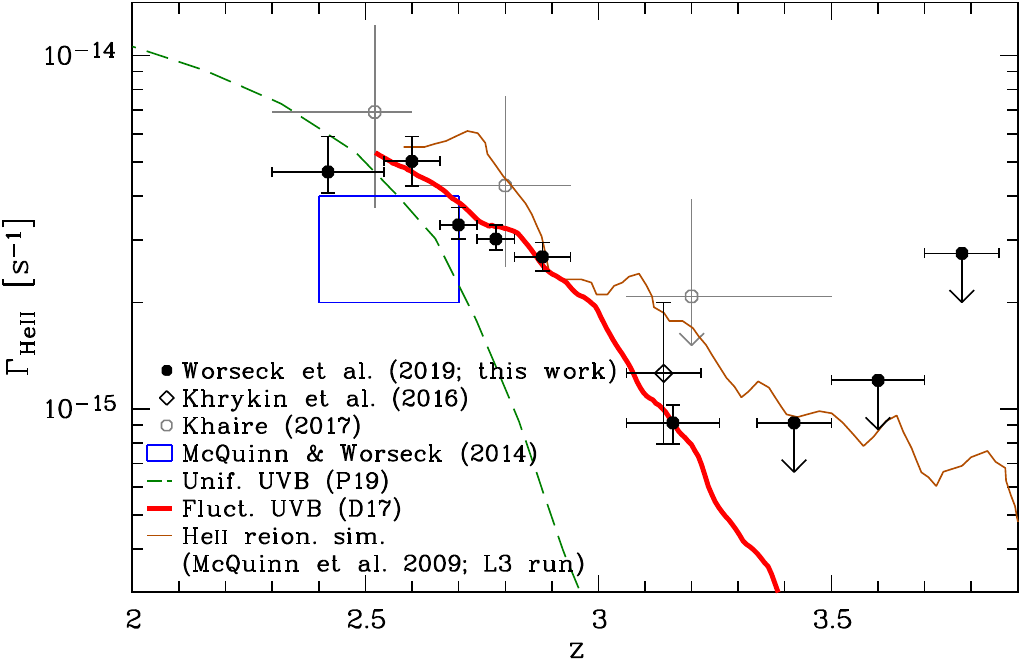}
\caption{\label{fig:he2gammacomp}
\ion{He}{2} photoionization rate as a function of redshift. Filled circles show the \gheii\ values inferred from
the median \teff\ in the indicated redshift ranges (Table~\ref{tab:gammahe2}).
Error bars are $1\sigma$ derived from the posterior of \gheii\ given the measured median \teff,
and arrows indicate $1\sigma$ upper limits. We also show previous estimates based on subsets
of our data \citep{khrykin16,khaire17} or inferred from the number density ratio
$n_\mathrm{HeII}/n_\mathrm{HI}$ \citep{mcquinn14}.
Lines show predictions based on a 1D uniform UV background synthesis model \citep{puchwein19},
3D semianalytic calculations with a fluctuating UV background \citep{davies17},
and a 3D numerical radiative transfer simulation of \ion{He}{2} reionization \citep{mcquinn09a}.
}
\end{figure}

In Figure~\ref{fig:he2gammacomp} we compare the redshift evolution of \gheii\ inferred from our
\teff\ sample to several published estimates and models. We reiterate that our values approximately
correspond to the median values for a range of plausible \gheii\ distributions (Figure~\ref{fig:he2gammavalid}),
but our \teff\ measurements cannot well constrain the \gheii\ distribution or the spatial scale of \gheii\ fluctuations.
The \ion{He}{2} photoionization rate drops by a factor of $\simeq 5$ between $z\simeq 2.6$ and $z\simeq 3.1$.
The decrease likely continues to $z>3.3$ where our constraints on \gheii\ are limited by saturation,
instrument sensitivity and sample size.

At $z\simeq 3.1$ our \gheii\ value is consistent with the estimate $\mgheii=10^{-14.9\pm 0.2}$\,s$^{-1}$
by \citet{khrykin16} that was based on a rough comparison of a subset of our \teff\ data to a 25\,$h^{-1}$\,cMpc
smoothed particle hydrodynamics simulation without forward-modeling to the actual data quality.
\citet{khaire17} matched the median \teff\ values from \citetalias{worseck16} to predictions for an optically thin IGM
in photoionization equilibrium 
by applying a model for the \ion{He}{2}-to-\ion{H}{1} number density ratio $n_\mathrm{HeII}/n_\mathrm{HI}$
to the \ion{H}{1} column density distribution $f\left(N_\mathrm{HI},z\right)$
 which yields \gheii\ for an assumed value of \ghi.
The estimates critically depend on the assumed $f\left(N_\mathrm{HI},z\right)$ and the minimum
$N_\mathrm{HI}$ for Ly$\alpha$ forest lines, so the systematic offset between our values and the ones by
\citet{khaire17} is not surprising\footnote{
\citet{khaire17} incorrectly estimated the error on the median \teff\ from
the distribution of errors of individual \teff\ measurements. In Figure~\ref{fig:he2gammacomp} we show his
overestimated propagated errors for \gheii, noting that there are additional systematic uncertainties due to the
assumed $f\left(N_\mathrm{HI},z\right)$.
}.
\citet{mcquinn14} analyzed the coeval \ion{H}{1} and \ion{He}{2} Ly$\alpha$ absorption in two
post-reionization ($2.4<z<2.7$) \ion{He}{2} sightlines calibrated to a smoothed particle hydrodynamics simulation,
finding small (factor $<2$) variations of the number density ratio $n_\mathrm{HeII}/n_\mathrm{HI}$ around a value
of $\approx 100$ on scales of a few cMpc. With $\mghi=0.5$--$1\times 10^{-12}$\,s$^{-1}$ \citep{faucher08b,becker13b}
they estimated $\mgheii=2$--$4\times 10^{-15}$\,s$^{-1}$, with the true value being probably closer to the upper end
of this range (see \citealt{becker13b} for a discussion on \ghi). The remaining small differences to our \gheii\ values
are probably caused by the different thermal history employed in the simulation used by \citet{mcquinn14}.

Figure~\ref{fig:he2gammacomp} also shows model predictions for $\mgheii\left(z\right)$. The recent uniform
UV background synthesis model by \citet{puchwein19} is in broad agreement with our inferred \gheii\ values at $z\la 2.7$.
Since our inferences assume photoionization equilibrium, we plot their equivalent-equilibrium $\mgheii\left(z\right)$.
At $z>2.7$ the \citet{puchwein19} \ion{He}{2} photoionization rate drops much more rapidly than our measurements.
This is primarily due to rapid evolution in the \ion{He}{3} volume filling factor during \ion{He}{2} reionization that
completes at $z\simeq 2.8$ in the \citet{puchwein19} model for their particular choice of quasar emissivity and
IGM clumping factor. During reionization the \citet{puchwein19} \gheii\ represents an average over the
inhomogeneously ionized Universe captured by the \ion{He}{3} volume filling factor.
Since our values incorporate fully and partially ionized regions characterized by low and high \teff, respectively,
they are comparable to \citet{puchwein19} if the \gheii\ distribution is not highly skewed or bimodal.
 The observed shallow redshift evolution of the \ion{He}{2} effective optical depth
(Figure~\ref{fig:he2tau_model}) and the inferred \ion{He}{2} photoionization rate both indicate that \ion{He}{2} reionization
proceeded more gradually or ended earlier than predicted by \citet{puchwein19}.
Indeed, the \ion{He}{3} volume filling factor is expected to evolve gradually at the end of reionization when
the opacity to ionizing photons is dominated by Lyman limit systems instead of the diffuse IGM
\citep[e.g.][]{furlanetto05,furlanetto08,bolton09,madau17}. \citet{puchwein19} did not model \ion{He}{2} Lyman
limit systems explicitly, which likely resolves the discrepancy to our median \teff\ values and the inferred \gheii\ at $z>2.7$.

The median $\mgheii\left(z\right)$ from the \citetalias{davies17} fluctuating UV background model
(thick red line in Figure~\ref{fig:he2gammacomp}) matches our inferred values very closely at $z<3.3$
where our data set has sufficient sensitivity. This directly follows from the good agreement of the \teff\ distributions
(Figure~\ref{fig:he2fluctcomp}).
The mean \gheii\ from \citetalias{davies17} (not shown) is higher than the median
at all redshifts, in particular at $z>3$, since it is dominated by ionized regions.
\citetalias{davies17} predict a strong redshift evolution of the
median \gheii\ due to the evolving quasar number density and the fluctuating mean free path in the post-reionization IGM. 

Finally, the brown curve shows the volume-averaged $\mgheii\left(z\right)$ from a numerical simulation of
\ion{He}{2} reionization \citep{mcquinn09a}. Their L3 simulation, which includes a filtering prescription for low-column
density systems, broadly reproduces the observed scatter in \teff\ at $2.7\la z\la 3$ \citepalias{worseck16}.
At $z>3$ their volume-averaged \gheii\ is dominated by \ion{He}{3} regions, and the jaggedness of the curve
is due to variance in the number of active quasars captured in the (186\,Mpc)$^3$ simulation volume.
Accounting for these differences and uncertainties, our inferred \gheii\ values agree quite well with the
\citet{mcquinn09a} radiative transfer simulation.

\subsection{The Median \ion{He}{2} Fraction}

By combining our measurements of the median \gheii\ with the Nyx simulation skewers, 
we can estimate the median fraction of \ion{He}{2}, \xheii, in the IGM, which we list in Table~\ref{tab:gammahe2}.
While the median IGM has been fully reionized with $\mxheii<1\%$ at $z<3$, at $z>3$ the median IGM has 
a substantial residual \ion{He}{2} fraction of $\mxheii\gtrsim2.5\%$, indicating that we are probing
the tail end of the helium reionization process.
A caveat here is that we are mainly sensitive to \gheii\ and hence \xheii\ in ionized regions, and that
for highly bimodal distributions our estimate of the median value is strongly impacted by the limited sample size.
As such, it is not yet clear how to relate our \xheii\ value to the \ion{He}{3} volume filling factor used in
simple estimates of the \ion{He}{2} reionization history \citep[e.g.][]{haardt12,madau17,puchwein19}.

\subsection{Uncertainties}
\label{sect:uncertainties}

In addition to the statistical errors computed in Section~\ref{sect:he2ionrate} our inferred \gheii\ values may be
affected by systematic uncertainties, primarily due to our approximation to use optically thin simulations to represent
the immediate post-reionization IGM with potentially large variations in \gheii\ and residual patches of \ion{He}{2}.
On the other hand, it is fair to assume photoionization equilibrium that allows a simple rescaling of optical depths
in these simulations, since our inferred $\mgheii\ga 10^{-15}$\,s$^{-1}$ corresponds to an equilibration timescale
$t_\mathrm{eq}\approx\mgheii^{-1}\la 3\times 10^7$ years, which is short compared to any cosmological effect
\citep{mcquinn09b,khrykin16}.

A further systematic uncertainty stems from the specific thermal history assumed in the Nyx simulation,
which depends on the amount of heat injected into the IGM during \ion{He}{2} reionization.
We estimate the magnitude of this effect by imposing two different IGM thermal states onto the Nyx
skewers which bracket the $\sim1\sigma$ range of the measurements by \citet{walther18} at $z=3$:
$\left(T_0,\gamma\right) = \left(10700\,\mathrm{K}, 1.80\right)$ and $\left(15000\,\mathrm{K}, 1.36\right)$,
where the IGM temperature-density relation is defined to be a power law $T\left(\rho\right)=T_0\left(\rho/\bar{\rho}\right)^{\gamma-1}$
and $\bar{\rho}$ is the mean density of the Universe \citep{walther19}.
Due to the temperature dependence of the \ion{He}{2} recombination rate, these two thermal states require more or less
photoionization by the UV background to match the same \teff. Compared to our fiducial simulation, the relatively cold
and hot models (at mean density) would shift our inferred \gheii\ values by approximately $+0.1$ and $-0.2$\,dex, respectively.
By comparison, the modest \teff\ convergence issues of our fiducial Nyx simulation result in a modest bias of our
inferred \gheii\ values by $+10$\% (Appendix~\ref{sect:convapp}).

A final caveat is that our statistical errors on \gheii\ are possibly underestimated, because our inference
just includes the variance in \teff\ due to density fluctuations and data quality, but not the variance of
a fluctuating UV background. However, this is relevant when the UV background fluctuations are largest
and occur on large scales, i.e.\ at $z>3.3$ where we just obtain upper limits on \gheii\ due to
the sensitivity limit of our data. As our upper limits on \gheii\ also depend on the assumed prior,
we estimate that mild UV background fluctuations do not significantly increase our statistical errors.

\section{Summary and Conclusions}
\label{sect:conclusions}

We have conducted a systematic survey to characterize the ionization state of intergalactic helium at $2.3<z<3.8$
with \textit{HST} \ion{He}{2} Ly$\alpha$ absorption spectra. Building on earlier results \citepalias{worseck16},
we have analyzed \textit{HST}/COS spectra of eight additional \ion{He}{2}-transparent $z_\mathrm{em}>3$ quasars,
six of which had been discovered in our dedicated survey for FUV-bright high-redshift quasars (dubbed HE2QS,
Figure~\ref{fig:he2newspc}). These spectra increase the redshift pathlength sensitive to high \ion{He}{2} effective
optical depths $\mteff\sim 5$ by more than a factor of two at $3<z<3.5$, and provide the first statistically
meaningful sample of six \ion{He}{2} sightlines at $z>3.5$. Adding archival higher-quality high-resolution spectra
of four known \ion{He}{2}-transparent quasars, we have constructed a total sample of 25 science-grade
(S/N$\ga 3$) \textit{HST} \ion{He}{2} Ly$\alpha$ absorption spectra, available from our survey data
repository
\footnote{\url{https://archive.stsci.edu/prepds/hers/}}.
Our main results can be summarized as follows:

\begin{enumerate}

\item
At $z>3.5$ \ion{He}{2} Ly$\alpha$ absorption is predominantly saturated with some isolated narrow
($\Delta v<650$\,km\,s$^{-1}$) transmission spikes, most of which are unresolved and/or impacted by
Poisson noise in our \textit{HST}/COS G140L spectra (Figure~\ref{fig:he2spczoom2}).
At $z<3.5$ these features become more numerous and broader (Figure~\ref{fig:he2spczoom1}),
but the \ion{He}{2} Ly$\alpha$ absorption is still patchy at $2.7<z<3.3$ with significant sightline-to sightline variance.

\item
The \ion{He}{2} effective optical depth on a scale $\Delta z=0.04$ ($\approx 40$\,cMpc at $z=3$) increases
from $\mteff\simeq 2$ at $z=2.7$ to a sensitivity limit $\mteff\ga 5$ at $z>3$, but with significant
sightline-to-sightline variance at $z>2.7$ that increases with redshift (Figure~\ref{fig:he2tau_model}).
At $z>3$ regions with $\mteff<5$ gradually disappear as a result of the diminishing number and significance
of \ion{He}{2} transmission spikes. At $z\simeq 3.4$ our larger sample yields a lower fraction of statistically
significant $\mteff\la 5$ values ($26.1$\%) than in \citetalias{worseck16} (50\%), bringing the data into better
agreement with numerical models of the fluctuating \ion{He}{2}-ionizing radiation field at the tail end of
\ion{He}{2} reionization (\citealt{compostella14}; \citetalias{davies17}). Still, 6/38 $\Delta z=0.04$ regions at $z>3.5$
show statistically significant \ion{He}{2} transmission.

\item
We have compared our \teff\ measurements to predictions for a range of UV background models applied
to outputs of a large ($100h^{-1}$\,cMpc) high-resolution ($25h^{-1}$\,ckpc) hydrodynamical simulation,
forward-modeling variations in data quality and sample size.
At $z>2.74$ the observed variance in \teff\ cannot be reproduced by a spatially uniform redshift-dependent
\ion{He}{2}-ionizing background (Figure~\ref{fig:he2gammavar1}), strictly confining the applicability of
common UV background synthesis models to $z<2.74$ \citep[e.g.][]{faucher09,haardt12,puchwein19}.
Instead, the observed \teff\ distributions closely agree with predictions of the \citetalias{davies17} fluctuating
post-reionization UV background that is due to the varying quasar number density and the mean free path
to \ion{He}{2}-ionizing photons. This suggests an extended overlap epoch of \ion{He}{3} zones around
quasars at $2.7\la z\la 3.3$ that is captured by the \citetalias{davies17} model without deliberately tuning its parameters.
However, at $z>3.3$ we cannot distinguish between this scenario and ongoing \ion{He}{2} reionization due to
limited sensitivity to $\mteff\sim 5$ and model assumptions (\citealt{davies14}; \citetalias{davies17}).
 
\item 
We have developed a method to infer the characteristic \ion{He}{2} photoionization rate by matching the median
\teff\ of observed and mock data, respectively. Tests with mock data confirmed that our procedure approximately
recovers the median value of a \gheii\ distribution with sightline-to-sightline variance but spatial coherence
on the adopted scale $\Delta z=0.04$. The inferred \gheii\ decreases by a factor $\simeq 5$ between $z\simeq 2.6$
and $z\simeq 3.1$ (Figure~\ref{fig:he2gammacomp}), in very good agreement with the median $\mgheii\left(z\right)$
by \citetalias{davies17}. At $3.06<z<3.26$ our inferred
$\mgheii=\left[9.1^{+1.1}_{-1.2}\,\mathrm{(stat.)}\,^{+2.4}_{-3.4}\,\mathrm{(sys.)}\right]\times 10^{-16}$\,s$^{-1}$
translates to a median \ion{He}{2} fraction $\mxheii\simeq 0.025$, confirming that our sample of
\ion{He}{2} sightlines probes the tail end of \ion{He}{2} reionization that is well approximated by the \citetalias{davies17}
fluctuating UV background model. At $z>3.3$ our constraints are limited by saturation in \ion{He}{2} \lya\ due
to IGM density evolution, decreasing instrument sensitivity, and a small sample size.

\end{enumerate}

In summary, our sample of \ion{He}{2} sightlines probes the extended end phase of \ion{He}{2} reionization and the
build-up of the \ion{He}{2}-ionizing background with gradually diminishing fluctuations, in good agreement with
recent models (\citealt{compostella14}; \citetalias{davies17}). At $z>3$ the observed \teff\ distributions are consistent with models of
\ion{He}{2} reionization primarily driven by the observed quasar population at $z>4$ \citep{compostella14}.
At face value, our six $\mteff\la 4$ values at $z>3.5$ indicate mild tension with these models, which may be due to
the few specific quasar models run on limited simulation volumes \citep[e.g.][]{daloisio17}.
Future progress requires large-volume \ion{He}{2} reionization simulations with accurate radiative transfer
(see \citealt{laplante17} for a recent effort)
but also more and higher-quality data. Our forthcoming \textit{HST} program (PID~15356, PI Worseck) aims at
resolving the \ion{He}{2} Ly$\alpha$ absorption toward the two UV-brightest $z>3.5$ \ion{He}{2}-transmitting quasars
discovered in our dedicated survey, to verify and resolve isolated narrow \ion{He}{2} transmission spikes. 

In the near future, optical slitless prism spectroscopy currently obtained by the \textit{Gaia} satellite will
provide a complete census of bright quasars on the full sky \citep[e.g.][]{proft15} that does not exhibit
the specific bias against UV-bright $z_\mathrm{em}>2.7$ quasars in optical color-selected samples \citep{worseck11}.
Correlation of the \textit{Gaia} quasar catalog to \textit{GALEX} photometry will reveal further likely
\ion{He}{2}-transparent quasars to complete \textit{HST}'s legacy on \ion{He}{2} absorption spectroscopy.
The physical interpretation of these data will require dedicated efforts to run large-volume radiative transfer
simulations of \ion{He}{2} reionization that use updated quasar luminosity functions \citep{kulkarni18}
and the latest constraints on the distribution of quasar lifetimes \citep{eilers17,schmidt18,khrykin19}.
These simulations will also make detailed predictions for the early stages of \ion{He}{2} reionization at $z>4$
that may be studied with sensitive \ion{He}{2} absorption spectroscopy obtained with
a next-generation large-aperture UV space telescope.

\acknowledgments

We thank Zarija Luki\'c for providing the Nyx simulation, Jose O\~norbe for help with the
convergence test simulations, and Jo Bovy for adapting his algorithm
to specifically select FUV-bright quasar candidates.
We thank the referee for their helpful and constructive report.

Support for Programs GO\,13013 and GO\,13875 was provided by NASA through a grant from
the Space Telescope Science Institute, which is operated by the Association of
Universities for Research in Astronomy, Inc., under NASA contract NAS5-26555.
G.W. has been partially supported by the Deutsches Zentrum f\"ur Luft- und
Raumfahrt (DLR) through grant numbers 50\,OR\,1317 and 50\,OR\,1512.
F.B.D. has been partially supported by NASA through grant number HST-AR-15014
from the Space Telescope Science Institute.

This work is based on observations collected at the Centro Astron\'omico Hispano Alem\'an (CAHA) at Calar Alto,
operated jointly by the Max-Planck-Institut f\"ur Astronomie and the Instituto de Astrof\'isica de Andaluc\'ia (CSIC).
It also uses data obtained from Lick Observatory, owned and operated by the University of California.

Funding for SDSS-III has been provided by the Alfred P. Sloan Foundation, the Participating Institutions,
the National Science Foundation, and the U.S. Department of Energy Office of Science.
The SDSS-III web site is \url{http://www.sdss3.org/}.
SDSS-III is managed by the Astrophysical Research Consortium for the Participating Institutions of the
SDSS-III Collaboration including the University of Arizona, the Brazilian Participation Group, Brookhaven National Laboratory,
Carnegie Mellon University, University of Florida, the French Participation Group, the German Participation Group,
Harvard University, the Instituto de Astrof\'isica de Canarias, the Michigan State/Notre Dame/JINA Participation Group,
Johns Hopkins University, Lawrence Berkeley National Laboratory, Max Planck Institute for Astrophysics,
Max Planck Institute for Extraterrestrial Physics, New Mexico State University, New York University, Ohio State University,
Pennsylvania State University, University of Portsmouth, Princeton University, the Spanish Participation Group,
University of Tokyo, University of Utah, Vanderbilt University, University of Virginia, University of Washington, and Yale University.

The Pan-STARRS1 Surveys (PS1) and the PS1 public science archive have been made possible through contributions by
the Institute for Astronomy, the University of Hawaii, the Pan-STARRS Project Office, the Max-Planck Society and its participating institutes
(the Max Planck Institute for Astronomy, Heidelberg and the Max Planck Institute for Extraterrestrial Physics, Garching), 
The Johns Hopkins University, Durham University, the University of Edinburgh, the Queen's University Belfast,
the Harvard-Smithsonian Center for Astrophysics, the Las Cumbres Observatory Global Telescope Network Incorporated,
the National Central University of Taiwan, the Space Telescope Science Institute, the National Aeronautics and Space Administration under
Grant No. NNX08AR22G issued through the Planetary Science Division of the NASA Science Mission Directorate,
the National Science Foundation Grant No. AST-1238877, the University of Maryland, Eotvos Lorand University (ELTE),
the Los Alamos National Laboratory, and the Gordon and Betty Moore Foundation.

This publication makes use of data products from the Wide-field Infrared Survey Explorer, which is a joint project of
the University of California, Los Angeles, and the Jet Propulsion Laboratory/California Institute of Technology,
funded by the National Aeronautics and Space Administration.

\vspace{5mm}
\facilities{
HST (STIS, COS),
CAO:2.2m (Calar Alto Faint Object Spectrograph),
Shane (Kast Double Spectrograph),
GALEX,
PS1,
Sloan,
WISE
}

\bibliographystyle{aasjournal}
\bibliography{he2cos3}

\begin{thebibliography}{}
\expandafter\ifx\csname natexlab\endcsname\relax\def\natexlab#1{#1}\fi
\providecommand{\url}[1]{\href{#1}{#1}}
\providecommand{\dodoi}[1]{doi:~\href{http://doi.org/#1}{\nolinkurl{#1}}}
\providecommand{\doeprint}[1]{\href{http://ascl.net/#1}{\nolinkurl{http://ascl.net/#1}}}
\providecommand{\doarXiv}[1]{\href{https://arxiv.org/abs/#1}{\nolinkurl{https://arxiv.org/abs/#1}}}

\bibitem[{{Aihara} {et~al.}(2011){Aihara}, {Allende Prieto}, {An}, {Anderson},
  {Aubourg}, {Balbinot}, {Beers}, {Berlind}, {Bickerton}, {Bizyaev}, {Blanton},
  {Bochanski}, {Bolton}, {Bovy}, {Brandt}, {Brinkmann}, {Brown}, {Brownstein},
  {Busca}, {Campbell}, {Carr}, {Chen}, {Chiappini}, {Comparat}, {Connolly},
  {Cortes}, {Croft}, {Cuesta}, {da Costa}, {Davenport}, {Dawson}, {Dhital},
  {Ealet}, {Ebelke}, {Edmondson}, {Eisenstein}, {Escoffier}, {Esposito},
  {Evans}, {Fan}, {Femen{\'{\i}}a Castell{\'a}}, {Font-Ribera}, {Frinchaboy},
  {Ge}, {Gillespie}, {Gilmore}, {Gonz{\'a}lez Hern{\'a}ndez}, {Gott}, {Gould},
  {Grebel}, {Gunn}, {Hamilton}, {Harding}, {Harris}, {Hawley}, {Hearty}, {Ho},
  {Hogg}, {Holtzman}, {Honscheid}, {Inada}, {Ivans}, {Jiang}, {Johnson},
  {Jordan}, {Jordan}, {Kazin}, {Kirkby}, {Klaene}, {Knapp}, {Kneib},
  {Kochanek}, {Koesterke}, {Kollmeier}, {Kron}, {Lampeitl}, {Lang}, {Le Goff},
  {Lee}, {Lin}, {Long}, {Loomis}, {Lucatello}, {Lundgren}, {Lupton}, {Ma},
  {MacDonald}, {Mahadevan}, {Maia}, {Makler}, {Malanushenko}, {Malanushenko},
  {Mandelbaum}, {Maraston}, {Margala}, {Masters}, {McBride}, {McGehee},
  {McGreer}, {M{\'e}nard}, {Miralda-Escud{\'e}}, {Morrison}, {Mullally},
  {Muna}, {Munn}, {Murayama}, {Myers}, {Naugle}, {Neto}, {Nguyen}, {Nichol},
  {O'Connell}, {Ogando}, {Olmstead}, {Oravetz}, {Padmanabhan},
  {Palanque-Delabrouille}, {Pan}, {Pandey}, {P{\^a}ris}, {Percival},
  {Petitjean}, {Pfaffenberger}, {Pforr}, {Phleps}, {Pichon}, {Pieri}, {Prada},
  {Price-Whelan}, {Raddick}, {Ramos}, {Reyl{\'e}}, {Rich}, {Richards}, {Rix},
  {Robin}, {Rocha-Pinto}, {Rockosi}, {Roe}, {Rollinde}, {Ross}, {Ross},
  {Rossetto}, {S{\'a}nchez}, {Sayres}, {Schlegel}, {Schlesinger}, {Schmidt},
  {Schneider}, {Sheldon}, {Shu}, {Simmerer}, {Simmons}, {Sivarani}, {Snedden},
  {Sobeck}, {Steinmetz}, {Strauss}, {Szalay}, {Tanaka}, {Thakar}, {Thomas},
  {Tinker}, {Tofflemire}, {Tojeiro}, {Tremonti}, {Vandenberg}, {Vargas
  Maga{\~n}a}, {Verde}, {Vogt}, {Wake}, {Wang}, {Weaver}, {Weinberg}, {White},
  {White}, {Yanny}, {Yasuda}, {Yeche}, \& {Zehavi}}]{aihara11}
{Aihara}, H., {Allende Prieto}, C., {An}, D., {et~al.} 2011, ApJS, 193, 29,
  \dodoi{10.1088/0067-0049/193/2/29}

\bibitem[{{Almgren} {et~al.}(2013){Almgren}, {Bell}, {Lijewski}, {Luki{\'c}},
  \& {Van Andel}}]{almgren13}
{Almgren}, A.~S., {Bell}, J.~B., {Lijewski}, M.~J., {Luki{\'c}}, Z., \& {Van
  Andel}, E. 2013, \apj, 765, 39, \dodoi{10.1088/0004-637X/765/1/39}

\bibitem[{Anderson {et~al.}(1999)Anderson, Hogan, Williams, \&
  Carswell}]{anderson99}
Anderson, S.~F., Hogan, C.~J., Williams, B.~F., \& Carswell, R.~F. 1999, AJ,
  117, 56

\bibitem[{{Bahcall}(1979)}]{bahcall79}
{Bahcall}, J.~N. 1979, in NASA Conference Publication, Vol. 2111, Scientific
  Research with the Space Telescope: IAU Colloquium No. 54, ed. M.~S. {Longair}
  \& J.~W. {Warner}, 215--240

\bibitem[{Becker \& Bolton(2013)}]{becker13b}
Becker, G.~D., \& Bolton, J.~S. 2013, MNRAS, 436, 1023

\bibitem[{Becker {et~al.}(2011)Becker, Bolton, Haehnelt, \& Sargent}]{becker11}
Becker, G.~D., Bolton, J.~S., Haehnelt, M.~G., \& Sargent, W.~L.~W. 2011,
  MNRAS, 410, 1096

\bibitem[{{Becker} {et~al.}(2015){Becker}, {Bolton}, {Madau}, {Pettini},
  {Ryan-Weber}, \& {Venemans}}]{becker15}
{Becker}, G.~D., {Bolton}, J.~S., {Madau}, P., {et~al.} 2015, \mnras, 447,
  3402, \dodoi{10.1093/mnras/stu2646}

\bibitem[{{Becker} {et~al.}(2018){Becker}, {Davies}, {Furlanetto}, {Malkan},
  {Boera}, \& {Douglass}}]{becker18}
{Becker}, G.~D., {Davies}, F.~B., {Furlanetto}, S.~R., {et~al.} 2018, \apj,
  863, 92, \dodoi{10.3847/1538-4357/aacc73}

\bibitem[{{Boera} {et~al.}(2014){Boera}, {Murphy}, {Becker}, \&
  {Bolton}}]{boera14}
{Boera}, E., {Murphy}, M.~T., {Becker}, G.~D., \& {Bolton}, J.~S. 2014, \mnras,
  441, 1916, \dodoi{10.1093/mnras/stu660}

\bibitem[{Bolton \& Becker(2009)}]{bolton09c}
Bolton, J.~S., \& Becker, G. 2009, MNRAS, 398, L26

\bibitem[{{Bolton} {et~al.}(2014){Bolton}, {Becker}, {Haehnelt}, \&
  {Viel}}]{bolton14}
{Bolton}, J.~S., {Becker}, G.~D., {Haehnelt}, M.~G., \& {Viel}, M. 2014,
  \mnras, 438, 2499, \dodoi{10.1093/mnras/stt2374}

\bibitem[{{Bolton} {et~al.}(2010){Bolton}, {Becker}, {Wyithe}, {Haehnelt}, \&
  {Sargent}}]{bolton10}
{Bolton}, J.~S., {Becker}, G.~D., {Wyithe}, J.~S.~B., {Haehnelt}, M.~G., \&
  {Sargent}, W.~L.~W. 2010, \mnras, 406, 612,
  \dodoi{10.1111/j.1365-2966.2010.16701.x}

\bibitem[{Bolton {et~al.}(2009{\natexlab{a}})Bolton, Oh, \&
  Furlanetto}]{bolton09}
Bolton, J.~S., Oh, S.~P., \& Furlanetto, S.~R. 2009{\natexlab{a}}, MNRAS, 395,
  736

\bibitem[{Bolton {et~al.}(2009{\natexlab{b}})Bolton, Oh, \&
  Furlanetto}]{bolton09b}
---. 2009{\natexlab{b}}, MNRAS, 396, 2405

\bibitem[{{Bolton} {et~al.}(2008){Bolton}, {Viel}, {Kim}, {Haehnelt}, \&
  {Carswell}}]{bolton08}
{Bolton}, J.~S., {Viel}, M., {Kim}, T.-S., {Haehnelt}, M.~G., \& {Carswell},
  R.~F. 2008, \mnras, 386, 1131, \dodoi{10.1111/j.1365-2966.2008.13114.x}

\bibitem[{Bolton {et~al.}(2012)}]{bolton12}
Bolton, J.~S., {et~al.} 2012, MNRAS, 419, 2880

\bibitem[{{Bosman} {et~al.}(2018){Bosman}, {Fan}, {Jiang}, {Reed}, {Matsuoka},
  {Becker}, \& {Haehnelt}}]{bosman18}
{Bosman}, S.~E.~I., {Fan}, X., {Jiang}, L., {et~al.} 2018, \mnras,
  \dodoi{10.1093/mnras/sty1344}

\bibitem[{{Bovy} {et~al.}(2011{\natexlab{a}}){Bovy}, {Hogg}, \&
  {Roweis}}]{bovy11b}
{Bovy}, J., {Hogg}, D.~W., \& {Roweis}, S.~T. 2011{\natexlab{a}}, Annals of
  Applied Statistics, 5, \dodoi{10.1214/10-AOAS439}

\bibitem[{{Bovy} {et~al.}(2011{\natexlab{b}}){Bovy}, {Hennawi}, {Hogg},
  {Myers}, {Kirkpatrick}, {Schlegel}, {Ross}, {Sheldon}, {McGreer},
  {Schneider}, \& {Weaver}}]{bovy11}
{Bovy}, J., {Hennawi}, J.~F., {Hogg}, D.~W., {et~al.} 2011{\natexlab{b}}, \apj,
  729, 141, \dodoi{10.1088/0004-637X/729/2/141}

\bibitem[{{Bryan} \& {Machacek}(2000)}]{bryan00}
{Bryan}, G.~L., \& {Machacek}, M.~E. 2000, \apj, 534, 57,
  \dodoi{10.1086/308735}

\bibitem[{Calura {et~al.}(2012)Calura, Tescari, D'Odorico, Viel, Cristiani,
  Kim, \& Bolton}]{calura12}
Calura, F., Tescari, E., D'Odorico, V., {et~al.} 2012, MNRAS, 422, 3019

\bibitem[{Cardelli {et~al.}(1989)Cardelli, Clayton, \& Mathis}]{cardelli89}
Cardelli, J.~A., Clayton, G.~C., \& Mathis, J.~S. 1989, ApJ, 345, 245

\bibitem[{{Chambers} {et~al.}(2016){Chambers}, {Magnier}, {Metcalfe},
  {Flewelling}, {Huber}, {Waters}, {Denneau}, {Draper}, {Farrow}, {Finkbeiner},
  {Holmberg}, {Koppenhoefer}, {Price}, {Saglia}, {Schlafly}, {Smartt},
  {Sweeney}, {Wainscoat}, {Burgett}, {Grav}, {Heasley}, {Hodapp}, {Jedicke},
  {Kaiser}, {Kudritzki}, {Luppino}, {Lupton}, {Monet}, {Morgan}, {Onaka},
  {Stubbs}, {Tonry}, {Banados}, {Bell}, {Bender}, {Bernard}, {Botticella},
  {Casertano}, {Chastel}, {Chen}, {Chen}, {Cole}, {Deacon}, {Frenk},
  {Fitzsimmons}, {Gezari}, {Goessl}, {Goggia}, {Goldman}, {Grebel}, {Hambly},
  {Hasinger}, {Heavens}, {Heckman}, {Henderson}, {Henning}, {Holman}, {Hopp},
  {Ip}, {Isani}, {Keyes}, {Koekemoer}, {Kotak}, {Long}, {Lucey}, {Liu},
  {Martin}, {McLean}, {Morganson}, {Murphy}, {Nieto-Santisteban}, {Norberg},
  {Peacock}, {Pier}, {Postman}, {Primak}, {Rae}, {Rest}, {Riess}, {Riffeser},
  {Rix}, {Roser}, {Schilbach}, {Schultz}, {Scolnic}, {Szalay}, {Seitz},
  {Shiao}, {Small}, {Smith}, {Soderblom}, {Taylor}, {Thakar}, {Thiel},
  {Thilker}, {Urata}, {Valenti}, {Walter}, {Watters}, {Werner}, {White},
  {Wood-Vasey}, \& {Wyse}}]{chambers16}
{Chambers}, K.~C., {Magnier}, E.~A., {Metcalfe}, N., {et~al.} 2016, ArXiv
  e-prints.
\newblock \doarXiv{1612.05560}

\bibitem[{{Chardin} {et~al.}(2015){Chardin}, {Haehnelt}, {Aubert}, \&
  {Puchwein}}]{chardin15}
{Chardin}, J., {Haehnelt}, M.~G., {Aubert}, D., \& {Puchwein}, E. 2015, \mnras,
  453, 2943, \dodoi{10.1093/mnras/stv1786}

\bibitem[{{Chardin} {et~al.}(2017){Chardin}, {Puchwein}, \&
  {Haehnelt}}]{chardin17}
{Chardin}, J., {Puchwein}, E., \& {Haehnelt}, M.~G. 2017, \mnras, 465, 3429,
  \dodoi{10.1093/mnras/stw2943}

\bibitem[{{Compostella} {et~al.}(2013){Compostella}, {Cantalupo}, \&
  {Porciani}}]{compostella13}
{Compostella}, M., {Cantalupo}, S., \& {Porciani}, C. 2013, \mnras, 435, 3169,
  \dodoi{10.1093/mnras/stt1510}

\bibitem[{{Compostella} {et~al.}(2014){Compostella}, {Cantalupo}, \&
  {Porciani}}]{compostella14}
---. 2014, \mnras, 445, 4186, \dodoi{10.1093/mnras/stu2035}

\bibitem[{Croft {et~al.}(1997)Croft, Weinberg, Katz, \& Hernquist}]{croft97}
Croft, R.~A.~C., Weinberg, D.~H., Katz, N., \& Hernquist, L. 1997, ApJ, 488,
  532

\bibitem[{{D'Aloisio} {et~al.}(2018){D'Aloisio}, {McQuinn}, {Davies}, \&
  {Furlanetto}}]{daloisio18}
{D'Aloisio}, A., {McQuinn}, M., {Davies}, F.~B., \& {Furlanetto}, S.~R. 2018,
  \mnras, 473, 560, \dodoi{10.1093/mnras/stx2341}

\bibitem[{{D'Aloisio} {et~al.}(2015){D'Aloisio}, {McQuinn}, \&
  {Trac}}]{daloisio15}
{D'Aloisio}, A., {McQuinn}, M., \& {Trac}, H. 2015, \apjl, 813, L38,
  \dodoi{10.1088/2041-8205/813/2/L38}

\bibitem[{{D'Aloisio} {et~al.}(2017){D'Aloisio}, {Upton Sanderbeck}, {McQuinn},
  {Trac}, \& {Shapiro}}]{daloisio17}
{D'Aloisio}, A., {Upton Sanderbeck}, P.~R., {McQuinn}, M., {Trac}, H., \&
  {Shapiro}, P.~R. 2017, \mnras, 468, 4691, \dodoi{10.1093/mnras/stx711}

\bibitem[{{Davies} {et~al.}(2018{\natexlab{a}}){Davies}, {Becker}, \&
  {Furlanetto}}]{davies18a}
{Davies}, F.~B., {Becker}, G.~D., \& {Furlanetto}, S.~R. 2018{\natexlab{a}},
  \apj, 860, 155, \dodoi{10.3847/1538-4357/aac2d6}

\bibitem[{Davies \& Furlanetto(2014)}]{davies14}
Davies, F.~B., \& Furlanetto, S.~R. 2014, MNRAS, 437, 1141

\bibitem[{{Davies} \& {Furlanetto}(2016)}]{davies16}
{Davies}, F.~B., \& {Furlanetto}, S.~R. 2016, \mnras, 460, 1328,
  \dodoi{10.1093/mnras/stw931}

\bibitem[{Davies {et~al.}(2017)Davies, Furlanetto, \& Dixon}]{davies17}
Davies, F.~B., Furlanetto, S.~R., \& Dixon, K.~L. 2017, MNRAS, 465, 2886

\bibitem[{{Davies} {et~al.}(2018{\natexlab{b}}){Davies}, {Hennawi},
  {Ba{\~n}ados}, {Luki{\'c}}, {Decarli}, {Fan}, {Farina}, {Mazzucchelli},
  {Rix}, {Venemans}, {Walter}, {Wang}, \& {Yang}}]{davies18b}
{Davies}, F.~B., {Hennawi}, J.~F., {Ba{\~n}ados}, E., {et~al.}
  2018{\natexlab{b}}, \apj, 864, 142, \dodoi{10.3847/1538-4357/aad6dc}

\bibitem[{{Eilers} {et~al.}(2018){Eilers}, {Davies}, \& {Hennawi}}]{eilers18}
{Eilers}, A.-C., {Davies}, F.~B., \& {Hennawi}, J.~F. 2018, \apj, 864, 53,
  \dodoi{10.3847/1538-4357/aad4fd}

\bibitem[{{Eilers} {et~al.}(2017){Eilers}, {Davies}, {Hennawi}, {Prochaska},
  {Luki{\'c}}, \& {Mazzucchelli}}]{eilers17}
{Eilers}, A.-C., {Davies}, F.~B., {Hennawi}, J.~F., {et~al.} 2017, \apj, 840,
  24, \dodoi{10.3847/1538-4357/aa6c60}

\bibitem[{Fan {et~al.}(2006)}]{fan06}
Fan, X., {et~al.} 2006, AJ, 132, 117

\bibitem[{Fardal {et~al.}(1998)Fardal, Giroux, \& Shull}]{fardal98}
Fardal, M.~A., Giroux, M.~L., \& Shull, J.~M. 1998, AJ, 115, 2206

\bibitem[{Faucher-Gigu\`{e}re {et~al.}(2008)Faucher-Gigu\`{e}re, Lidz,
  Hernquist, \& Zaldarriaga}]{faucher08b}
Faucher-Gigu\`{e}re, C.-A., Lidz, A., Hernquist, L., \& Zaldarriaga, M. 2008,
  ApJ, 688, 85

\bibitem[{Faucher-Gigu\`{e}re {et~al.}(2009)Faucher-Gigu\`{e}re, Lidz,
  Zaldarriaga, \& Hernquist}]{faucher09}
Faucher-Gigu\`{e}re, C.-A., Lidz, A., Zaldarriaga, M., \& Hernquist, L. 2009,
  ApJ, 703, 1416

\bibitem[{Fechner \& Reimers(2007)}]{fechner07}
Fechner, C., \& Reimers, D. 2007, A\&A, 461, 847

\bibitem[{Fechner {et~al.}(2006)}]{fechner06}
Fechner, C., {et~al.} 2006, A\&A, 455, 91

\bibitem[{Feldman \& Cousins(1998)}]{feldman98}
Feldman, G.~J., \& Cousins, R.~D. 1998, Phys. Rev. D, 57, 3873

\bibitem[{Furlanetto(2009)}]{furlanetto09}
Furlanetto, S.~R. 2009, ApJ, 703, 702

\bibitem[{Furlanetto \& Dixon(2010)}]{furlanetto10}
Furlanetto, S.~R., \& Dixon, K. 2010, ApJ, 714, 355

\bibitem[{{Furlanetto} \& {Oh}(2005)}]{furlanetto05}
{Furlanetto}, S.~R., \& {Oh}, S.~P. 2005, \mnras, 363, 1031,
  \dodoi{10.1111/j.1365-2966.2005.09505.x}

\bibitem[{Furlanetto \& Oh(2008{\natexlab{a}})}]{furlanetto08}
Furlanetto, S.~R., \& Oh, S.~P. 2008{\natexlab{a}}, ApJ, 681, 1

\bibitem[{Furlanetto \& Oh(2008{\natexlab{b}})}]{furlanetto08b}
---. 2008{\natexlab{b}}, ApJ, 682, 14

\bibitem[{{Garaldi} {et~al.}(2019){Garaldi}, {Compostella}, \&
  {Porciani}}]{garaldi19}
{Garaldi}, E., {Compostella}, M., \& {Porciani}, C. 2019, \mnras, 483, 5301,
  \dodoi{10.1093/mnras/sty3414}

\bibitem[{{Garzilli} {et~al.}(2012){Garzilli}, {Bolton}, {Kim}, {Leach}, \&
  {Viel}}]{garzilli12}
{Garzilli}, A., {Bolton}, J.~S., {Kim}, T.-S., {Leach}, S., \& {Viel}, M. 2012,
  \mnras, 424, 1723, \dodoi{10.1111/j.1365-2966.2012.21223.x}

\bibitem[{{Giallongo} {et~al.}(2015){Giallongo}, {Grazian}, {Fiore}, {Fontana},
  {Pentericci}, {Vanzella}, {Dickinson}, {Kocevski}, {Castellano}, {Cristiani},
  {Ferguson}, {Finkelstein}, {Grogin}, {Hathi}, {Koekemoer}, {Newman}, \&
  {Salvato}}]{giallongo15}
{Giallongo}, E., {Grazian}, A., {Fiore}, F., {et~al.} 2015, A\&A, 578, A83,
  \dodoi{10.1051/0004-6361/201425334}

\bibitem[{Gleser {et~al.}(2005)Gleser, Nusser, Benson, Ohno, \&
  Sugiyama}]{gleser05}
Gleser, L., Nusser, A., Benson, A.~J., Ohno, H., \& Sugiyama, N. 2005, MNRAS,
  361, 1399

\bibitem[{Green {et~al.}(2012)}]{green12}
Green, J.~C., {et~al.} 2012, ApJ, 744, 60

\bibitem[{Haardt \& Madau(2012)}]{haardt12}
Haardt, F., \& Madau, P. 2012, ApJ, 746, 125

\bibitem[{{Haehnelt} \& {Steinmetz}(1998)}]{haehnelt98b}
{Haehnelt}, M.~G., \& {Steinmetz}, M. 1998, \mnras, 298, L21,
  \dodoi{10.1046/j.1365-8711.1998.01879.x}

\bibitem[{Heap {et~al.}(2000)}]{heap00}
Heap, S.~R., {et~al.} 2000, ApJ, 534, 69

\bibitem[{{Hiss} {et~al.}(2018){Hiss}, {Walther}, {Hennawi}, {O{\~n}orbe},
  {O'Meara}, {Rorai}, \& {Luki{\'c}}}]{hiss18}
{Hiss}, H., {Walther}, M., {Hennawi}, J.~F., {et~al.} 2018, \apj, 865, 42,
  \dodoi{10.3847/1538-4357/aada86}

\bibitem[{Hogan {et~al.}(1997)Hogan, Anderson, \& Rugers}]{hogan97}
Hogan, C.~J., Anderson, S.~F., \& Rugers, M.~H. 1997, AJ, 113, 1495

\bibitem[{{Hopkins} {et~al.}(2007){Hopkins}, {Richards}, \&
  {Hernquist}}]{hopkins07}
{Hopkins}, P.~F., {Richards}, G.~T., \& {Hernquist}, L. 2007, \apj, 654, 731,
  \dodoi{10.1086/509629}

\bibitem[{Hui \& Gnedin(1997)}]{hui97}
Hui, L., \& Gnedin, N.~Y. 1997, MNRAS, 292, 27

\bibitem[{{Inoue} {et~al.}(2014){Inoue}, {Shimizu}, {Iwata}, \&
  {Tanaka}}]{inoue14}
{Inoue}, A.~K., {Shimizu}, I., {Iwata}, I., \& {Tanaka}, M. 2014, \mnras, 442,
  1805, \dodoi{10.1093/mnras/stu936}

\bibitem[{Jakobsen {et~al.}(1994)}]{jakobsen94}
Jakobsen, P., {et~al.} 1994, Nature, 370, 35

\bibitem[{{Jiang} {et~al.}(2016){Jiang}, {McGreer}, {Fan}, {Strauss},
  {Ba{\~n}ados}, {Becker}, {Bian}, {Farnsworth}, {Shen}, {Wang}, {Wang},
  {Wang}, {White}, {Wu}, {Wu}, {Yang}, \& {Yang}}]{jiang16}
{Jiang}, L., {McGreer}, I.~D., {Fan}, X., {et~al.} 2016, \apj, 833, 222,
  \dodoi{10.3847/1538-4357/833/2/222}

\bibitem[{{Keating} {et~al.}(2018){Keating}, {Puchwein}, \&
  {Haehnelt}}]{keating18}
{Keating}, L.~C., {Puchwein}, E., \& {Haehnelt}, M.~G. 2018, \mnras, 477, 5501,
  \dodoi{10.1093/mnras/sty968}

\bibitem[{{Khaire}(2017)}]{khaire17}
{Khaire}, V. 2017, \mnras, 471, 255, \dodoi{10.1093/mnras/stx1487}

\bibitem[{{Khrykin} {et~al.}(2016){Khrykin}, {Hennawi}, {McQuinn}, \&
  {Worseck}}]{khrykin16}
{Khrykin}, I.~S., {Hennawi}, J.~F., {McQuinn}, M., \& {Worseck}, G. 2016, ApJ,
  824, 133, \dodoi{10.3847/0004-637X/824/2/133}

\bibitem[{{Khrykin} {et~al.}(2019){Khrykin}, {Hennawi}, \&
  {Worseck}}]{khrykin19}
{Khrykin}, I.~S., {Hennawi}, J.~F., \& {Worseck}, G. 2019, \mnras, 484, 3897,
  \dodoi{10.1093/mnras/stz135}

\bibitem[{Kriss {et~al.}(2001)}]{kriss01}
Kriss, G.~A., {et~al.} 2001, Sci, 293, 1112

\bibitem[{{Kulkarni} {et~al.}(2018{\natexlab{a}}){Kulkarni}, {Keating},
  {Haehnelt}, {Bosman}, {Puchwein}, {Chardin}, \& {Aubert}}]{kulkarni18b}
{Kulkarni}, G., {Keating}, L.~C., {Haehnelt}, M.~G., {et~al.}
  2018{\natexlab{a}}, arXiv e-prints.
\newblock \doarXiv{1809.06374}

\bibitem[{{Kulkarni} {et~al.}(2018{\natexlab{b}}){Kulkarni}, {Worseck}, \&
  {Hennawi}}]{kulkarni18}
{Kulkarni}, G., {Worseck}, G., \& {Hennawi}, J.~F. 2018{\natexlab{b}}, ArXiv
  e-prints.
\newblock \doarXiv{1807.09774}

\bibitem[{{La Plante} \& {Trac}(2016)}]{laplante16}
{La Plante}, P., \& {Trac}, H. 2016, \apj, 828, 90,
  \dodoi{10.3847/0004-637X/828/2/90}

\bibitem[{{La Plante} {et~al.}(2017){La Plante}, {Trac}, {Croft}, \&
  {Cen}}]{laplante17}
{La Plante}, P., {Trac}, H., {Croft}, R., \& {Cen}, R. 2017, \apj, 841, 87,
  \dodoi{10.3847/1538-4357/aa7136}

\bibitem[{{Lee} {et~al.}(2015){Lee}, {Hennawi}, {Spergel}, {Weinberg}, {Hogg},
  {Viel}, {Bolton}, {Bailey}, {Pieri}, {Carithers}, {Schlegel}, {Lundgren},
  {Palanque-Delabrouille}, {Suzuki}, {Schneider}, \& {Y{\`e}che}}]{lee15}
{Lee}, K.-G., {Hennawi}, J.~F., {Spergel}, D.~N., {et~al.} 2015, \apj, 799,
  196, \dodoi{10.1088/0004-637X/799/2/196}

\bibitem[{Lidz {et~al.}(2010)Lidz, Faucher-Gigu\`{e}re, Dall'Aglio, McQuinn,
  Fechner, Zaldarriaga, Hernquist, \& Dutta}]{lidz10}
Lidz, A., Faucher-Gigu\`{e}re, C.-A., Dall'Aglio, A., {et~al.} 2010, ApJ, 718,
  199

\bibitem[{{Luki{\'c}} {et~al.}(2015){Luki{\'c}}, {Stark}, {Nugent}, {White},
  {Meiksin}, \& {Almgren}}]{lukic15}
{Luki{\'c}}, Z., {Stark}, C.~W., {Nugent}, P., {et~al.} 2015, \mnras, 446,
  3697, \dodoi{10.1093/mnras/stu2377}

\bibitem[{{Madau}(2017)}]{madau17}
{Madau}, P. 2017, \apj, 851, 50, \dodoi{10.3847/1538-4357/aa9715}

\bibitem[{{Madau} \& {Haardt}(2015)}]{madau15}
{Madau}, P., \& {Haardt}, F. 2015, \apjl, 813, L8,
  \dodoi{10.1088/2041-8205/813/1/L8}

\bibitem[{Madau {et~al.}(1999)Madau, Haardt, \& Rees}]{madau99}
Madau, P., Haardt, F., \& Rees, M.~J. 1999, ApJ, 514, 648

\bibitem[{Madau \& Meiksin(1994)}]{madau94}
Madau, P., \& Meiksin, A. 1994, ApJ, 433, L53

\bibitem[{Martin {et~al.}(2005)}]{martin05}
Martin, D.~C., {et~al.} 2005, ApJ, 619, L1

\bibitem[{{Matsuoka} {et~al.}(2018){Matsuoka}, {Strauss}, {Kashikawa}, {Onoue},
  {Iwasawa}, {Tang}, {Lee}, {Imanishi}, {Nagao}, {Akiyama}, {Asami}, {Bosch},
  {Furusawa}, {Goto}, {Gunn}, {Harikane}, {Ikeda}, {Izumi}, {Kawaguchi},
  {Kato}, {Kikuta}, {Kohno}, {Komiyama}, {Lupton}, {Minezaki}, {Miyazaki},
  {Murayama}, {Niida}, {Nishizawa}, {Noboriguchi}, {Oguri}, {Ono}, {Ouchi},
  {Price}, {Sameshima}, {Schulze}, {Shirakata}, {Silverman}, {Sugiyama},
  {Tait}, {Takada}, {Takata}, {Tanaka}, {Toba}, {Utsumi}, {Wang}, \&
  {Yamashita}}]{matsuoka18}
{Matsuoka}, Y., {Strauss}, M.~A., {Kashikawa}, N., {et~al.} 2018, \apj, 869,
  150, \dodoi{10.3847/1538-4357/aaee7a}

\bibitem[{McDonald {et~al.}(2001)McDonald, Miralda-Escud\'{e}, Rauch, Sargent,
  Barlow, \& Cen}]{mcdonald01b}
McDonald, P., Miralda-Escud\'{e}, J., Rauch, M., {et~al.} 2001, ApJ, 562, 52

\bibitem[{{McGreer} {et~al.}(2018){McGreer}, {Fan}, {Jiang}, \&
  {Cai}}]{mcgreer18}
{McGreer}, I.~D., {Fan}, X., {Jiang}, L., \& {Cai}, Z. 2018, \aj, 155, 131,
  \dodoi{10.3847/1538-3881/aaaab4}

\bibitem[{{McGreer} {et~al.}(2013){McGreer}, {Jiang}, {Fan}, {Richards},
  {Strauss}, {Ross}, {White}, {Shen}, {Schneider}, {Myers}, {Brandt}, {DeGraf},
  {Glikman}, {Ge}, \& {Streblyanska}}]{mcgreer13}
{McGreer}, I.~D., {Jiang}, L., {Fan}, X., {et~al.} 2013, \apj, 768, 105,
  \dodoi{10.1088/0004-637X/768/2/105}

\bibitem[{McQuinn(2009)}]{mcquinn09b}
McQuinn, M. 2009, ApJ, 704, L89

\bibitem[{{McQuinn} \& {Upton Sanderbeck}(2016)}]{mcquinn16}
{McQuinn}, M., \& {Upton Sanderbeck}, P.~R. 2016, \mnras, 456, 47,
  \dodoi{10.1093/mnras/stv2675}

\bibitem[{{McQuinn} \& {Worseck}(2014)}]{mcquinn14}
{McQuinn}, M., \& {Worseck}, G. 2014, \mnras, 440, 2406,
  \dodoi{10.1093/mnras/stu242}

\bibitem[{McQuinn {et~al.}(2009)}]{mcquinn09a}
McQuinn, M., {et~al.} 2009, ApJ, 694, 842

\bibitem[{{Meiksin} \& {Tittley}(2012)}]{tittley12}
{Meiksin}, A., \& {Tittley}, E.~R. 2012, \mnras, 423, 7,
  \dodoi{10.1111/j.1365-2966.2011.20380.x}

\bibitem[{Miralda-Escud\'{e} {et~al.}(2000)Miralda-Escud\'{e}, Haehnelt, \&
  Rees}]{miralda-escude00}
Miralda-Escud\'{e}, J., Haehnelt, M., \& Rees, M.~J. 2000, ApJ, 530, 1

\bibitem[{{Mitra} {et~al.}(2018){Mitra}, {Choudhury}, \& {Ferrara}}]{mitra18}
{Mitra}, S., {Choudhury}, T.~R., \& {Ferrara}, A. 2018, \mnras, 473, 1416,
  \dodoi{10.1093/mnras/stx2443}

\bibitem[{Morrissey {et~al.}(2007)}]{morrissey07}
Morrissey, P., {et~al.} 2007, ApJS, 173, 682

\bibitem[{{Murthy}(2014)}]{murthy14}
{Murthy}, J. 2014, \apjs, 213, 32, \dodoi{10.1088/0067-0049/213/2/32}

\bibitem[{{O{\~n}orbe} {et~al.}(2017){O{\~n}orbe}, {Hennawi}, \&
  {Luki{\'c}}}]{onorbe17}
{O{\~n}orbe}, J., {Hennawi}, J.~F., \& {Luki{\'c}}, Z. 2017, \apj, 837, 106,
  \dodoi{10.3847/1538-4357/aa6031}

\bibitem[{{Parsa} {et~al.}(2018){Parsa}, {Dunlop}, \& {McLure}}]{parsa18}
{Parsa}, S., {Dunlop}, J.~S., \& {McLure}, R.~J. 2018, \mnras, 474, 2904,
  \dodoi{10.1093/mnras/stx2887}

\bibitem[{Picard \& Jakobsen(1993)}]{picard93}
Picard, A., \& Jakobsen, P. 1993, A\&A, 276, 331

\bibitem[{{Planck Collaboration} {et~al.}(2018){Planck Collaboration},
  {Aghanim}, {Akrami}, {Ashdown}, {Aumont}, {Baccigalupi}, {Ballardini},
  {Banday}, {Barreiro}, {Bartolo}, {Basak}, {Battye}, {Benabed}, {Bernard},
  {Bersanelli}, {Bielewicz}, {Bock}, {Bond}, {Borrill}, {Bouchet}, {Boulanger},
  {Bucher}, {Burigana}, {Butler}, {Calabrese}, {Cardoso}, {Carron},
  {Challinor}, {Chiang}, {Chluba}, {Colombo}, {Combet}, {Contreras}, {Crill},
  {Cuttaia}, {de Bernardis}, {de Zotti}, {Delabrouille}, {Delouis}, {Di
  Valentino}, {Diego}, {Dor{\'e}}, {Douspis}, {Ducout}, {Dupac}, {Dusini},
  {Efstathiou}, {Elsner}, {En{\ss}lin}, {Eriksen}, {Fantaye}, {Farhang},
  {Fergusson}, {Fernandez-Cobos}, {Finelli}, {Forastieri}, {Frailis},
  {Franceschi}, {Frolov}, {Galeotta}, {Galli}, {Ganga}, {G{\'e}nova-Santos},
  {Gerbino}, {Ghosh}, {Gonz{\'a}lez-Nuevo}, {G{\'o}rski}, {Gratton},
  {Gruppuso}, {Gudmundsson}, {Hamann}, {Handley}, {Herranz}, {Hivon}, {Huang},
  {Jaffe}, {Jones}, {Karakci}, {Keih{\"a}nen}, {Keskitalo}, {Kiiveri}, {Kim},
  {Kisner}, {Knox}, {Krachmalnicoff}, {Kunz}, {Kurki-Suonio}, {Lagache},
  {Lamarre}, {Lasenby}, {Lattanzi}, {Lawrence}, {Le Jeune}, {Lemos},
  {Lesgourgues}, {Levrier}, {Lewis}, {Liguori}, {Lilje}, {Lilley}, {Lindholm},
  {L{\'o}pez-Caniego}, {Lubin}, {Ma}, {Mac{\'{\i}}as-P{\'e}rez}, {Maggio},
  {Maino}, {Mandolesi}, {Mangilli}, {Marcos-Caballero}, {Maris}, {Martin},
  {Martinelli}, {Mart{\'{\i}}nez-Gonz{\'a}lez}, {Matarrese}, {Mauri}, {McEwen},
  {Meinhold}, {Melchiorri}, {Mennella}, {Migliaccio}, {Millea}, {Mitra},
  {Miville-Desch{\^e}nes}, {Molinari}, {Montier}, {Morgante}, {Moss}, {Natoli},
  {N{\o}rgaard-Nielsen}, {Pagano}, {Paoletti}, {Partridge}, {Patanchon},
  {Peiris}, {Perrotta}, {Pettorino}, {Piacentini}, {Polastri}, {Polenta},
  {Puget}, {Rachen}, {Reinecke}, {Remazeilles}, {Renzi}, {Rocha}, {Rosset},
  {Roudier}, {Rubi{\~n}o-Mart{\'{\i}}n}, {Ruiz-Granados}, {Salvati}, {Sandri},
  {Savelainen}, {Scott}, {Shellard}, {Sirignano}, {Sirri}, {Spencer},
  {Sunyaev}, {Suur-Uski}, {Tauber}, {Tavagnacco}, {Tenti}, {Toffolatti},
  {Tomasi}, {Trombetti}, {Valenziano}, {Valiviita}, {Van Tent}, {Vibert},
  {Vielva}, {Villa}, {Vittorio}, {Wandelt}, {Wehus}, {White}, {White},
  {Zacchei}, \& {Zonca}}]{planckcollab18}
{Planck Collaboration}, {Aghanim}, N., {Akrami}, Y., {et~al.} 2018, ArXiv
  e-prints.
\newblock \doarXiv{1807.06209}

\bibitem[{{Prochaska} {et~al.}(2014){Prochaska}, {Madau}, {O'Meara}, \&
  {Fumagalli}}]{prochaska14}
{Prochaska}, J.~X., {Madau}, P., {O'Meara}, J.~M., \& {Fumagalli}, M. 2014,
  \mnras, 438, 476, \dodoi{10.1093/mnras/stt2218}

\bibitem[{{Proft} \& {Wambsganss}(2015)}]{proft15}
{Proft}, S., \& {Wambsganss}, J. 2015, \aap, 574, A46,
  \dodoi{10.1051/0004-6361/201323280}

\bibitem[{{Puchwein} {et~al.}(2015){Puchwein}, {Bolton}, {Haehnelt}, {Madau},
  {Becker}, \& {Haardt}}]{puchwein15}
{Puchwein}, E., {Bolton}, J.~S., {Haehnelt}, M.~G., {et~al.} 2015, \mnras, 450,
  4081, \dodoi{10.1093/mnras/stv773}

\bibitem[{{Puchwein} {et~al.}(2019){Puchwein}, {Haardt}, {Haehnelt}, \&
  {Madau}}]{puchwein19}
{Puchwein}, E., {Haardt}, F., {Haehnelt}, M.~G., \& {Madau}, P. 2019, \mnras,
  \dodoi{10.1093/mnras/stz222}

\bibitem[{Reimers {et~al.}(1997)}]{reimers97}
Reimers, D., {et~al.} 1997, A\&A, 327, 890

\bibitem[{Reimers {et~al.}(2005)}]{reimers05}
---. 2005, A\&A, 442, 63

\bibitem[{{Ricci} {et~al.}(2017){Ricci}, {Marchesi}, {Shankar}, {La Franca}, \&
  {Civano}}]{ricci17}
{Ricci}, F., {Marchesi}, S., {Shankar}, F., {La Franca}, F., \& {Civano}, F.
  2017, \mnras, 465, 1915, \dodoi{10.1093/mnras/stw2909}

\bibitem[{Ricotti {et~al.}(2000)Ricotti, Gnedin, \& Shull}]{ricotti00}
Ricotti, M., Gnedin, N.~Y., \& Shull, J.~M. 2000, ApJ, 534, 41

\bibitem[{{Rorai} {et~al.}(2018){Rorai}, {Carswell}, {Haehnelt}, {Becker},
  {Bolton}, \& {Murphy}}]{rorai18}
{Rorai}, A., {Carswell}, R.~F., {Haehnelt}, M.~G., {et~al.} 2018, \mnras, 474,
  2871, \dodoi{10.1093/mnras/stx2862}

\bibitem[{{Rorai} {et~al.}(2017{\natexlab{a}}){Rorai}, {Becker}, {Haehnelt},
  {Carswell}, {Bolton}, {Cristiani}, {D'Odorico}, {Cupani}, {Barai}, {Calura},
  {Kim}, {Pomante}, {Tescari}, \& {Viel}}]{rorai17a}
{Rorai}, A., {Becker}, G.~D., {Haehnelt}, M.~G., {et~al.} 2017{\natexlab{a}},
  \mnras, 466, 2690, \dodoi{10.1093/mnras/stw2917}

\bibitem[{{Rorai} {et~al.}(2017{\natexlab{b}}){Rorai}, {Hennawi}, {O{\~n}orbe},
  {White}, {Prochaska}, {Kulkarni}, {Walther}, {Luki{\'c}}, \&
  {Lee}}]{rorai17b}
{Rorai}, A., {Hennawi}, J.~F., {O{\~n}orbe}, J., {et~al.} 2017{\natexlab{b}},
  Science, 356, 418, \dodoi{10.1126/science.aaf9346}

\bibitem[{{Rudie} {et~al.}(2012){Rudie}, {Steidel}, \& {Pettini}}]{rudie12}
{Rudie}, G.~C., {Steidel}, C.~C., \& {Pettini}, M. 2012, \apjl, 757, L30,
  \dodoi{10.1088/2041-8205/757/2/L30}

\bibitem[{{Sahnow} {et~al.}(2016){Sahnow}, {Penton}, {Ake}, {DeRosa}, {Ely},
  {Lockwood}, {Oliveira}, {Plesha}, {Proffitt}, {Roman-Duval}, {Sonnentrucker},
  {Taylor}, \& {White}}]{sahnow16}
{Sahnow}, D.~J., {Penton}, S., {Ake}, T., {et~al.} 2016, in \procspie, Vol.
  9905, Society of Photo-Optical Instrumentation Engineers (SPIE) Conference
  Series, 99052T

\bibitem[{Schaye {et~al.}(2000)Schaye, Theuns, Rauch, Efstathiou, \&
  Sargent}]{schaye00}
Schaye, J., Theuns, T., Rauch, M., Efstathiou, G., \& Sargent, W.~L.~W. 2000,
  MNRAS, 318, 817

\bibitem[{Schlegel {et~al.}(1998)Schlegel, Finkbeiner, \& Davis}]{schlegel98}
Schlegel, D.~J., Finkbeiner, D.~P., \& Davis, M. 1998, ApJ, 500, 525

\bibitem[{{Schmidt} {et~al.}(2018){Schmidt}, {Hennawi}, {Worseck}, {Davies},
  {Luki{\'c}}, \& {O{\~n}orbe}}]{schmidt18}
{Schmidt}, T.~M., {Hennawi}, J.~F., {Worseck}, G., {et~al.} 2018, \apj, 861,
  122, \dodoi{10.3847/1538-4357/aac8e4}

\bibitem[{Shull {et~al.}(2004)Shull, Tumlinson, Giroux, Kriss, \&
  Reimers}]{shull04}
Shull, J.~M., Tumlinson, J., Giroux, M.~L., Kriss, G.~A., \& Reimers, D. 2004,
  ApJ, 600, 570

\bibitem[{Shull {et~al.}(2010)}]{shull10}
Shull, J.~M., {et~al.} 2010, ApJ, 722, 1312

\bibitem[{Smette {et~al.}(2002)Smette, Heap, Williger, Tripp, Jenkins, \&
  Songaila}]{smette02}
Smette, A., Heap, S.~R., Williger, G.~M., {et~al.} 2002, ApJ, 564, 542

\bibitem[{Sokasian {et~al.}(2002)Sokasian, Abel, \& Hernquist}]{sokasian02}
Sokasian, A., Abel, T., \& Hernquist, L. 2002, MNRAS, 332, 601

\bibitem[{Syphers {et~al.}(2012)Syphers, Anderson, Zheng, Meiksin, Schneider,
  \& York}]{syphers12}
Syphers, D., Anderson, S.~F., Zheng, W., {et~al.} 2012, AJ, 143, 100

\bibitem[{Syphers \& Shull(2013)}]{syphers13}
Syphers, D., \& Shull, J.~M. 2013, ApJ, 765, 119

\bibitem[{{Syphers} \& {Shull}(2014)}]{syphers14}
{Syphers}, D., \& {Shull}, J.~M. 2014, \apj, 784, 42,
  \dodoi{10.1088/0004-637X/784/1/42}

\bibitem[{Syphers {et~al.}(2009{\natexlab{a}})}]{syphers09a}
Syphers, D., {et~al.} 2009{\natexlab{a}}, ApJ, 690, 1181

\bibitem[{Syphers {et~al.}(2009{\natexlab{b}})}]{syphers09b}
---. 2009{\natexlab{b}}, ApJS, 185, 20

\bibitem[{{Theuns} {et~al.}(1998){Theuns}, {Leonard}, {Efstathiou}, {Pearce},
  \& {Thomas}}]{theuns98}
{Theuns}, T., {Leonard}, A., {Efstathiou}, G., {Pearce}, F.~R., \& {Thomas},
  P.~A. 1998, \mnras, 301, 478, \dodoi{10.1046/j.1365-8711.1998.02040.x}

\bibitem[{{Theuns} {et~al.}(2002){Theuns}, {Schaye}, {Zaroubi}, {Kim},
  {Tzanavaris}, \& {Carswell}}]{theuns02c}
{Theuns}, T., {Schaye}, J., {Zaroubi}, S., {et~al.} 2002, \apjl, 567, L103,
  \dodoi{10.1086/339998}

\bibitem[{Theuns {et~al.}(2002)Theuns, Zharoubi, Kim, Tzanavaris, \&
  Carswell}]{theuns02a}
Theuns, T., Zharoubi, S., Kim, T.-S., Tzanavaris, P., \& Carswell, R.~F. 2002,
  MNRAS, 332, 367

\bibitem[{Tittley \& Meiksin(2007)}]{tittley07}
Tittley, E.~R., \& Meiksin, A. 2007, MNRAS, 380, 1369

\bibitem[{{Viel} {et~al.}(2009){Viel}, {Bolton}, \& {Haehnelt}}]{viel09}
{Viel}, M., {Bolton}, J.~S., \& {Haehnelt}, M.~G. 2009, \mnras, 399, L39,
  \dodoi{10.1111/j.1745-3933.2009.00720.x}

\bibitem[{{Wakker} {et~al.}(2015){Wakker}, {Hernandez}, {French}, {Kim},
  {Oppenheimer}, \& {Savage}}]{wakker15}
{Wakker}, B.~P., {Hernandez}, A.~K., {French}, D.~M., {et~al.} 2015, ApJ, 814,
  40, \dodoi{10.1088/0004-637X/814/1/40}

\bibitem[{{Walther} {et~al.}(2018){Walther}, {Hennawi}, {Hiss}, {O{\~n}orbe},
  {Lee}, {Rorai}, \& {O'Meara}}]{walther18}
{Walther}, M., {Hennawi}, J.~F., {Hiss}, H., {et~al.} 2018, \apj, 852, 22,
  \dodoi{10.3847/1538-4357/aa9c81}

\bibitem[{{Walther} {et~al.}(2019){Walther}, {O{\~n}orbe}, {Hennawi}, \&
  {Luki{\'c}}}]{walther19}
{Walther}, M., {O{\~n}orbe}, J., {Hennawi}, J.~F., \& {Luki{\'c}}, Z. 2019,
  \apj, 872, 13, \dodoi{10.3847/1538-4357/aafad1}

\bibitem[{{Wang} {et~al.}(2018){Wang}, {Yang}, {Fan}, {Wu}, {Yue}, {Li},
  {Bian}, {Jiang}, {Ba{\~n}ados}, {Schindler}, {Findlay}, {Davies}, {Decarli},
  {Farina}, {Green}, {Hennawi}, {Huang}, {Mazzuccheli}, {McGreer}, {Venemans},
  {Walter}, {Dye}, {Lyke}, {Myers}, \& {Haze Nunez}}]{wang18}
{Wang}, F., {Yang}, J., {Fan}, X., {et~al.} 2018, arXiv e-prints.
\newblock \doarXiv{1810.11926}

\bibitem[{Worseck \& Prochaska(2011)}]{worseck11}
Worseck, G., \& Prochaska, J.~X. 2011, ApJ, 728, 23

\bibitem[{{Worseck} {et~al.}(2016){Worseck}, {Prochaska}, {Hennawi}, \&
  {McQuinn}}]{worseck16}
{Worseck}, G., {Prochaska}, J.~X., {Hennawi}, J.~F., \& {McQuinn}, M. 2016,
  \apj, 825, 144, \dodoi{10.3847/0004-637X/825/2/144}

\bibitem[{Worseck {et~al.}(2011)}]{worseck11b}
Worseck, G., {et~al.} 2011, ApJ, 733, L24

\bibitem[{{Wright} {et~al.}(2010){Wright}, {Eisenhardt}, {Mainzer}, {Ressler},
  {Cutri}, {Jarrett}, {Kirkpatrick}, {Padgett}, {McMillan}, {Skrutskie},
  {Stanford}, {Cohen}, {Walker}, {Mather}, {Leisawitz}, {Gautier}, {McLean},
  {Benford}, {Lonsdale}, {Blain}, {Mendez}, {Irace}, {Duval}, {Liu}, {Royer},
  {Heinrichsen}, {Howard}, {Shannon}, {Kendall}, {Walsh}, {Larsen}, {Cardon},
  {Schick}, {Schwalm}, {Abid}, {Fabinsky}, {Naes}, \& {Tsai}}]{wright10}
{Wright}, E.~L., {Eisenhardt}, P.~R.~M., {Mainzer}, A.~K., {et~al.} 2010, AJ,
  140, 1868, \dodoi{10.1088/0004-6256/140/6/1868}

\bibitem[{Wyithe \& Loeb(2003)}]{wyithe03}
Wyithe, J.~S.~B., \& Loeb, A. 2003, ApJ, 586, 693

\bibitem[{{Zaldarriaga} {et~al.}(2001){Zaldarriaga}, {Hui}, \&
  {Tegmark}}]{zaldarriaga01}
{Zaldarriaga}, M., {Hui}, L., \& {Tegmark}, M. 2001, \apj, 557, 519,
  \dodoi{10.1086/321652}

\bibitem[{{Zheng} {et~al.}(2015){Zheng}, {Syphers}, {Meiksin}, {Kriss},
  {Schneider}, {York}, \& {Anderson}}]{zheng15}
{Zheng}, W., {Syphers}, D., {Meiksin}, A., {et~al.} 2015, \apj, 806, 142,
  \dodoi{10.1088/0004-637X/806/1/142}

\bibitem[{Zheng {et~al.}(2004{\natexlab{a}})}]{zheng04b}
Zheng, W., {et~al.} 2004{\natexlab{a}}, AJ, 127, 656

\bibitem[{Zheng {et~al.}(2004{\natexlab{b}})}]{zheng04}
---. 2004{\natexlab{b}}, ApJ, 605, 631

\bibitem[{Zheng {et~al.}(2005)}]{zheng05}
Zheng, W., {et~al.} 2005, in IAU Colloquium 199: Probing Galaxies through
  Quasar Absorption Lines, ed. P.~R. Williams, C.~Shu, \& B.~M\'{e}nard
  (Cambridge University Press), 484

\bibitem[{Zheng {et~al.}(2008)}]{zheng08}
---. 2008, ApJ, 686, 195

\end{thebibliography}

\appendix\twocolumngrid

\section{Measured \ion{He}{2} effective optical depths}
\label{sect:taueffapp}
\startlongtable
\begin{deluxetable}{lCRrr}
\tabletypesize{\footnotesize}
\tablewidth{0pt}
\renewcommand{\arraystretch}{1.0}
\tablecaption{\label{tab:he2tau}Measured \teff\ in new or reanalyzed spectra (Table~\ref{tab:qsosample})}
\tablehead{
\colhead{Quasar}&\colhead{$z$}&\colhead{\teff}&\colhead{stat. 1$\sigma$ error}&\colhead{sys. error}
}
\startdata
HS~1024$+$1849      &2.60  &1.69  &$^{+0.61}_{-0.01}$    &$^{+0.00}_{-0.21}$\\
                    &2.64  &1.12  &$^{+0.10}_{-0.03}$    &$^{+0.00}_{-0.04}$\\
                    &2.68  &2.19  &$^{+0.08}_{-0.11}$    &$^{+0.03}_{-0.00}$\\
                    &2.72  &2.73  &$^{+0.15}_{-0.09}$    &$^{+0.00}_{-0.03}$\\
                    &2.76  &3.04  &$^{+0.18}_{-0.09}$    &$^{+0.00}_{-0.04}$\\
Q~1602$+$576        &2.56  &1.37  &$^{+0.24}_{-0.15}$    &$^{+0.00}_{-0.04}$\\
                    &2.60  &1.43  &$^{+0.04}_{-0.11}$    &$^{+0.05}_{-0.00}$\\
                    &2.64  &1.09  &$^{+0.04}_{-0.03}$    &$^{+0.00}_{-0.00}$\\
                    &2.68  &1.58  &$^{+0.03}_{-0.03}$    &$^{+0.00}_{-0.00}$\\
                    &2.72  &1.75  &$^{+0.02}_{-0.04}$    &$^{+0.00}_{-0.00}$\\
                    &2.76  &1.55  &$^{+0.02}_{-0.02}$    &$^{+0.00}_{-0.00}$\\
                    &2.80  &2.29  &$^{+0.02}_{-0.05}$    &$^{+0.02}_{-0.00}$\\
HE2QS~J2157$+$2330  &2.68  &2.42  &$^{+0.18}_{-0.19}$    &$^{+0.03}_{-0.00}$\\
                    &2.72  &2.24  &$^{+0.13}_{-0.12}$    &$^{+0.00}_{-0.00}$\\
                    &2.76  &2.01  &$^{+0.09}_{-0.09}$    &$^{+0.00}_{-0.00}$\\
                    &2.80  &1.68  &$^{+0.07}_{-0.07}$    &$^{+0.00}_{-0.00}$\\
                    &2.84  &2.26  &$^{+0.09}_{-0.09}$    &$^{+0.00}_{-0.00}$\\
                    &2.88  &2.20  &$^{+0.10}_{-0.09}$    &$^{+0.00}_{-0.00}$\\
HE2QS~J2149$-$0859  &2.72  &1.17  &$^{+0.16}_{-0.13}$    &$^{+0.01}_{-0.01}$\\
                    &2.76  &1.52  &$^{+0.13}_{-0.13}$    &$^{+0.02}_{-0.00}$\\
                    &2.80  &4.19  &$^{+\infty}_{-0.00}$  &$^{+0.00}_{-0.00}$\\
                    &2.84  &1.83  &$^{+0.14}_{-0.12}$    &$^{+0.02}_{-0.01}$\\
                    &2.88  &1.91  &$^{+0.15}_{-0.14}$    &$^{+0.02}_{-0.00}$\\
                    &3.12  &4.50  &$^{+\infty}_{-0.00}$  &$^{+0.00}_{-0.00}$\\
                    &3.16  &4.39  &$^{+\infty}_{-0.00}$  &$^{+0.00}_{-0.00}$\\
                    &3.20  &4.41  &$^{+\infty}_{-0.00}$  &$^{+0.00}_{-0.00}$\\
Q~0302$-$003        &2.76  &1.75  &$^{+0.06}_{-0.06}$    &$^{+0.00}_{-0.00}$\\
                    &2.80  &2.17  &$^{+0.07}_{-0.07}$    &$^{+0.00}_{-0.00}$\\
                    &2.84  &2.07  &$^{+0.05}_{-0.05}$    &$^{+0.00}_{-0.00}$\\
                    &2.88  &3.74  &$^{+0.22}_{-0.18}$    &$^{+0.04}_{-0.02}$\\
                    &2.92  &4.34  &$^{+0.45}_{-0.27}$    &$^{+0.04}_{-0.06}$\\
                    &3.08  &5.29  &$^{+0.93}_{-0.54}$    &$^{+0.60}_{-0.04}$\\
                    &3.12  &5.73  &$^{+\infty}_{-0.00}$  &$^{+0.00}_{-0.00}$\\
                    &3.16  &4.41  &$^{+0.30}_{-0.24}$    &$^{+0.07}_{-0.03}$\\
HE2QS~J0233$-$0149  &2.68  &2.12  &$^{+0.32}_{-0.22}$    &$^{+0.00}_{-0.01}$\\
                    &2.72  &1.93  &$^{+0.15}_{-0.13}$    &$^{+0.00}_{-0.00}$\\
                    &2.76  &2.32  &$^{+0.13}_{-0.12}$    &$^{+0.00}_{-0.00}$\\
                    &2.80  &1.84  &$^{+0.09}_{-0.09}$    &$^{+0.00}_{-0.00}$\\
                    &2.84  &1.47  &$^{+0.07}_{-0.06}$    &$^{+0.00}_{-0.00}$\\
                    &2.88  &3.16  &$^{+0.21}_{-0.18}$    &$^{+0.02}_{-0.00}$\\
                    &3.08  &2.14  &$^{+0.08}_{-0.09}$    &$^{+0.00}_{-0.00}$\\
                    &3.12  &2.89  &$^{+0.14}_{-0.13}$    &$^{+0.00}_{-0.00}$\\
                    &3.16  &5.21  &$^{+\infty}_{-0.00}$  &$^{+0.00}_{-0.00}$\\
                    &3.20  &5.37  &$^{+\infty}_{-0.00}$  &$^{+0.00}_{-0.00}$\\
                    &3.24  &4.45  &$^{+0.64}_{-0.40}$    &$^{+0.06}_{-0.00}$\\
HS~0911$+$4809      &2.72  &2.12  &$^{+0.08}_{-0.07}$    &$^{+0.00}_{-0.00}$\\
                    &2.76  &2.74  &$^{+0.08}_{-0.12}$    &$^{+0.03}_{-0.00}$\\
                    &2.80  &2.06  &$^{+0.09}_{-0.01}$    &$^{+0.00}_{-0.04}$\\
                    &2.84  &2.63  &$^{+0.07}_{-0.07}$    &$^{+0.00}_{-0.00}$\\
                    &2.88  &5.02  &$^{+0.66}_{-0.49}$    &$^{+0.23}_{-0.00}$\\
                    &2.92  &4.66  &$^{+0.45}_{-0.28}$    &$^{+0.00}_{-0.03}$\\
                    &3.08  &4.47  &$^{+0.43}_{-0.28}$    &$^{+0.01}_{-0.03}$\\
                    &3.12  &5.36  &$^{+1.03}_{-0.52}$    &$^{+0.38}_{-0.05}$\\
                    &3.16  &3.30  &$^{+0.09}_{-0.10}$    &$^{+0.01}_{-0.00}$\\
                    &3.20  &4.07  &$^{+0.22}_{-0.16}$    &$^{+0.00}_{-0.03}$\\
                    &3.24  &5.70  &$^{+\infty}_{-0.00}$  &$^{+0.00}_{-0.00}$\\
HE2QS~J0916$+$2408  &2.80  &1.56  &$^{+0.06}_{-0.06}$    &$^{+0.00}_{-0.00}$\\
                    &2.84  &1.47  &$^{+0.05}_{-0.05}$    &$^{+0.00}_{-0.00}$\\
                    &2.88  &1.93  &$^{+0.08}_{-0.07}$    &$^{+0.00}_{-0.00}$\\
                    &3.12  &5.57  &$^{+\infty}_{-0.00}$  &$^{+0.00}_{-0.00}$\\
                    &3.16  &5.42  &$^{+\infty}_{-0.00}$  &$^{+0.00}_{-0.00}$\\
                    &3.20  &5.56  &$^{+\infty}_{-0.00}$  &$^{+0.00}_{-0.00}$\\
                    &3.36  &5.29  &$^{+\infty}_{-0.00}$  &$^{+0.00}_{-0.00}$\\
SDSS~J1253$+$6817   &2.84  &2.95  &$^{+0.15}_{-0.13}$    &$^{+0.01}_{-0.00}$\\
                    &2.88  &2.62  &$^{+0.10}_{-0.11}$    &$^{+0.01}_{-0.00}$\\
                    &3.12  &3.38  &$^{+0.18}_{-0.16}$    &$^{+0.01}_{-0.00}$\\
                    &3.16  &3.02  &$^{+0.14}_{-0.13}$    &$^{+0.00}_{-0.00}$\\
                    &3.20  &5.42  &$^{+\infty}_{-0.00}$  &$^{+0.00}_{-0.00}$\\
                    &3.24  &5.54  &$^{+\infty}_{-0.00}$  &$^{+0.00}_{-0.00}$\\
                    &3.36  &5.57  &$^{+\infty}_{-0.00}$  &$^{+0.00}_{-0.00}$\\
SDSS~J2346$-$0016   &2.84  &3.13  &$^{+0.14}_{-0.13}$    &$^{+0.00}_{-0.00}$\\
                    &2.88  &2.56  &$^{+0.09}_{-0.08}$    &$^{+0.00}_{-0.00}$\\
                    &3.12  &5.36  &$^{+1.05}_{-0.55}$    &$^{+0.36}_{-0.00}$\\
                    &3.16  &5.74  &$^{+\infty}_{-0.00}$  &$^{+0.00}_{-0.00}$\\
                    &3.20  &5.76  &$^{+\infty}_{-0.00}$  &$^{+0.00}_{-0.00}$\\
                    &3.36  &5.88  &$^{+\infty}_{-0.00}$  &$^{+0.00}_{-0.00}$\\
                    &3.40  &5.94  &$^{+\infty}_{-0.00}$  &$^{+0.00}_{-0.00}$\\
                    &3.44  &5.81  &$^{+\infty}_{-0.00}$  &$^{+0.00}_{-0.00}$\\
HE2QS~J2311$-$1417  &3.08  &3.25  &$^{+0.14}_{-0.13}$    &$^{+0.00}_{-0.00}$\\
                    &3.12  &3.91  &$^{+0.24}_{-0.21}$    &$^{+0.02}_{-0.00}$\\
                    &3.16  &5.22  &$^{+1.04}_{-0.56}$    &$^{+0.27}_{-0.00}$\\
                    &3.20  &5.60  &$^{+\infty}_{-0.00}$  &$^{+0.00}_{-0.00}$\\
                    &3.24  &5.56  &$^{+\infty}_{-0.00}$  &$^{+0.00}_{-0.00}$\\
                    &3.36  &5.40  &$^{+\infty}_{-0.00}$  &$^{+0.00}_{-0.00}$\\
                    &3.40  &5.39  &$^{+\infty}_{-0.00}$  &$^{+0.00}_{-0.00}$\\
                    &3.44  &5.24  &$^{+\infty}_{-0.00}$  &$^{+0.00}_{-0.00}$\\
                    &3.48  &5.17  &$^{+\infty}_{-0.00}$  &$^{+0.00}_{-0.00}$\\
                    &3.52  &5.12  &$^{+\infty}_{-0.00}$  &$^{+0.00}_{-0.00}$\\
                    &3.56  &5.19  &$^{+\infty}_{-0.00}$  &$^{+0.00}_{-0.00}$\\
                    &3.60  &5.12  &$^{+\infty}_{-0.00}$  &$^{+0.00}_{-0.00}$\\
                    &3.64  &5.01  &$^{+\infty}_{-0.00}$  &$^{+0.00}_{-0.00}$\\
SDSS~J1137$+$6237   &3.16  &5.23  &$^{+\infty}_{-0.00}$  &$^{+0.00}_{-0.00}$\\
                    &3.20  &5.21  &$^{+\infty}_{-0.00}$  &$^{+0.00}_{-0.00}$\\
                    &3.24  &5.13  &$^{+\infty}_{-0.00}$  &$^{+0.00}_{-0.00}$\\
                    &3.36  &5.01  &$^{+\infty}_{-0.00}$  &$^{+0.00}_{-0.00}$\\
                    &3.40  &5.02  &$^{+\infty}_{-0.00}$  &$^{+0.00}_{-0.00}$\\
                    &3.52  &4.80  &$^{+\infty}_{-0.00}$  &$^{+0.00}_{-0.00}$\\
                    &3.56  &4.80  &$^{+\infty}_{-0.00}$  &$^{+0.00}_{-0.00}$\\
                    &3.60  &4.77  &$^{+\infty}_{-0.00}$  &$^{+0.00}_{-0.00}$\\
                    &3.64  &4.71  &$^{+\infty}_{-0.00}$  &$^{+0.00}_{-0.00}$\\
                    &3.68  &4.68  &$^{+\infty}_{-0.00}$  &$^{+0.00}_{-0.00}$\\
                    &3.72  &3.96  &$^{+0.84}_{-0.46}$    &$^{+0.40}_{-0.08}$\\
HE2QS~J1630$+$0435  &3.08  &1.63  &$^{+0.06}_{-0.06}$    &$^{+0.00}_{-0.00}$\\
                    &3.12  &4.14  &$^{+0.36}_{-0.27}$    &$^{+0.02}_{-0.00}$\\
                    &3.16  &5.44  &$^{+\infty}_{-0.00}$  &$^{+0.00}_{-0.00}$\\
                    &3.20  &5.51  &$^{+\infty}_{-0.00}$  &$^{+0.00}_{-0.00}$\\
                    &3.24  &5.46  &$^{+\infty}_{-0.00}$  &$^{+0.00}_{-0.00}$\\
                    &3.36  &5.28  &$^{+\infty}_{-0.00}$  &$^{+0.00}_{-0.00}$\\
                    &3.40  &5.39  &$^{+\infty}_{-0.00}$  &$^{+0.00}_{-0.00}$\\
                    &3.44  &5.35  &$^{+\infty}_{-0.00}$  &$^{+0.00}_{-0.00}$\\
                    &3.48  &4.89  &$^{+1.05}_{-0.59}$    &$^{+0.30}_{-0.00}$\\
                    &3.52  &5.32  &$^{+\infty}_{-0.00}$  &$^{+0.00}_{-0.00}$\\
                    &3.56  &5.37  &$^{+\infty}_{-0.00}$  &$^{+0.00}_{-0.00}$\\
                    &3.60  &3.86  &$^{+0.42}_{-0.31}$    &$^{+0.00}_{-0.00}$\\
                    &3.64  &5.27  &$^{+\infty}_{-0.00}$  &$^{+0.00}_{-0.00}$\\
                    &3.68  &5.30  &$^{+\infty}_{-0.00}$  &$^{+0.00}_{-0.00}$\\
                    &3.72  &5.29  &$^{+\infty}_{-0.00}$  &$^{+0.00}_{-0.00}$\\
SDSS~J1614$+$4859   &3.20  &4.86  &$^{+\infty}_{-0.00}$  &$^{+0.00}_{-0.00}$\\
                    &3.36  &4.68  &$^{+\infty}_{-0.00}$  &$^{+0.00}_{-0.00}$\\
                    &3.40  &4.73  &$^{+\infty}_{-0.00}$  &$^{+0.00}_{-0.00}$\\
                    &3.52  &4.56  &$^{+\infty}_{-0.00}$  &$^{+0.00}_{-0.00}$\\
                    &3.56  &4.58  &$^{+\infty}_{-0.00}$  &$^{+0.00}_{-0.00}$\\
                    &3.60  &4.52  &$^{+\infty}_{-0.00}$  &$^{+0.00}_{-0.00}$\\
                    &3.64  &4.47  &$^{+\infty}_{-0.00}$  &$^{+0.00}_{-0.00}$\\
                    &3.68  &4.46  &$^{+\infty}_{-0.00}$  &$^{+0.00}_{-0.00}$\\
                    &3.72  &4.38  &$^{+\infty}_{-0.00}$  &$^{+0.00}_{-0.00}$\\
SDSS~J1711$+$6052   &3.36  &5.38  &$^{+1.16}_{-0.59}$    &$^{+1.30}_{-0.07}$\\
                    &3.40  &4.43  &$^{+0.40}_{-0.28}$    &$^{+0.08}_{-0.05}$\\
                    &3.52  &5.45  &$^{+\infty}_{-0.00}$  &$^{+0.00}_{-0.00}$\\
                    &3.56  &5.45  &$^{+\infty}_{-0.00}$  &$^{+0.00}_{-0.00}$\\
                    &3.60  &5.39  &$^{+\infty}_{-0.00}$  &$^{+0.00}_{-0.00}$\\
                    &3.64  &5.34  &$^{+\infty}_{-0.00}$  &$^{+0.00}_{-0.00}$\\
                    &3.68  &5.29  &$^{+\infty}_{-0.00}$  &$^{+0.00}_{-0.00}$\\
                    &3.72  &5.19  &$^{+\infty}_{-0.00}$  &$^{+0.00}_{-0.00}$\\
                    &3.76  &5.15  &$^{+\infty}_{-0.00}$  &$^{+0.00}_{-0.00}$\\
SDSS~J1319$+$5202   &3.20  &5.40  &$^{+\infty}_{-0.00}$  &$^{+0.00}_{-0.00}$\\
                    &3.36  &5.06  &$^{+\infty}_{-0.00}$  &$^{+0.00}_{-0.00}$\\
                    &3.40  &4.85  &$^{+1.18}_{-0.65}$    &$^{+1.32}_{-0.00}$\\
                    &3.44  &2.39  &$^{+0.15}_{-0.11}$    &$^{+0.00}_{-0.01}$\\
                    &3.48  &3.97  &$^{+0.76}_{-0.40}$    &$^{+0.00}_{-0.02}$\\
                    &3.52  &4.72  &$^{+\infty}_{-0.00}$  &$^{+0.00}_{-0.00}$\\
                    &3.56  &4.65  &$^{+\infty}_{-0.00}$  &$^{+0.00}_{-0.00}$\\
                    &3.60  &4.53  &$^{+\infty}_{-0.00}$  &$^{+0.00}_{-0.00}$\\
                    &3.64  &4.46  &$^{+\infty}_{-0.00}$  &$^{+0.00}_{-0.00}$\\
                    &3.68  &3.98  &$^{+1.00}_{-0.54}$    &$^{+0.34}_{-0.00}$\\
                    &3.72  &4.28  &$^{+\infty}_{-0.00}$  &$^{+0.00}_{-0.00}$\\
                    &3.76  &3.04  &$^{+0.44}_{-0.34}$    &$^{+0.04}_{-0.00}$\\
                    &3.80  &3.59  &$^{+1.00}_{-0.54}$    &$^{+0.23}_{-0.00}$\\
                    &3.84  &3.38  &$^{+0.84}_{-0.47}$    &$^{+0.08}_{-0.00}$\\
\enddata
\tablecomments{Sensitivity lower limits on \teff\ are marked with infinite upper error.}
\end{deluxetable}

\section{Convergence of the \ion{He}{2} Effective Optical Depths in the Nyx Simulation}
\label{sect:convapp}

Here we assess the convergence of the \ion{He}{2} Ly$\alpha$ forest transmission in our hydrodynamical simulation.
Similar to the $z\gtrsim5$ \ion{H}{1} Ly$\alpha$ forest \citep{bolton09c}, the high optical depth of the \ion{He}{2}
Ly$\alpha$ forest at $z>3$ is difficult to fully resolve in simulations \citep{compostella13,compostella14}.

\begin{figure}[h!]
\includegraphics[width=\linewidth]{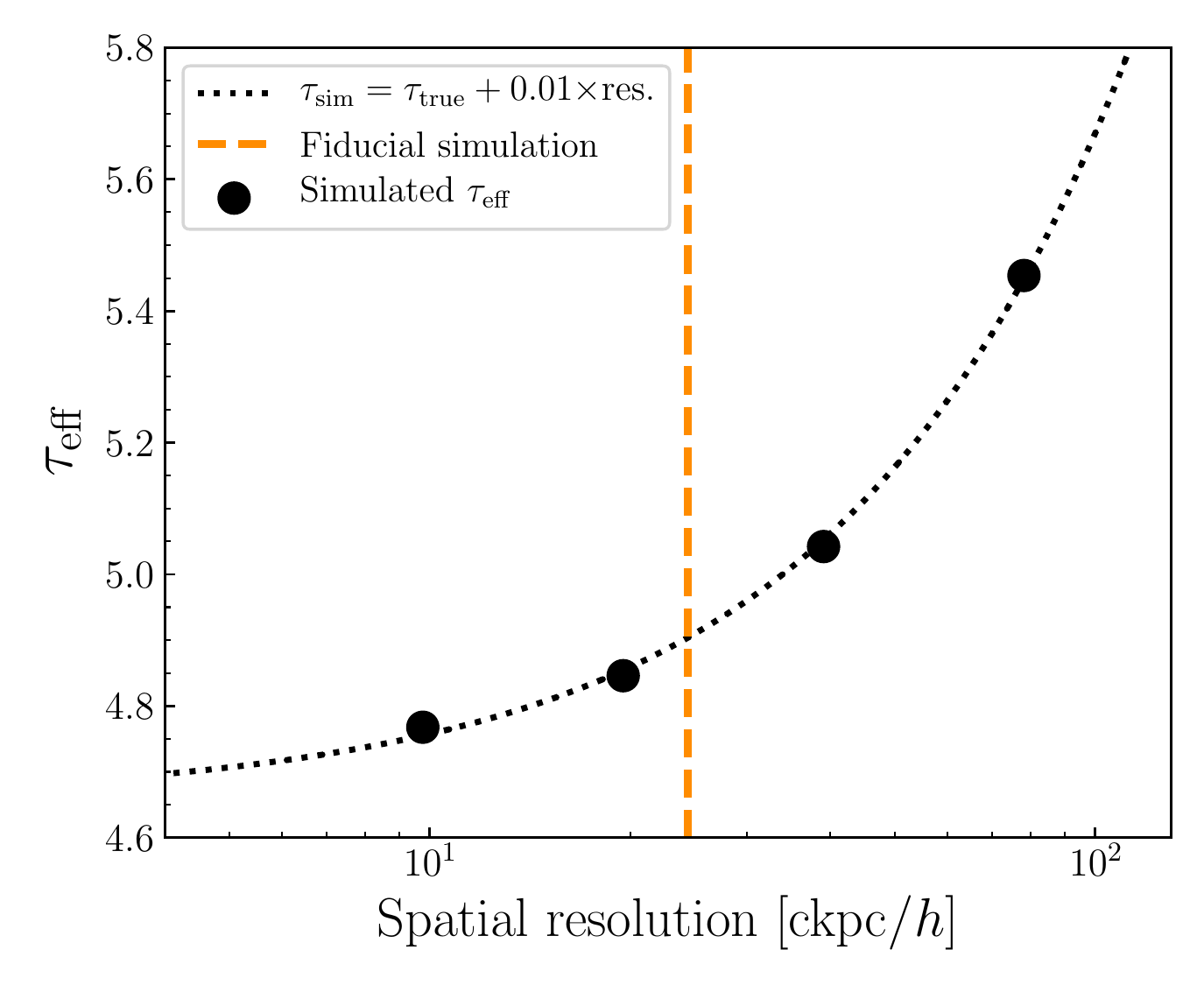}
\caption{\label{fig:converge}
\ion{He}{2} effective optical depth as a function of the spatial resolution of four ($20h^{-1}$\,cMpc)$^3$ volumes at $z=3$,
assuming $\mgheii=10^{-15}$\,s$^{-1}$. The dotted line shows a linear fit,
while the dashed line indicates the spatial resolution of our fiducial simulation.
}
\end{figure}

Our convergence test simulations consist of a suite of ($20h^{-1}$\,cMpc)$^3$ volumes at $z=3$ from \citet{lukic15} on a series
of grids: $256^3$, $512^3$, $1024^3$, and $2048^3$. The fiducial simulation used in this work
is most similar to the $1024^3$ simulation.
In Figure~\ref{fig:converge} we show the \ion{He}{2} Ly$\alpha$ effective optical depth at a fixed
$\mgheii=10^{-15}$\,s$^{-1}$ as a function of the spatial resolution in the simulation, with the resolution of our
fiducial simulation shown by the vertical dashed line. Similar to the case of \ion{H}{1} Ly$\alpha$,
we see a smooth trend of convergence towards higher resolution. We find that a linear fit to \teff\ as a function
of spatial resolution, shown by the dotted line in Figure~\ref{fig:converge}, very closely matches the trend.
Extrapolating this curve to infinite resolution (similar to \citealt{lukic15}, although our functional form differs),
we find that our fiducial simulation is unconverged at the $\sim5\%$ level in \teff. Based on the relationship
between \teff\ and \gheii, this implies that our \gheii\ measurements may be biased high by $\sim10\%$.

\end{document}